\documentclass[twocolumn]{aastex631}

\newcommand{\nb}{n_\mathrm{b}}
\newcommand{\nel}{n_\mathrm{e}}
\newcommand{\el}{\mathrm{e}}
\newcommand{\nue}{{\nu_\mathrm{e}}}
\newcommand{\nua}{{\bar{\nu}_\mathrm{e}}}
\newcommand{\nux}{{\nu_x}}

\newcommand{\nui}{{\nu_i}}

\def\lesssim{\mathrel{\hbox{\rlap{\hbox{\lower4pt\hbox{$\sim$}}}\hbox{$<$}}}}
\def\gtrsim{\mathrel{\hbox{\rlap{\hbox{\lower4pt\hbox{$\sim$}}}\hbox{$>$}}}}
\newcommand{\bea}{\begin{eqnarray}}
\newcommand{\eea}{\end{eqnarray}}

\newcommand{\msun}{{M_{\odot}}}
\newcommand{\apr}{{$\mathtt{APR}$}}
\newcommand{\sfho}{{$\mathtt{SFHo}$}}
\newcommand{\lattimer}{{$\mathtt{LS220}$}}

\newcommand{\mej}{{M_{\rm ej}}}
\newcommand{\mejn}{{M_{{\rm ej}, n}}}

\DeclareRobustCommand{\review1}[1]{ {{#1}} }

\usepackage{amsmath}

\shorttitle{GRMHD simulations of BNS mergers with weak interactions}
\shortauthors{Combi \& Siegel}
\graphicspath{{./}{./figures/}}

\begin{document}

\title{GRMHD simulations of neutron-star mergers with weak interactions: \\ r-process nucleosynthesis and electromagnetic signatures of dynamical ejecta}

\author[0000-0002-5427-1207]{Luciano Combi}
\altaffiliation{CITA National Fellow}
\affiliation{Instituto Argentino de Radioastronom\'ia (IAR, CCT La Plata, CONICET/CIC)\\
C.C.5, (1984) Villa Elisa, Buenos Aires, Argentina}
\affiliation{Perimeter Institute for Theoretical Physics, Waterloo, Ontario N2L 2Y5, Canada}
\affiliation{Department of Physics, University of Guelph, Guelph, Ontario N1G 2W1, Canada}

\author[0000-0001-6374-6465]{Daniel M.~Siegel}
\affiliation{Institute of Physics, University of Greifswald, D-17489 Greifswald, Germany}
\affiliation{Perimeter Institute for Theoretical Physics, Waterloo, Ontario N2L 2Y5, Canada}
\affiliation{Department of Physics, University of Guelph, Guelph, Ontario N1G 2W1, Canada}

\begin{abstract}

Fast neutron-rich material ejected dynamically over $\lesssim\!10$\,ms during the merger of a binary neutron-star (BNS) system can give rise to distinctive electromagnetic counterparts to the system's gravitational-wave emission that can serve as a `smoking gun' to distinguish between a BNS and a NS---black-hole merger. We present novel ab-initio modeling of the associated kilonova precursor and kilonova afterglow based on three-dimensional general-relativistic magneto-hydrodynamic (GRMHD) simulations of BNS mergers with tabulated, composition-dependent, finite-temperature equations of state (EOSs), weak interactions, and approximate neutrino transport. We analyze dynamical mass ejection from 1.35--1.35\,$M_\odot$ binaries, typical of the observed Galactic double--NS systems and consistent with inferred properties of the first observed BNS merger GW170817, using three nuclear EOSs that span the range of allowed compactness of $1.35\,M_\odot$-neutron stars. Nuclear reaction network calculations yield a robust 2nd-to-3rd-peak r-process. We find $\text{few}\times 10^{-6}M_\odot$ of fast ($v>0.6c$) ejecta that give rise to broad-band synchrotron emission on $\sim\!\text{yr}$ timescales, consistent with recent tentative evidence for excess X-ray/radio emission following GW170817. We find $\approx\!2\times10^{-5}\,M_\odot$ of free neutrons that power a kilonova precursor on $\lesssim\!\text{h}$ timescale. A boost in early UV/optical brightness by a factor of a few due to previously neglected relativistic effects, with appreciable enhancements up to $\lesssim\!10$\,h post-merger, provides promising prospects for future detection with UV/optical telescopes such as \emph{Swift} or ULTRASAT out to $\lesssim\!250\,\text{Mpc}$. We find that a recently predicted opacity boost due to highly ionized lanthanides at $\gtrsim\!70000$\,K is unlikely to affect the early kilonova lightcurve based on the obtained ejecta structures. Azimuthal inhomogeneities in dynamical ejecta composition for soft EOSs found here (``lanthanide/actinide pockets'') may have observable consequences for both early kilonova and late-time nebular emission.

\end{abstract}


\section{Introduction}
\label{sec:introduction}

The first detection of gravitational waves from the merger of a binary neutron star (BNS) system, called GW170817 \citep{abbott2017GW170817}, was associated with fireworks of electromagnetic (EM) counterparts across the electromagnetic spectrum. The largest EM follow-up campaign ever conducted \citep{abbott2017Multimessenger} revealed an associated short gamma-ray burst (GRB) GRB170817, followed by broad-band GRB afterglow emission, as well as the quasi-thermal counterpart AT2017gfo---the first unambiguous detection of a kilonova \citep{li1998Transient,metzger2010electromagnetic,barnes2013Effect,metzger2020Kilonovae}, powered by radioactive heating from the production of heavy elements via the rapid neutron-capture process (r-process; \citealt{burbidge1957Synthesis,cameron1957nuclear}) in dense, neutron-rich plasma ejected from the merger site.

The GRB afterglow of GW170817 showed several interesting and unusual properties and has been followed up for several years until today. The faintness of the prompt GRB gamma-ray emission given the proximity of only 40\,Mpc, the unusual rising X-ray flux appearing after nine days \citep{troja2017Xray}, peaking at $\approx\!160$ days after the merger, followed by a steep decay \citep{haggard2017Deep, hallinan2017Radio, d2018evolution, lyman2018Optical, margutti2017Electromagnetic, mooley2018Mildly, troja2018outflow, ruan2018Brightening}, as well as the observation of superluminal motion of the radio centroid, are interpreted as a structured jet expanding into the interstellar medium (ISM) viewed at an angle of $\theta_{\rm obs} \sim 20^{o}-30^{o}$ from the jet core \citep{lamb2017Electromagnetic, lamb2018Latetime, alexander2017Electromagnetic, hotokezaka2018Synchrotron, wu2019gw170817, fong2017Electromagnetic, ghirlanda2019Compact, hajela2019Two, lamb2019Optical,troja2020Thousand}. After several years, the broad-band synchrotron afterglow originating in the decelerating external shock driven into the ISM has faded to levels at which other possible emission components may be revealed, with first observational indications \citep{hajela2022Evidence,troja2021accurate}. One robust additional emission component expected on timescales of years after the event is the broad-band synchrotron emission from a transrelativistic shock of fast dynamical ejecta expanding into the ISM (e.g., \citealt{nakar2011Radio,hotokezaka2015Mass,hotokezaka2018Synchrotron,nedora2021Dynamical,balasubramanian2022}), which we refer to here as the `kilonova afterglow'. Late-time fall-back accretion onto a remnant black hole offers a possible alternative explanation to a rebrightening \citep{metzger2021neutrino}. The existence of a rebrightening in GW170817 is still a matter of debate \citep{troja2021accurate,connor2022continued, balasubramanian2022}.

The astrophysical sites that give rise to the synthesis of heavy elements via the r-process remain a topic of active debate (see \citealt{horowitz2019r,cowan2021Origin,siegel2022r} for recent reviews). Several lines of evidence, including measurements of radioactive isotopes in deposits on the deep seafloor (e.g.,~\citealt{wallner2015Abundance,hotokezaka2015Shortlived}), abundances of metal-poor stars formed in the smallest dwarf galaxies (e.g., \citealt{ji2016Rprocess,tsujimoto2017enrichment}), in the halo of the Milky Way \citep{macias2018stringent}, as well as in globular clusters \citep{zevin2019can,kirby2020stars}, point to a low-rate, high-yield source, much rarer than ordinary core-collapse supernovae both in early and recent Galactic history (see Fig.~3 of \citealt{siegel2019GW170817} for a compilation of observational constraints). From the most promising candidates, which include neutron-star mergers \citep{lattimer1974Blackholeneutronstar,symbalisty1982neutron}, rare types of core-collapse supernovae producing rapidly spinning, strongly magnetized proto-neutron stars (magnetorotational supernovae; \citealt{winteler2012Magnetorotationally, symbalisty1985expanding,nishimura2006r}), and hyper-accreting black holes (``collapsars''; \citealt{pruet2003nucleosynthesis, surman2006nucleosynthesis, siegel2019collapsars, siegel2021super}; see also \citealt{grichener2019common}), definitive evidence for the production of r-process elements has only been obtained in the case of neutron-star mergers through the GW170817 kilonova so far. 

\begin{figure*}[ht]
  \centering
  \includegraphics[width=2\columnwidth]{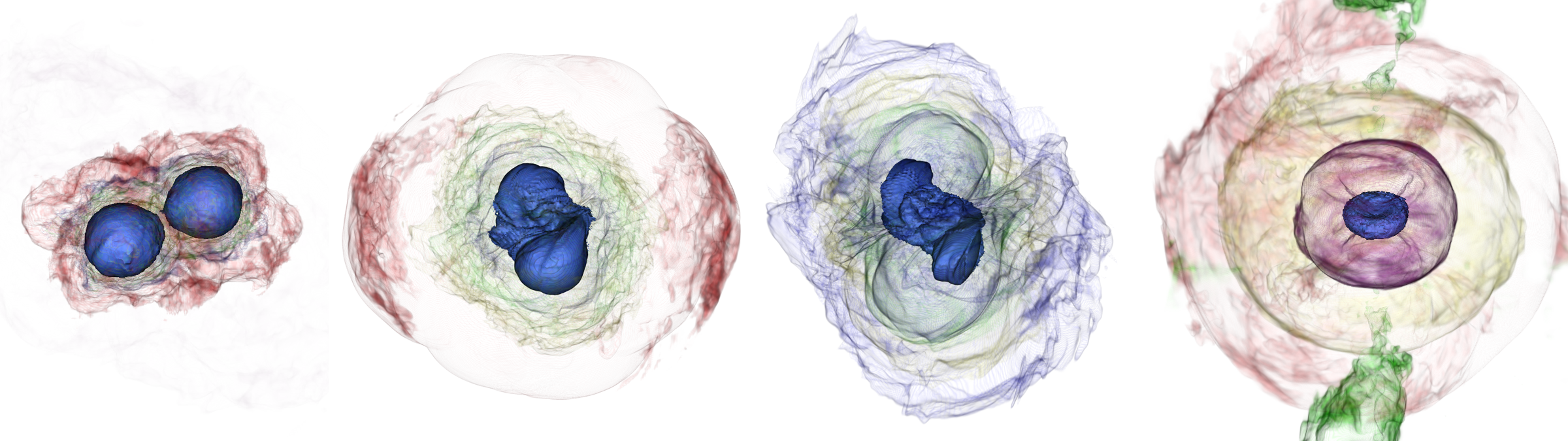}
  \caption{Volume renderings of rest-mass density layers from our APR4 BNS simulation illustrating the different merger stages and associated ejecta components. From left to right: (i) The two neutron stars reach high angular speeds before the collision and tidally deform; (ii) At merger, the neutron star cores bounce off of each other, generating shock-heated material that is ejected at high speeds, represented here as a red density layer; (iii) Due to tidal forces during merger, `tidal tails' develop that eject cold and highly neutron-rich material, represented here by a blue density layer; (iv) After merger much of the debris material circularizes around the remnant neutron star to form an accretion disk. Winds launched from this disk eject a significant fraction of the total unbound material of this BNS merger. The hot remnant neutron star additionally launches a neutrino and magnetically driven wind in the polar direction, highlighted here by a green density layer.}
  \label{fig-merger}
\end{figure*}

Several mechanisms can give rise to the ejection of neutron-rich material conducive to undergoing r-process nucleosynthesis and powering kilonovae in BNS mergers\footnote{Provided the total mass of the BNS system is below a certain threshold, typically $\approx\!1.3-1.7\,M_{\rm TOV}$, where $M_{\rm TOV}$ is the maximum mass for non-rotating NSs (e.g., \citealt{bauswein2013Prompt}). Above this threshold the merging BNS system promptly collapses into a black hole, resulting in negligible amounts of ejecta.}. In the dynamical merger phase (typically lasting $\lesssim\!10$ ms), unbound matter is ejected from the system by tidal forces (sometimes forming `tidal tails'; \citealt{rosswog1999Mass,rosswog2013Dynamic, korobkin2012Astrophysical, goriely2011RProcess}), by shocks generated at the collision interface of the two stars \citep{hotokezaka2013Mass, oechslin2007relativistic, sekiguchi2016Dynamical}, and by a sequence of quasi-radial oscillations (`bounces'; \citealt{bauswein2013Systematics, radice2018Binary}) of a double-core structure of the remnant that forms immediately upon merger (see Fig.~\ref{fig-merger}; `dynamical ejecta'). 

After merger, a remnant NS may be formed, temporarily supported from collapsing into a black hole by solid-body or differential rotation \citep{duez2006Evolution, kaplan2014Influence, siegel2014Magnetically}, which can unbind material in a combination of neutrino-driven \citep{dessart2009Neutrino,perego2014Neutrinodriven,desai2022three} and magnetically driven winds \citep{siegel2014Magnetically,ciolfi2020Magnetically,metzger2018Magnetar,mosta2020Magnetar,curtis2021Process} over the lifetime and Kelvin-Helmholtz cooling time of the remnant ($\sim10\,\text{ms}-1\,\text{s}$). Bound material surrounding the remnant circularizes and forms a neutrino-cooled accretion disk \citep{fernandez2013Delayed,just2015Comprehensive,siegel2017ThreeDimensional,lippuner2017Signatures,fujibayashi2018Mass,fernandez2019Longterm,miller2019Full,nedora2021Numerical} with typical masses of $0.01–0.3\,\msun$. A combination of heating-cooling imbalance in the disk corona in the presence of MHD turbulence \citep{siegel2017ThreeDimensional,siegel2018Threedimensional} at early times and viscously-driven outflows at later evolution stages \citep{fernandez2019Longterm}, together with nuclear binding energy release as seed particles for the r-process form, can eject up to $\sim30-40\%$ of the initial disk mass. 

The GW170817 kilonova was likely generated by a combination of $\approx\! 0.01\,M_\odot$ of dynamical ejecta and winds from the remnant NS for the early blue emission, and by $\approx\!0.05\,M_\odot$ post-merger disk ejecta for the red emission (see, e.g., \citealt{siegel2019GW170817,metzger2020Kilonovae,radice2020Dynamics, margutti2021first, pian2021mergers} for reviews overseeing the interpretation of the event). The likely dominance of disk outflows over other ejecta channels may extend to the population of merging BNS systems in general if these systems follow a distribution similar to the Galactic double-neutron star systems observable as radio pulsars \citep{siegel2022r}. The population of merging BNS systems is a matter of active debate (e.g., \citealt{landry2020nonparametric}) and future combined gravitational wave and EM observations will be necessary to infer the observed population. In a population sense, the contribution of neutron-star--black-hole (NS--BH) systems to Galactic r-process nucleosynthesis is likely by far subdominant relative to BNS mergers \citep{chen2021relative}.

Although likely subdominant with respect to other ejecta channels in many scenarios, dynamical ejecta can give rise to distinctive EM signals associated with its most energetic component. The outermost $\sim\!10^{-8}\,M_{\odot}$ of dynamical ejecta might reach ultra-relativistic velocities and serve as a breakout medium for a relativistic jet and give rise to prompt gamma-ray emission similar to GRB 170817A as envisioned by \citet{beloborodov2020relativistic}. The transrelativistic tail of ejecta with velocity $\gtrsim (0.5-0.6)c$ drives a shock into the ISM, giving rise to the aforementioned broad-band `kilonova afterglow'. Furthermore, this transrelativistic component typically has a sufficiently rapid expansion timescale ($<5$\,ms) to produce free neutrons, preventing rapid neutron capture to occur in these fastest layers. Free-neutron decay on a timescale of $\approx\!15$\,min acts with a specific heating rate roughly an order of magnitude larger than typical r-process heating at a similar epoch in the outermost layers of the ejecta, giving rise to a short-lived thermal UV/optical kilonova ``precursor'' transient on a timescale of $\lesssim\!1$\,hr after merger \citep{kulkarni2005Modeling,metzger2015Neutronpowered,ishii2018free}. Such early blue emission might overlap (and be confused) with additional heating of the outermost ejecta layers by a cocoon of hot shock-heated material surrounding a jet propagating into the ejecta envelope (e.g., \citealt{kasliwal2017Illuminating,gottlieb2018cocoon}). The aforementioned signatures, if detected early on after merger in future gravitational-wave events, will help to distinguish between a BNS and an NS--BH merger, a distinction that is presently not possible based on gravitational-wave data alone (except when exclusions can be made based on plausible ranges for inferred component masses). This motivates a detailed investigation of the physical mechanisms related to dynamical ejecta, a characterization of its properties, and self-consistent modeling of its unique EM signatures based on first-principle numerical simulations, as attempted here.

In this paper, we present a set of general-relativistic magnetohydrodynamic (GRMHD) simulations of equal-mass BNSs typical of the Galactic double-neutron star systems and consistent with the inferred source parameters of GW170817. These simulations self-consistently combine several physical ingredients from inspiral to post-merger, including magnetic fields, tabulated (finite temperature, composition-dependent) equations of state (EOS), weak interactions, and approximate neutrino transport via a ray-by-ray scheme. We employ these simulations to explore in detail the physical ejection mechanisms of dynamical ejecta, the nucleosynthesis it gives rise to, and its distinctive EM signatures including neutron precursor emission and the kilonova afterglow. We focus here on building novel, self-consistent models for the EM signatures that are directly based on ab-initio numerical simulations of BNS mergers.

The paper is organized as follows. In Sec.~\ref{sec-simusetup}, we describe our simulation setup, including numerical methods, initial data, and simulation diagnostics. In Sec.~\ref{sec:simulation_results}, we discuss simulation results on dynamical ejecta, provide a comprehensive discussion of ejection mechanisms, and conclude with a brief description of the post-merger phase. Section~\ref{sec-nucleosyn} reflects on nucleosynthesis results of the dynamical ejecta. In Sec.~\ref{sec-emsig}, we present models of the kilonova emission including the contribution of free-neutron heating (Secs.~\ref{sec:kilonova_model} and \ref{sec:kilonova_lightcurves}), of the non-thermal kilonova afterglow (Sec.~\ref{sec:afterglows}), and discuss their application in the context of GW170817 (Secs.~\ref{sec:kilonova_GW170817} and \ref{sec:kilonova_afterglows_GW170817}). We also discuss the role of exceptionally high opacities associated with high ionization states of lanthanides recently calculated by \citet{banerjee2022opacity} with respect to early kilonova and precursor emission (Sec.~\ref{sec:ionization_lanthanides}). We also compute high-energy gamma-ray emission from inverse-Compton and synchrotron self-Compton processes associated with the kilonova afterglow (Sec.~\ref{sec:gamma_rays}). Conclusions are presented in Sec.~\ref{sec-conclusions}. Finally, two appendices elaborate on numerical details regarding convergence, ejecta analysis, and the use of passive tracer particles.

\section{Simulation details}
\label{sec-simusetup}

In the following subsections, we briefly discuss the analytic foundations (Secs.~\ref{sec:GRMHDnu} and \ref{sec:leakage}), numerical methods (Sec.~\ref{sec:numerical_setup}), initial data (Sec.~\ref{sec:initial_data}) as well as simulation dignostics (Sec.~\ref{sec:simulation_diagnostics}) employed here for simulating BNS mergers. Our methods closely follow \citet{siegel2018Threedimensional} for the GRMHD part with weak interactions and \citet{radice2018Binary} for modeling neutrino transport via a one-moment closure scheme (`M0 scheme').

\subsection{GRMHD with weak interactions}
\label{sec:GRMHDnu}

We model BNS mergers within ideal GRMHD coupled to weak interactions. The equations of ideal GRMHD comprise energy-momentum conservation, baryon and lepton number conservation, Maxwell's equations in the ideal MHD regime, and Einstein's equations:
\begin{eqnarray}
\nabla_\mu T^{\mu\nu} &=& Q u^\nu,
\label{eq-tmunucon}\\
\nabla_\mu (\nb u^\mu) &=& 0,
\label{eq-baryoncon}\\
\nabla_\mu (\nel u^\mu) &=& R,
\label{eq-electrondencon} \\
\nabla_\nu F^{*\mu\nu} &=& 0,\\
\quad u_\mu F^{\mu\nu} &=& 0,
\label{eq-maxwell}
\end{eqnarray}
and
\begin{equation}
G_{\mu \nu} = 8 \pi T_{\mu \nu}.
\label{eq-einstein}
\end{equation}
Here, $G_{\mu\nu}$ is the Einstein tensor, $u^\mu$ denotes the four-velocity of the fluid, $\nb$, and $\nel$ are the baryon and electron number density, $F^{*\mu\nu}$ is the dual of the Faraday electromagnetic
tensor, and the quantities $Q$ and $R$ represent source terms due to weak interaction and neutrino transport (neutrino absorption and emission). The energy-momentum tensor is given by
\begin{equation}
  T^{\mu\nu}= \left(\rho h + b^2\right) u^\mu u^\nu + \left(p + \frac{b^2}{2}\right) g^{\mu\nu} - b^\mu b^\nu, \label{eq:Tmunu}
\end{equation}
where $\rho = \nb m_\mathrm{b}$ is the rest-mass density and $m_\mathrm{b}$ is the baryon mass, $p$ is the pressure, $h=1+\epsilon + p/\rho$ denotes the specific enthalpy, $\epsilon$ is the specific internal energy, $b^\mu\equiv (4\pi)^{-1/2}F^{*\mu\nu}u_\nu$ is the magnetic field vector in the frame comoving with the fluid,  $b^2/2 \equiv b^\mu b_\mu/2$ is the energy density of the magnetic field, and $g_{\mu\nu}$ is the space-time metric. To close the equations, we assume a three-parameter, finite-temperature, composition-dependent EOS, i.e., we assume that every dependent thermodynamic variable $\mathcal{P}$ is a function of mass density, temperature, and electron fraction,  $\mathcal{P} \equiv \mathcal{P} (\rho, T, Y_\el)$. The electron or proton fraction is defined by $Y_\el = n_{\rm p}/(n_{\rm p} + n_{\rm n})$, where $n_{\rm p}$ and $n_{\rm n}$ denote the proton and neutron number density, respectively. In accordance with the tabulated EOS employed here (Sec.~\ref{sec:initial_data}), we set the baryon mass $m_{\rm b}$ to the free neutron mass for \apr{} and \lattimer{}, and to the atomic mass unit in the case of the \sfho{} EOS (Sec.~\ref{sec:initial_data}). 

Using a 3+1 decomposition of spacetime into non-intersecting space-like hypersurfaces of constant coordinate time $t$ and timelike unit normal $n^\mu$ \citep{lichnerowicz1944espaces,arnowitt2008Dynamics} , we can write Eqs.~\eqref{eq-tmunucon}--\eqref{eq-maxwell} in conservative form as
\begin{equation}
  \partial_t(\sqrt{\gamma}\mathbf{q})
  + \partial_i[\alpha\sqrt{\gamma}\mathbf{f}^{(i)}(\mathbf{p},\mathbf{q})]
  = \alpha\sqrt{\gamma} \mathbf{s}(\mathbf{p}), \label{eq:GRMHDeqns}
\end{equation}
where $\mathbf{f}^{(i)}(\mathbf{p},\mathbf{q})$ and $\mathbf{s}(\mathbf{p})$ represent the fluxes and the source terms, respectively (see Eqs.~(18) and (19) in \citealt{siegel2018Threedimensional}), $\alpha$ denotes the lapse function, $\gamma$ is the determinant of the spatial metric $\gamma_{ij}$ on the hypersurfaces,
\begin{equation}
  \mathbf{q} \equiv [D,S_i,\tau,B^i,DY_\el] \label{eq:q}
\end{equation}
denotes the vector of conserved (evolved) variables, and 
\begin{equation}
  \mathbf{p} \equiv [\rho, v^i, \epsilon, B^i, Y_\el] \label{eq:p} 
\end{equation}
is the vector of primitive (physical) variables. The conserved vector is composed of the conserved mass density $D=\rho W$, the conserved momenta $S_i=-n_\mu T^{\mu}_{\phantom{\mu}i}$, the conserved energy $\tau = n_\mu n_\nu T^{\mu\nu}-D$, the three-vector components of the magnetic field $B^\mu = (4\pi)^{-1/2}F^{*\mu\nu}n_\nu$, as well as the conserved electron fraction $DY_\el$---as measured by the Eulerian observer, defined as the normal observer moving with four velocity $n^\mu$ perpendicular to the spatial hypersurfaces. This observer measures a Lorentz factor of the fluid of $W = -u^\mu n_\mu$, which moves with three-velocity $v_i = u_i/W$ on the hypersurfaces.

\subsection{Neutrino leakage scheme and neutrino transport}
\label{sec:leakage}

The composition of the fluid as traced by $Y_\el$, as well as its internal energy and momentum, can change due to weak interactions (i.e., due to the emission and absorption of neutrinos). In the evolution equations, the effect of these interactions is represented by the source terms $Q$ and $R$ of the right-hand side of Eqs.~\eqref{eq-tmunucon} and \eqref{eq-electrondencon}, which correspond to the net energy loss/deposition and net lepton number emission/absorption rate per unit volume, respectively. We use an energy-averaged leakage scheme to treat neutrino emission at finite optical depth as described in \citet{siegel2018Threedimensional}, closely following the methods of \citet{galeazzi2013Implementation} and \citet{radice2016Dynamical}, which are, in turn, based on \citet{ruffert1996Coalescing}.

In more detail, we track the reactions involving electron neutrinos, $\nue$, electron anti-neutrinos, $\nua$, and the heavy-lepton neutrinos $\nu_\mu$ and $\nu_\tau$ summarized as $\nux$. In the rest frame of the fluid, the net neutrino heating/cooling rate per unit volume and the net lepton emission/absorption rate per unit volume are given as a local balance of absorption and emission of free-streaming neutrinos:
\begin{equation}
  Q = \sum_\nui \kappa_\nui n_\nui E_\nui
  - \sum_\nui Q_\nui^\mathrm{eff} \label{eq:Q}
\end{equation}
and
\begin{equation}
  R = \sum_{\nui}\kappa_\nui n_\nui
  - (R_\nue^\mathrm{eff} - R_\nua^\mathrm{eff}).
\end{equation}
Here, $\nui = \{\nue,\nua,\nux\}$ and $\kappa_\nui$, $n_\nui$, and
$E_\nui$ denote the corresponding absorption opacities, number densities, and mean energies of the free-streaming neutrinos in the rest frame of the fluid, respectively. The effective emission rates $R^\mathrm{eff}_{\nu_i}$ and $Q^\mathrm{eff}_{\nu_i}$ are calculated from intrinsic (free) emission rates $R_\nui$ and $Q_\nui$ by taking into account the finite optical depth due to the surrounding medium:
\begin{equation}
  R^\mathrm{eff}_\nui = \frac{R_\nui}{1 +
    t_{\mathrm{diff},\nui}/t^{\mathrm{em,R}}_\nui}, \mskip10mu Q^\mathrm{eff}_\nui = \frac{Q_\nui}{1 +
    t_{\mathrm{diff},\nui}/t^{\mathrm{em,Q}}_\nui}.
\end{equation}

For a given neutrino species $\nui$, we define $t_{\mathrm{diff},\nui} = D_{\mathrm{diff}} \tau^2_\nui k^{-1}_\nui$ as the local diffusion timescale, where $\tau_\nui$ is the optical depth, and $D_\mathrm{diff} \equiv 6$ is the diffusion normalization factor \citep{oconnor2010New}. We define the local neutrino number and energy emission timescales as
\begin{equation}
   t^{\mathrm{em},R}_\nui = \frac{R_\nui}{n^{\mathrm{eq}}_\nui}, \mskip20mu
   t^{\mathrm{em},Q}_\nui = \frac{Q_\nui}{e_\nui}, \label{eq:tem}
\end{equation}
where $n^{\mathrm{eq}}_\nui$ is the neutrino number density in chemical equilibrium, and $e_\nui$ denotes the corresponding neutrino energy densities. The intrinsic emission rates take into account charged-current $\beta-$processes, electron-position pair annihilation, and plasmon decay (see Eqs.~(25)--(32) in \citealt{siegel2018Threedimensional}). Neutrino opacities $\kappa_\nui$ include contributions from absorption of electron and anti-electron neutrinos and from scattering on heavy nuclei $A$ and on free nucleons (see Eqs.~(33)--(39) in \citealt{siegel2018Threedimensional} and Eq.~(8) of \citealt{radice2018Binary}). We compute optical depths $\tau_\nui$ using the effective local approach of \citet{neilsen2014Magnetized}.

Finally, using a zeroth-moment (M0) scheme to approximate the general-relativistic Boltzmann transport equation \citep{radice2016Dynamical}, we evolve the number density $n_\nui$ and average energy $E_\nui$ of free-streaming neutrinos. In this scheme, only one moment of the neutrino distribution functions is used, the neutrino number currents $J^\mu_\nui$. Neglecting scattering, the first moment of the Boltzmann equation yields an equation for lepton number conservation,
\begin{equation}
    \nabla_\mu J^\mu_\nui = R_\nui^{\rm eff} - \kappa_{\nui} n_\nui, \label{eq:M0_J}
\end{equation}
which is closed by assuming that free-streaming neutrinos propagate on null radial rays, $J^\mu_\nui = n_\nui k^\mu$, with $k^\mu$ \review1{a null four-vector proportional to the radial coordinate of the grid,} and normalized such that $k^\mu u_\mu = -1$. In a stationary spacetime with time-like Killing vector $\partial_t$ and in the absence of interactions with the fluid, the mean neutrino energy $\mathcal{E}_\nui\equiv -p_\mu \partial_t^\mu$ as seen by an observer with four-velocity $\partial_t^\mu$ is conserved along the neutrino worldline (parametrized by the affine parameter $s$), $d\mathcal{E}_\nui/ds = d(-p_\mu \partial_t^\mu)/ds = 0$. Here, $p_\mu = E_\nui k_\mu$ is the average neutrino four-momentum. In the presence of heating and cooling, this becomes
\begin{equation}
    \frac{d\mathcal{E}_\nui}{ds} = \mathcal{Q}_{\nui,{\rm h}} - \mathcal{Q}_{\nui,{\rm c}}, \label{eq:M0_E}
\end{equation}
where $\mathcal{Q}_{\nui,{\rm h}}=-k_\mu \partial_t^\mu (Q^\mathrm{eff}_\nui/n_\nui)$ is the local effective average energy deposition rate per neutrino due to neutrino absorptions and $\mathcal{Q}_{\nui,{\rm c}}=\mathcal{E}_{\nui} (R^\mathrm{eff}_\nui/n_\nui)$ is the average energy loss rate per neutrino due to neutrino emission. The assumption of a stationary spacetime is approximately satisfied once the neutron stars merge and the spacetime becomes approximately stationary. We switch on this transport scheme only as the initially cold neutron stars start to merge and matter is heated up to at least several MeV to copiously produce neutrinos. The neutrino number densities $n_\nui$ and mean energies $E_\nui$ are obtained by evolving Eqs.~\eqref{eq:M0_J} and \eqref{eq:M0_E} on radial rays (radial worldlines) using a separate spherical grid that requires interpolation at each timestep onto the Cartesian grid on which the GRMHD equations are evolved.

This combined neutrino leakage and M0 transport approach include general-relativistic and Doppler kinematic effects and it is computationally far cheaper than more complex schemes such as M1 schemes in GRMHD \citep{foucart2018Monte,li2021Neutrino,radice2021New} or Monte-Carlo based radiation transport schemes \citep{miller2019Full, foucart2020monte} and thus allow for longer evolution timescales and wider parameter space studies. While quantitative differences exist between these methods, it is reassuring that quantitative comparisons between the methods \citep{radice2021New, foucart2020monte} reveal remarkable agreement for most physical observables, in particular with respect to ejecta mass, velocity, and nucleosynthesis. Deviations in individual physical quantities are typically on the $\sim\!10\%$ level (except for $\nu_x$ neutrinos, which, however, are not the focus here), within the error budget of other assumptions and approximations.
\subsection{Numerical set-up}
\label{sec:numerical_setup}

We evolve Einstein's equations coupled to the general-relativistic MHD equations in conservative form using the open-source framework of the \texttt{EinsteinToolkit}\footnote{\url{http://einsteintoolkit.org}} \citep{goodale2003Cactus,schnetter2004Evolutions,thornburg2004Fast,loffler2012Einstein, babiuc2019einstein}.
Regarding the matter part, we use an enhanced version of the public GRMHD code \texttt{GRHydro} \citep{baiotti2003New, mosta2013GRHydro}  presented in \citet{siegel2018Threedimensional} and \citet{siegel2018Recovery}. We implement a finite-volume scheme using piecewise parabolic reconstruction \citep{colella1984Piecewisea} and the approximate HLLE Rieman solver \citep{harten1987Uniformly}. The magnetic field is evolved using the FluxCT method \citep{toth2000Constraint} to maintain the solenoidal constraint during evolution. The recovery of primitive variables is carried out using the framework presented in \citet{siegel2018Recovery}, which provides support for general finite-temperature, composition-dependent, three-parameter EOS. We set a static atmosphere at a rest-mass density of $5 \%$ above the minimum density of the tabulated EOS. For most of our models, this is $\rho_{\rm atmo} \simeq 6 \times 10^2$\,g\,cm$^{-3}$, which is sufficiently small to capture the most tenuous mass outflows from the system. Indeed, at a typical extraction radius of $r \approx 450$ km, the total mass of the atmosphere enclosed at that radius is $(4/3) \pi r^3 \rho_{\rm atmo} \approx 1\times 10^{-7}~\msun$, which is an order of magnitude smaller than the total amount of material in the fast ejecta tails ($v>0.6c$; see below).

Our neutrino leakage and M0 transport scheme is based on the implementation in \texttt{WhiskyTHC} \citep{radice2012THC, radice2016Dynamical}. For the M0 transport, we use a uniform spherical grid of $2048$ radial rays extending up to $\simeq 765$ km, and a radial resolution of $\Delta r=240$ m. We activate neutrino heating when the center of mass of the stars is at $\simeq 12$ km, i.e. just after the NSs touch. 

Spacetime is evolved using the BSSNOK formulation \citep{nakamura1987General, shibata1995Evolution, baumgarte1999Evolving} as implemented in the \texttt{McLachlan} code \citep{brown2009Turduckening}. We use the $1+\log$ gauge and the $\Gamma$-driver shift condition with the damping parameter set to $\eta = 0.75$. We use a fourth-order finite-difference scheme with fifth-order Kreiss-Oliger dissipation and dissipation parameter $\epsilon=0.1$. At the time of black-hole formation, we start to track the apparent horizon with the methods of the \texttt{AHFinderDirect} code \citep{thornburg2004Fast} and we set the hydrodynamic variables to atmosphere floor values well inside the horizon. We also excise the M0 transport grid within a radius close to the apparent horizon and we keep the origin of the spherical transport grid centered on the black hole and move it with the black hole if the latter receives a `kick' after merger due to non-axisymmetric effects.

Time integration of the coupled Einstein and GRMHD equations is performed with the method of lines, using a third-order Runge-Kutta scheme \citep{gottlieb2009High}. The Courant-Friedrichs-Lewy (CFL) condition is set using a CFL factor of 0.25. 

In all simulations, we use a fixed Cartesian grid hierarchy composed of six nested Berger-Oliger refinement boxes, doubling the resolution on each level, as provided by the \texttt{Carpet} infrastructure \citep{schnetter2004Evolutions}. The finest mesh grid covers a radius of $\simeq 38$ km during the entire evolution and initially contains both NSs. We use fixed refinement boxes and overlap zones between levels to keep violations of the magnetic solenoidal constraints under control \citep{mosta2014Magnetorotational} and confined to the refinement boundaries without impacting the dynamics. The outer boundary is placed at a radius of $\simeq 1528$ km. For each EOS (see below), we perform two sets of simulations: $(i)$ a set of high-resolution simulations, setting the finest spatial resolution to $\Delta x=180$\,m and imposing reflection symmetry across the equatorial (orbital) plane for computational efficiency, and $(ii)$ a set of mid-resolution simulations with finest spatial resolution of $\Delta x=220$\,m without imposing any symmetries. This allows us to test how ejecta properties change with resolution and to analyze the influence of reflection symmetry with regard to the properties of the magnetic field.

\subsection{Models: EOS and initial data}
\label{sec:initial_data}

The properties of BNS mergers can be very sensitive to the choice of the EOS (e.g., \citealt{bauswein2013Systematics, radice2016Dynamical, radice2018Binary, hotokezaka2013Exploring, nedora2021Numerical}). The compactness of the star, as well as finite-temperature and composition-dependent effects, impact the lifetime of a merger remnant, the properties of matter outflows, the emitted gravitational waves, etc. We use three different nuclear EOSs in our set of simulations as tabulated by \citet{oconnor2010New} and \cite{schneider2017Opensource}, available at \url{stellarcollapse.org}. These tables tabulate all necessary thermodynamic quantities as a function of the independent variables $(\rho, T, Y_\el)$:
\begin{itemize}
\item LS220: We use the liquid-drop model with Skyrme interactions as implemented in \citet{schneider2017Opensource}, using a parametrization corresponding to \citet{lattimer1991Generalized}. At low densities, the EOS transitions from a single-nucleus approximation to nuclear statistical equilibrium (NSE) using 3335 nuclides \citep{schneider2017Opensource}.
\item SFHo: This EOS parametrizes nucleonic matter with a relativistic mean field model \citep{steiner2013Corecollapse}. The EOS describes nuclei and nonuniform nuclear matter using the statistical model of \citet{hempel2010Statistical}. 
\item APR: This EOS is based on the potential model of \citet{akmal1998Equation} built to fit results from nucleon-nucleon scattering and properties of light nuclei. We use the new 3D, finite-temperature implementation presented in \citet{schneider2019AkmalPandharipandeRavenhall}.
\end{itemize}

All three EOS transition to a nuclear statistical equilibrium including heavy nuclei in a thermodynamically consistent way, thus including the release of binding energy as alpha-particles and heavier nuclei form. This recombination energy of $\sim\!7$\,MeV per baryon can be an important source of energy to accelerate outflows. This effect is important, in particular, in the context of post-merger accretion disks \citep{fernandez2013Delayed,siegel2017ThreeDimensional,siegel2018Threedimensional}.

The full 3D tabulated LS220 and SFHo EOSs have been used in general-relativistic hydrodynamic (GRHD) simulations of BNS mergers \citep{sekiguchi2015Dynamical, nedora2021Numerical, radice2018Binary} and GRMHD simulations of the post-merger phase \citep{miller2019Full,li2021Neutrino,mosta2020Magnetar}, including weak interactions. APR-type EOSs have been used in both GRHD and GRMHD simulations but mostly using an ideal-gas approximation for the finite-temperature part \citep{ciolfi2017General} or without weak interactions \citep{hammond2021Thermal}. In this work, we present the first simulations of BNS mergers and their post-merger evolution with the APR EOS using the full 3D tabulated EOS and weak interactions. 

\begin{figure}[tb!]
  \centering
  \includegraphics[width=\columnwidth]{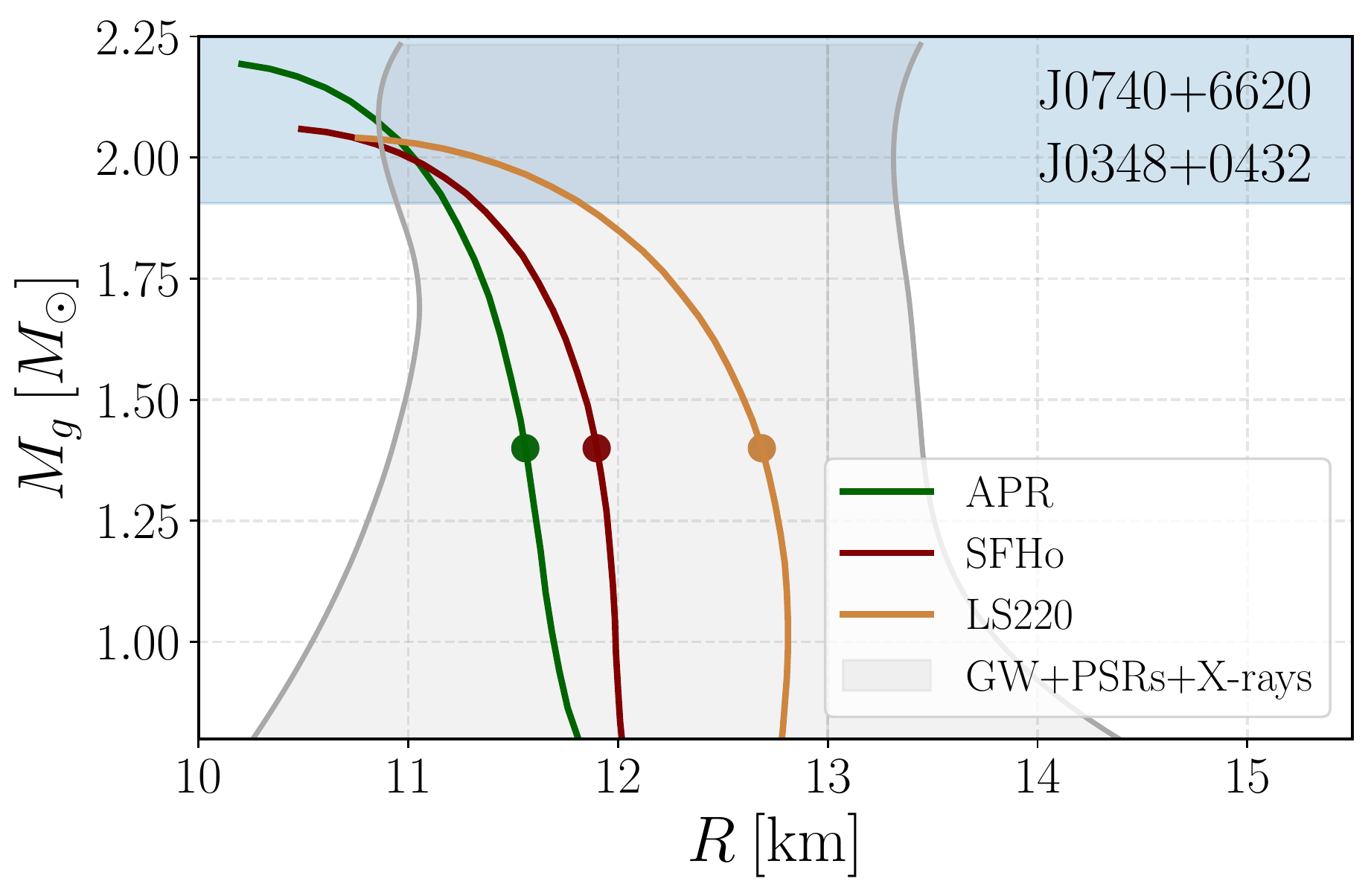}
  \caption{Gravitational mass as a function of stellar radius of non-rotating neutron stars in $\beta$-equilibrium at zero temperature for different EOSs. Dots mark models with $1.4\,M_\odot$ and corresponding radii 11.6,km, 11.9\,km, and 12.7\,km for APR, SFHo, and LS220, respectively.  The grey shaded area represents mass-radius constraints (90\% credible intervals) from current non-parametric inferences using heavy pulsar measurements, gravitational-wave observations of GW170817 and GW190425, and NICER data on J0030$+$0451 \citep{landry2020nonparametric}. The blue shaded area indicates the 68\% credible mass estimates from the heaviest pulsars measured so far \citep{cromartie2020relativistic, demorest2010two}}
  \label{fig-tov}
\end{figure}

The choice of EOSs roughly spans the current range of allowed neutron-star radii of $\approx\!11-13$\,km \citep{de2018Tidal, miller2019PSR,riley2019NICER, capano2020stringent, landry2020nonparametric, dietrich2020multimessenger} for a $1.4M_\odot$ neutron star (Fig.~\ref{fig-tov}), thus covering the range of different allowed compactness scenarios at given mass. The radii of a non-rotating, cold $1.4M_\odot$ neutron star in $\beta$-equilibrium are 11.6, km, 11.9\,km, and 12.7\,km for APR, SFHo, and LS220, respectively; the maximum masses of non-rotating neutron stars for these EOS are 2.2 $\msun$, 2.06 $\msun$ and 2.04 $\msun$, respectively. With differences in neutron-star compactness and maximum mass, these three EOSs give rise to a range of merger and post-merger phenomena. First, the `softer' APR and SFHo EOSs are expected to produce more violent collisions, owing to the faster relative velocities of the stars at merger \citep{bauswein2013Systematics}, and thus faster ejecta emanating from the collision interface than in the case of the `stiffer' LS220 EOS. Furthermore, the APR EOS leads to long-lived remnants for a typical $1.35M_\odot-1.35M_\odot$ binary \citep{ciolfi2019First, ciolfi2020Magnetically}, owing to the relatively high maximum TOV mass, while, for similar parameters, the SFHo and LS220 EOSs lead to short-lived remnants with light and heavy post-merger disks, respectively \citep{nedora2021Numerical, mosta2020Magnetar}.

We simulate a set of equal-mass binary neutron star systems on quasi-circular orbits starting at initial separations of $40$\,km. \review1{At this separation, the stars inspiral for ${\approx}3$ orbits before they merge. The relatively small separation introduces a small eccentricity to the quasi-circular orbit, affecting the GW waveforms; this, however, has a small impact on the hydrodynamics of the merger, compared to other properties, such as initial compactness of the stars (which is the focus of the present paper).}
Individual stars have ADM masses of $M_{\infty} = 1.35 \msun$ (at infinite separation), typical of galactic merging double neutron-star systems \citep{ozel2016masses}. Our selection of binary models are compatible with the inferred source parameters of GW170817, with chirp mass of $M_c = 1.188^{+0.004}_{-0.002} \msun$ and total gravitational mass of $2.74^{+0.04}_{-0.01}$ (low-spin prior) and mass ratio of 0.7--1.0 at 90\% confidence \citep{abbott2017GW170817}. Moreover, our simulations do not result in prompt collapse upon merger, which is strongly disfavored in the case of GW170817 by electromagnetic observations of the GW170817 kilonova (e.g., \citealt{margalit2017Constraining,bauswein2017Neutronstar,siegel2019GW170817}).

The initial data is built using the open-source code \texttt{LORENE} \citep{gourgoulhon2001Quasiequilibrium}\footnote{We use the secant fix suggested in \url{https://ccrgpages.rit.edu/~jfaber/BNSID/}} assuming that the stars are at zero temperature, $\beta$-equilibrated, and non-rotating. For this construction, we slice the 3D EOS table assuming the aforementioned conditions, subtracting the pressure contribution of photons \citep{radice2016Dynamical}, and generate a high-resolution table as a function of rest-mass density (i.e. an effectively barotropic table). The initial data is imported into the evolution code adapting the methods in \texttt{WhiskyTHC}\footnote{\url{https://bitbucket.org/FreeTHC}}.

After setting the hydrodynamical variables, we initialize the magnetic field as a (dynamically) weak poloidal seed field buried inside the stars. For this purpose, we specify the vector potential at each star according to
\begin{equation}
    A_{\phi} = A_b \: r^2_{\mathrm{cyl}} \left( 1 - \frac{\rho}{\rho_{\rm max}} \right)^{n_p} \mathrm{max}(p-p_{\rm cut},0)^{n_s},
\end{equation}
where $r_{\mathrm{cyl}}$ is the cylindrical radius relative to the star's center, $\rho_{\rm max}$ is the initial maximum rest-mass density, and $ \lbrace A_b, n_p, n_s, p_{\rm cut} \rbrace$ are free parameters \citep{liu2008General}. We choose $n_p = 0$ to set the maximum strength of the field at the center of the star, $n_s = 3$, and $p_{\rm cut}$ is set to $p_{\rm cut} = 0.4 p_{\rm max}$ in order to keep the field within the stellar interior and well off the stellar surface. Here, $p_{\rm max}$ is the initial maximum pressure within the star. Finally, we set $A_b$ such that the initial maximum field strength is $B_{\rm max} = 5 \times 10^{15}$\,G for all BNS models. This high value of the magnetic field strength is unlikely present in typical BNS, in which surface values of the poloidal component are expected to be closer to $B_{\rm pole} \sim 10^{12}$\,G, as observed in radio pulsars \citep{tauris2017formation}, while in our case, the strength of the magnetic field near the surface (where $p=0.4 p_{\rm max}$) is $\approx\! 3 \times 10^{14}$ G.

Our initial magnetic fields anticipate field strengths similar to those expected promptly after merger without the need to resolve the computationally extremely costly amplification process. Magnetic fields are amplified during the merger process to an equipartition level of $\gtrsim\!10^{15}-10^{16}$\,G via the Kevin-Helmholtz instability (e.g., \citealt{price2006Producing,kiuchi2014High, anderson2008magnetized,zrake2013magnetic} and the magnetorotational instability (e.g., \citealt{balbus1991Powerful,duez2006Collapse,siegel2013Magnetorotational}) on ms timescales. Specifying an initial poloidal seed field is an acceptable assumption motivated by the fact that the topology is `reset' during merger due to global dynamical effects (see below) as well as by the aforementioned small-scale amplification effects \citep{palenzuela2021Turbulent, aguilera2020turbulent}. Although large at absolute value, the initial magnetic fields are sufficiently weak to be dynamically insignificant and they do not alter the inspiral dynamics of the stars.

\review1{Finally, our initial magnetic field configuration neglects the external magnetosphere and large-scale fields likely present in the inspiral phase of a BNS system. Simulating regions of extremely high magnetic pressure in ideal MHD is challenging; various strategies have been pursued, such as imposing high-density atmospheres \citep{paschalidis2015relativistic, ruiz2016Binary} to reduce the plasma $\beta$ parameter and to maintain stability, improving the conservative-to-primitive scheme \citep{kastaun2021robust}, or using non-ideal (resistitive) MHD approximations \citep{andersson2022physics}. Our conservative-to-primitive scheme has already been significantly optimized to handle large plasma-$\beta$. However, in order to handle magnetospheric field strengths of $\sim10^{14}$\,G using the first approach, one would still need to increase the atmospheric density to levels at which it absorbs much of the high-velocity ejecta. This would thus preclude a detailed analysis of dynamical ejecta, which is one of the foci of the present paper. While the large-scale structure of the pre-merger field could have an important role in launching a short GRB after merger \citep{mosta2020Magnetar, ruiz2016Binary}, the properties of the magnetosphere in BNS are still largely unknown. For the purposes of analyzing the dynamical ejecta, we expect magnetospheric effects to be at least of second-order importance, compared to other hydrodynamical effects.}

\subsection{Simulation diagnostics and analysis}
\label{sec:simulation_diagnostics}

In the following, we briefly describe some key simulation diagnostic tools.

\subsubsection{Ejecta properties}
\label{sec:ejecta_properties}

This work focuses on dynamical ejecta from the merger process itself and the electromagnetic radiation it can give rise to. There are various proposed criteria to determine whether a fluid element becomes unbound from the merging system. Assuming a fluid element moves along a geodesic in a stationary, asymptotically flat spacetime, neglecting fluid pressure and self-gravity, a fluid element can reach spatial infinity if the specific kinetic energy at spatial infinity is  $E_\infty := -u_t + 1 > 0$, where $-u_t$ is conserved along the geodesic. The fluid then has an asymptotic escape Lorentz factor given by
\begin{equation}
 W_{\infty}= -u_t
\end{equation}
and the escape velocity is $v_{\infty} = \sqrt{1-1/W_{\infty}^2}$. This is known as the \textit{geodesic criterion}. Because conversion of thermal energy into kinetic energy by pressure gradients can occur in outflows, one guaranteed source of internal energy being the recombination of nucleons into alpha-particles and heavier nuclei, this criterion imposes a lower limit on the total unbound mass.

Assuming a stationary fluid, a stationary spacetime, and an asymptotic value of the specific enthalpy of $h_{\infty}$, an unbound fluid element fulfills the \textit{Bernoulli criterion}, if the specific energy at infinity satisfies $- hu_t > h_{\infty}$. Typically, the asymptotic enthalpy is $h_{\infty}=1$, if the EOS is independent of composition (e.g., for a polytropic EOS or ideal gas). In general, however, the asymptotic enthalpy depends on $Y_{\rm e}$. The asymptotic value of $Y_{\rm e}$ varies from fluid element to fluid element if one takes into account neutrino interactions and r-process nucleosynthesis. The former necessarily plays an important role for most ejecta elements due to the high ($\sim\!10$\,MeV) temperatures reached during the merger process, and the latter necessarily sets in as the ejected fluid element decompresses and/or moves out of nuclear statistical equilibrium (NSE) as it is being ejected from the merger site. We calculate $h_{\infty}$ taking into account the contribution of r-process heating as $h_{\infty}= 1+ \epsilon_{\infty}$, where $\epsilon_{\infty}$ is the average binding energy of the nuclei formed by the r-process (see also \citealt{foucart2021estimating,fujibayashi2020postmerger}). This number depends on how the binding energy is defined in the EOS table, and it is approximately independent of $Y_{\rm e}$ if we ignore neutrino cooling during the r-process. For our SFHo table, where the reference mass is the atomic mass unit, we have $h_{\infty} \simeq 1$, while LS220 and APR use the free neutron mass, which corresponds to $h_{\infty} \simeq 0.992$. Although this difference is small, it has a non-negligible effect on the measured kinetic energy of the ejecta. Using this criterion, the asymptotic Lorentz factor is:
\begin{equation}
    W_{\infty} = -\frac{h u_t}{h_\infty}.
\end{equation}
We note that the Bernoulli condition might overestimate the total unbound mass if the system is not stationary \citep{kastaun2015Properties}. 

We analyze the physical properties of the ejecta using two methods: discretizing the outflow on spherical detector surfaces and computing surface integrals, and by injecting passive tracer particles into the simulation domain that record plasma properties along their Lagrangian trajectories. We place different spherical surface detectors at radii $R_{\rm D} \approx \lbrace 300, 450, 600, 750, 900, 1050, 1350 \rbrace$\,km, and extract various quantities of the fluid on spherical surface grids with resolution $(n_\theta,n_\phi)=(56,96)$. The cumulative ejected mass across a detector (coordinate) sphere with radius $R_{\rm D}$ is given by
\begin{equation}
\mej = \int dt \oint_{R_{\rm D}} \alpha \sqrt{\gamma} \mathcal{W} \rho u^r dA,
\end{equation}
where $dA = R_{\rm D}^2 \sin\theta\, d\theta d\phi$, and $\mathcal{W}(\theta, \phi)$ is a function that is one (zero) if the fluid at a given grid point is unbound (bound). We also calculate mass-averaged properties of the fluid (such as $Y_\el$ and specific entropy) through the detectors as 
\begin{equation}
\langle \mathcal{P} \rangle \equiv \frac{1}{T}\sum_i dt_i \frac{1}{M_i} \sum_j dM_{i,j} \mathcal{P}(j), 
\end{equation}
where $dM_{i,j}$ is the amount of mass passing through the detector during a time interval $dt_i$ for the fluid component with property $\mathcal{P} \in [\mathcal{P}(j)-\delta,\mathcal{P}(j)+\delta]$, where $\delta$ refers to an appropriately chosen bin width; $M_i$ is the total mass crossing the detector surface in $dt_i$, and $T$ is the total time interval considered.  When using surface integration to calculate outflow properties, one ideally uses large surface radii for extraction to capture the entire unbound outflow component and to ensure that the fluid is approximately stationary. This requires, in particular, to evolve the system for sufficiently long timescales for outflows to cross the relevant detector spheres. Here, we evolve the systems for more than $30$\,ms post-merger, which is sufficient to yield approximate convergence of the unbound component at large radii using the geodesic criterion (see Appendix \ref{app:convergence_tests}).

\subsubsection{Tracers}
\label{sec:tracers}

Ejected material from BNS mergers undergoes rapid neutron-capture nucleosynthesis and produces heavy elements. We compute nucleosynthesis abundances arising from the simulation ejecta using a nuclear reaction network. Specifically, we sample thermodynamic properties such as $\rho, T, Y_\el,$ and $W$ using several families of passive tracer particles injected into the simulation domain and input the recorded Lagrangian fluid trajectory histories into the nuclear reaction network in a post-processing step (Sec.~\ref{sec:nucleosynthesis_setup}).

We place a total of $\sim\!10^5$ tracer particles into the crust of each star using a probability function proportional to rest-mass density. At the time of merger, but before ejection of material commences, we allocate $1/4$ of all tracers into a second family of tracers and resample regions of high specific entropy ($s >30 k_b$) using the same probability function, to ensure sufficient sampling of the shock-heated ejecta component that originates in the collision interface. Finally, after merger, when the system has settled into a quasi-stationary state ($\sim 20$ ms after merger), we allocate 1/4 of all tracers into a third family to adequately track the outflows from the post-merger accretion disks. These tracers are placed within a radius of $r \in (20, 100)$ km and a polar angle of $\theta \in (40,140)$ using the same probability function as before. This three-stage resampling method allows us to accurately identify, track, characterize, and distinguish between different components of outflows, which are ejected by different physical mechanisms. 

In order to compute mass distributions and mass-weighted averages of ejecta properties based on tracers, we (re)assign mass to each tracer once it becomes part of an outflow. For a set of tracers crossing a sphere of fixed radius ($R_{\rm D} = 440$\,km) at a given time $t$, we define the mass of those tracers integrating the mass-flux using the rest-mass density and velocity recorded by the tracer, $m \equiv \sum_i \alpha_i\sqrt{\gamma_i}\rho_i u_i^r (4 \pi R_{\rm D}^2) \Delta t$, where $\Delta t$ is the output frequency of the tracer data (see also \citealt{bovard2017use}). We find that this method of assigning mass very well reproduces the mass distribution of outflows as extracted by the grid-based spherical surface outflow detectors. In Appendix \ref{app:tracer_sampling}, we present a comparison between the $Y_{\rm e}$ and $v_{\infty}$ distributions as obtained by the grid-based and tracer methods, respectively.

\subsubsection{Nucleosynthesis setup}
\label{sec:nucleosynthesis_setup}

Using the full set of unbound tracers, we perform nuclear reaction network calculations with the open-source reaction network \texttt{SkyNet}\footnote{\url{https://bitbucket.org/jlippuner/skynet}} \citep{lippuner2017SkyNet} in post-processing. \texttt{SkyNet} includes 7852 isotopes from free neutrons to $^{336}$Cn, including 140000 nuclear reactions. Nuclear masses (experimental data where available, data from the finite range droplet macroscopic model (FRDM; \citealt{moller2016nuclear}) otherwise), partition functions, and forward strong reaction rates are taken from the JINA REACLIB database \citep{cyburt2010JINA}. Inverse strong reaction rates are derived from detailed balance. Where available, weak reaction rates are taken from \citet{fuller1982stellar}, \citet{oda1994rate}, \citet{langanke2000shell}, and from REACLIB otherwise. Spontaneous and neutron-induced fission rates are obtained from \citet{frankel1947calculations}, \citet{mamdouh2001fission}, \citet{panov2010neutron}, and \citet{wahl2002systematics}.

The Lagrangian trajectories of unbound tracers are followed through the grid until the end of each simulation or until they exit the grid. At that point, we smoothly transition and extrapolate their trajectories assuming homologous expansion ($\rho\propto t^{-3}$). We start evolving the composition with \texttt{SkyNet} at $t=t_{7 \rm GK}$, once the temperature drops below $7$\,GK. At this point, a fluid element represented by a tracer particle is still in NSE, and \texttt{SkyNet} initially evolves the composition in NSE accordingly until the temperature drops further to $5$\,GK. At $t=t_{5 \rm GK}$ NSE starts to break down and \texttt{SkyNet} automatically switches over to a full network evolution. As a result the $Y_{\rm e}$ of a fluid element then decouples from the $Y_{\rm e}$ evolution on the simulation grid. Reaction network calculations are performed up to $10^9$\,s, when a `stable' final abundance pattern has emerged. Using the full distribution of tracer particles instead of, e.g., feeding the outflow history recorded by spherical detectors into the nuclear reaction network (e.g., \citealt{radice2018Binary,nedora2021Numerical}) allows us to distinguish between different ejecta components and to determine the initial conditions for nucleosynthesis self-consistently for each ejecta component and fluid element of the flow.

\begin{deluxetable}{cccc}
\label{tab:sim_properties}
\tablenum{1}
\tablecaption{Properties of our suite of high-resolution BNS simulations. Here, $M_{1,2}$ denotes the gravitational mass of the neutron stars, $R_{\rm NS}$ their radius, $d_{12}$ is the initial binary separation, and $B_{\rm max}(0)$ is the initial maximum magnetic field strength; $t_{\rm BH}$ is the time of black-hole formation, $M_{\rm ej}$ is the total dynamical ejecta mass, $M_{\rm ej}(v>0.6)$ is the amount of fast moving dynamical ejecta, and $\mejn$ is the amount of ejecta mass in free neutrons. In brackets, we show several mass and time-averaged ejecta properties, such as the electron fraction ($Y_{\rm e}$), the specific entropy ($s$), the asymptotic velocity ($v_{\infty}$), the angular direction of the ejecta measured from the pole ($\theta$), and the electron fraction extracted at $T=5$\,GK by tracer particles ($Y_{\rm e}(T=5\,{\rm GK})$).}
\tablewidth{0pt}
\tablehead{
\colhead{Simulation} & \colhead{APR} &  \colhead{SFHo} & \colhead{LS220}
}
\startdata
$M_{1}$-$M_{2}$ [$\msun$] 		         &    $1.35-1.35$   & 	&    \\
$R_{\rm NS}$ [km]   	 		         &  11.6   &  11.9  &  12.7     \\
$d_{12}$ [km] 		         &    40   & 	& 	\\
$|B_{\rm max}(0)|$ [$10^{15} \:$ G]               & $5$ &  &  \\
\hline
$t_{\rm BH} - t_{\rm merger}$ [ms]               & $>50$ & $21$ &  $15$ \\
$\mej \: [10^{-3} \msun]$                       & $2.29$	   & $3.26$  &  $0.92$   \\
$\mej(v>0.6c) \: [10^{-6} \msun]$               & $4.9$	   & $3.8$   &  $0.6$  \\
$\mejn \: [10^{-5} \msun]$     & $2.03$	   & $1.57$  &  $2.44$   \\
$\langle Y_\el \rangle$						 & $0.25$	   & $0.28$  &  $0.24$   \\
$\langle s \rangle$ [$k_{\rm b}/\rm b$]      &  $13.1$      & $13.4$ &  $15$    \\
$\langle v_{\infty}\rangle$ [c]              & $0.206$  & $0.217$ &  $0.195$   \\
$\langle \theta\rangle$ [rad]                & $64.8$  & $63$ &  $64.6$ \\
$\langle Y_\el \: (T= 5 \textrm{GK}) \rangle$           & $0.17$  & $0.24$ &  $0.23$ \\
\enddata
\end{deluxetable}

\section{Simulation results}
\label{sec:simulation_results}

\subsection{Overview of simulations}
\label{sec:overview_simulations}

\begin{figure*}[tb!]
  \centering
  \includegraphics[width=2\columnwidth]{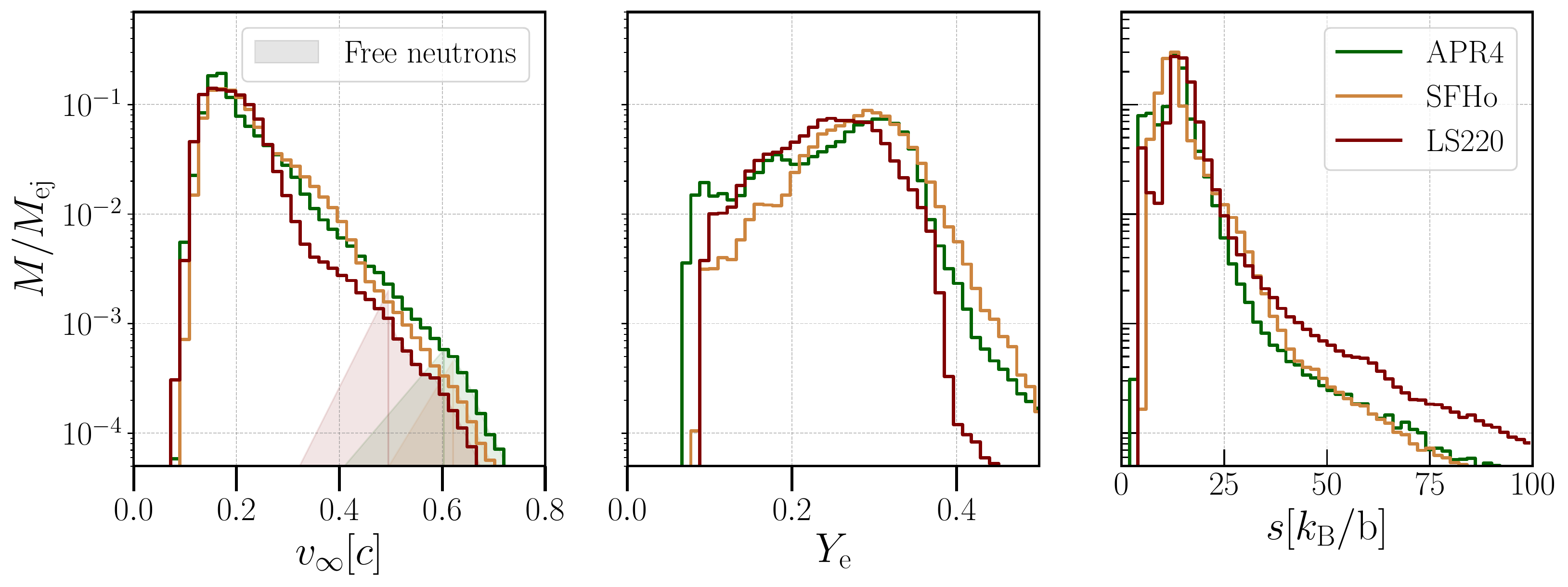}
  \caption{Key dynamical ejecta properties as measured at a radius of 440\,km according to the geodesic criterion: histograms of estimated asymptotic ejecta velocity ($v_{\infty}$), electron fraction ($Y_{e}$), and specific entropy ($s$), for each high-resolution simulation. We choose the geodesic criterion here to largely exclude secular ejecta from the remnant NS (not of interest here) and to focus on dynamical ejecta only. The high-velocity tails of the ejecta distributions that give rise to free-neutron decay and associated kilonova precursor emission (Secs.~\ref{sec:free_neutrons} and \ref{sec:kilonova_lightcurves}) are indicated as color-shaded areas.}
  \label{fig-histogram}
\end{figure*}

In each BNS simulation, the NSs inspiral for approximately three orbits, emitting GWs, before merging into a massive remnant neutron star. All remnants created in these runs are gravitationally unstable and eventually collapse into a black hole upon removal or redistribution of sufficient angular momentum due to magnetic stresses, spanning a wide range of lifetimes (cf.~Tab.~\ref{tab:sim_properties}). The remnant in the \apr{} case is a long-lived star with mass in the supramassive\footnote{Supramassive neutron stars refer to stars with mass above the maximum mass for nonrotating configurations and below the maximum mass for uniformly rotating configurations.} regime and a lifetime of likely more than a few hundred milliseconds \citep{ciolfi2019First}. In contrast, the \lattimer{} and \sfho{} binaries considered here lead to stars in the hypermassive\footnote{Configurations above the maximum mass for uniformly rotating neutron stars are referred to as hypermassive neutron stars, which can be temporarily stabilized against gravitational collapse by differential rotation.} regime with short lifetimes of $\approx\!15$\,ms and $\approx\!21$\,ms, respectively.

Our simulations self-consistently incorporate weak interactions and approximate neutrino transport, which is pivotal for accurately modeling ejecta properties, the compositional distribution represented by $Y_\el$, in particular. Furthermore, our simulations include magnetic fields, which play a key role in generating angular moment transport and outflows in the post-merger phase. In our setup, magnetic fields are initialized well inside the stars (cf.~Sec.~\ref{sec:initial_data}) and only `leak' out of the stars during the inspiral in an insignificant way. At merger, $\beta^{-1}\equiv b^2/p$ remains small, and the `buried' fields do not influence the ejection of dynamical ejecta. In this early stage of the merger process, our results resemble closely purely hydrodynamic simulations that include weak interactions and approximate neutrino absorption, but neglect magnetic fields (e.g., \citealt{sekiguchi2016Dynamical, radice2018Binary}).

In this paper, we focus on ejection mechanisms, ejecta properties, and observables of material ejected during the dynamical phase of the merger itself. We consider dynamical ejecta only, defined as material that is unbound by global dynamical processes. Table \ref{tab:sim_properties} provides an overview of the mass-averaged properties of the dynamical ejecta. Corresponding distributions of ejecta mass relevant for observables according to composition ($Y_\el$), asymptotic escape speed ($v_\infty$), and specific entropy ($s_\infty$) are summarized in Fig.~\ref{fig-histogram}. We extract physical quantities at a radius of $R= 300\,M \simeq 440$\,km, where $M$ is the total binary mass, using the geodesic criterion. We mainly focus on the geodesic criterion here, since at close separations of 440\,km it is somewhat insensitive to secular outflows such as neutrino-driven winds from the merger remnant (not of interest for the present study) and it thus acts as a filter for dynamical ejecta. The merger process during which dynamical ejecta is generated according to the geodesic criterion lasts approximately $10$\,ms in all our simulations (Sec.~\ref{sec:ejecta_dynamics} and Fig.~\ref{fig-unbmass}). We turn to a discussion of the details of mass ejection in the following subsections.

\begin{figure*}[tb!]
  \centering
  \includegraphics[width=2\columnwidth]{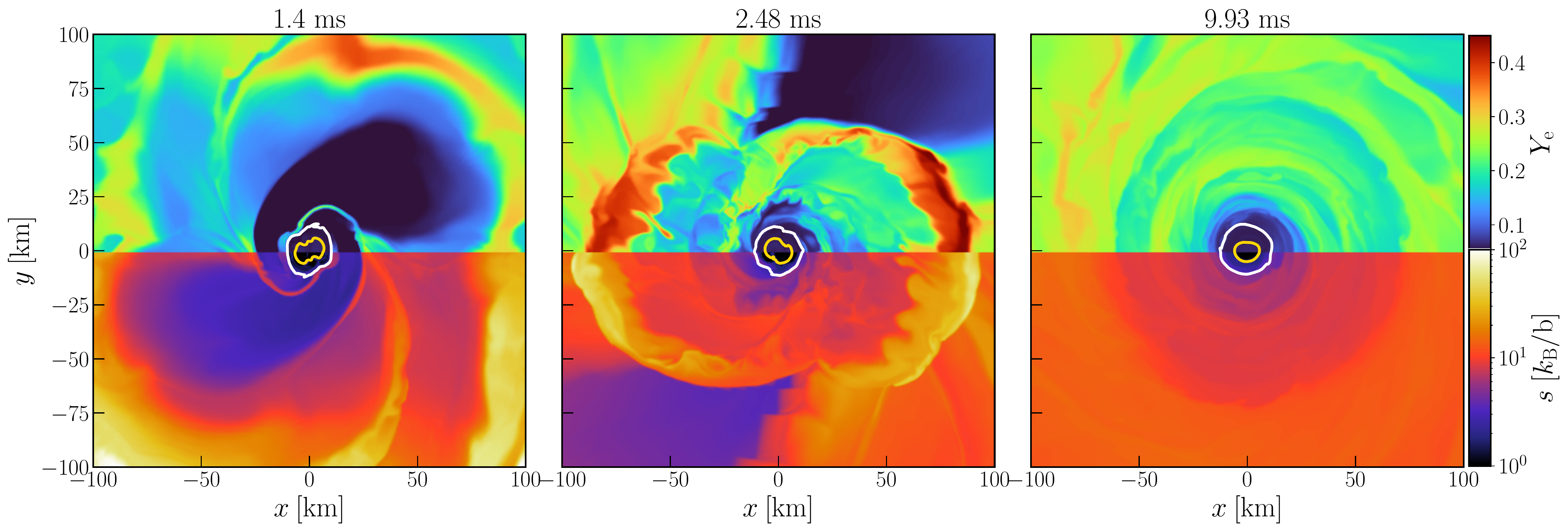}
  \caption{Evolution of the electron fraction (upper half-plane) and specific entropy (lower half-plane) in the equatorial plane during merger for the \sfho{} model. White and gold lines indicate rest-mass density contours at $\rho = 10^{13.5}, 10^{14.8}\, \rm g\,cm^{-3}$, respectively. 
  After the first bounce (Sec.~\ref{sec:ejecta_dynamics}, Fig.~\ref{fig-unbmass}), two neutron-rich tidal tails form (first panel). After the second quasi-radial bounce, a shock with associated heating and neutrino emission reprocesses part of the tidal tail to a higher electron fraction in a preferential direction defined by the phase of rotation (second panel). Finally, the remnant's oscillations are heavily damped, the ambient flow circularizes, and an accretion disk with $Y_e\approx 0.15-0.25$ and $s\approx 8-10\,k_B$ per baryon forms (third panel).}
  \label{fig-tri-yes}
\end{figure*}

\subsection{Ejecta dynamics and fast outflow}
\label{sec:ejecta_dynamics}

Two types of ejecta can be distinguished at merger: tidal and shock-heated ejecta. Tidal torques extract material from the surface of the stars during the final inspiral and merger process, creating spiral arms that expand into the orbital plane as they transport angular momentum outwards, expelling cold, neutron-rich material ($Y_\el \lesssim 0.1$) into the interstellar medium (see Fig.~\ref{fig-tri-yes}, first panel).
Because neutron stars are more compact in general relativity compared to Newtonian gravity, these tidal tails are not as prominent here as in Newtonian simulations (e.g., \citealt{rosswog1999Mass,korobkin2012Astrophysical,rosswog2013Dynamic}). Furthermore, for equal-mass binaries one expects a minimum of tidal ejecta: For a given EOS, tuning the binary mass ratio away from unity generally enhances the tidal torque on the lighter companion. This leads to increased tidal ejecta while reducing the shock-heated component originating in the collision interface \citep{hotokezaka2013Mass,bauswein2013Systematics,dietrich2015Numerical,lehner2016Unequal,sekiguchi2016Dynamical}. This is because the less massive companion becomes tidally elongated and seeks to `avoid' a (radial) collision. Finally, for a given binary mass ratio, changing the EOS from stiff (large NS radii) to soft (small NS radii) one expects the shock-heated component to be enhanced while reducing the tidal component \citep{hotokezaka2013Mass,bauswein2013Systematics,dietrich2015Numerical,lehner2016Unequal,sekiguchi2016Dynamical,palenzuela2015Effects}. This is because tidal forces are smaller for less extended objects, and NSs with smaller radii approach closer before merger, reaching higher orbital velocities at the collision, thus enhancing the shock power and associated ejecta mass.

\begin{figure}[tb!]
  \centering
  \includegraphics[width=1\columnwidth]{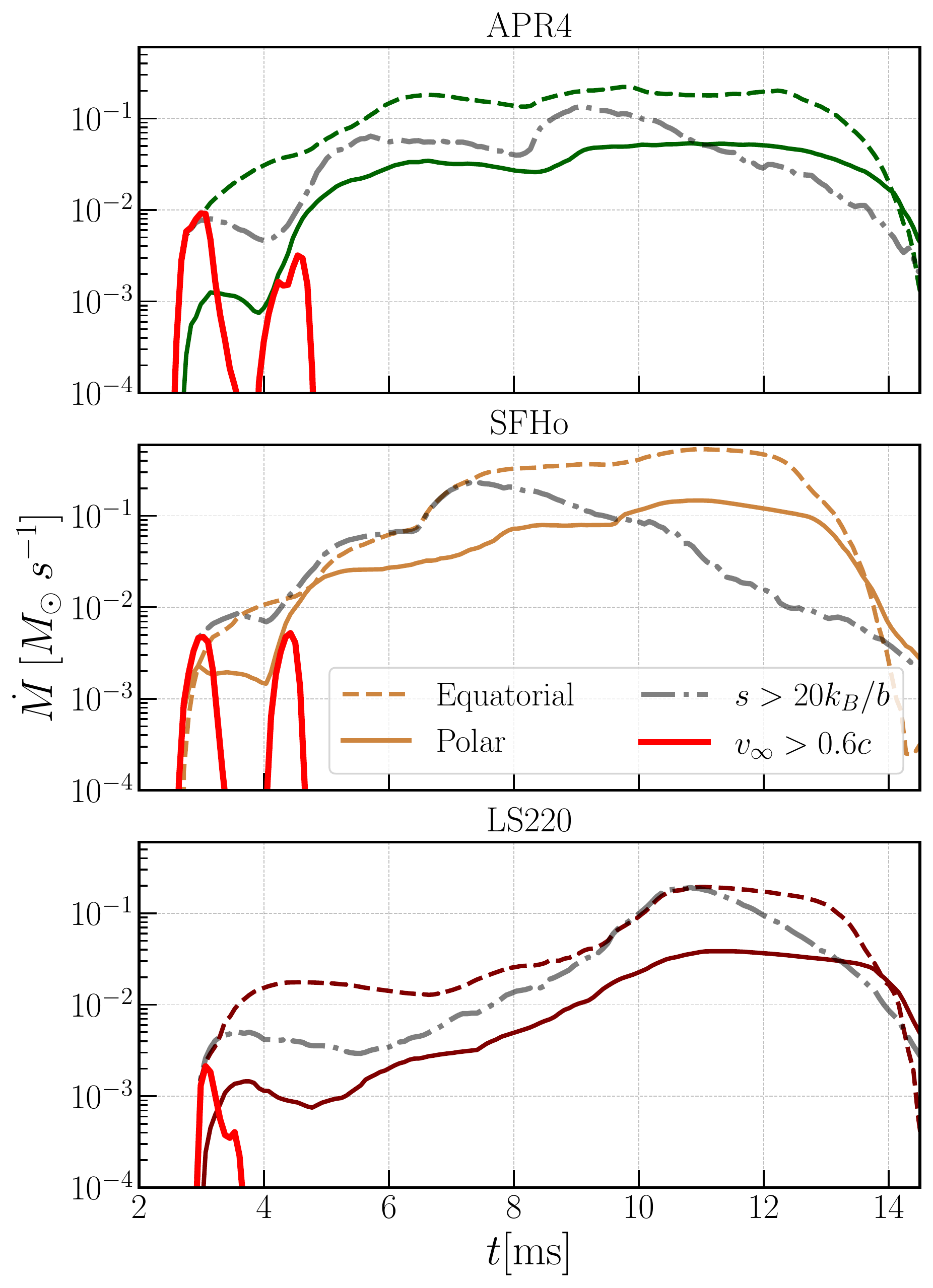}
  \caption{Total unbound mass flux according to the geodesic criterion through a coordinate sphere of radius $440$ km for the \apr{} (upper panel), \sfho{} (middle panel), and \lattimer{} (lower panel) models as a function of time after the onset of the merger. We distinguish between unbound ejecta with high specific entropy ($s>20 \: k_{\rm B}$ baryon $^{-1}$), fast velocities (asymptotic speed $v_{\infty}>0.6c$), as well as between polar (within $45^{\circ}$ from the polar axis) and equatorial (within $45^{\circ}$ from the equatorial plane) ejecta. We choose the geodesic criterion here to largely exclude secular wind ejecta from the remnant NS (not of interest here) and focus on dynamical ejecta only. The total amount of dynamical ejecta converges around 14 ms at this distance (see also Fig.~\ref{fig-conv-res}) and the mass flux drops.}
  \label{fig-massflux}
\end{figure}

With our NSs spanning the compactness range of currently allowed EOSs for typical galactic double neutron star masses, we find our runs span a range of dynamical mass ejection phenomena. A detailed analysis shows (see below) that for all systems considered here, by far most of the ejecta is expelled by shock waves produced in quasi-radial bounces of an oscillating double-core remnant structure that forms after the onset of the merger, with only a negligible amount of material being ejected by tidal tails (see Fig.~\ref{fig-massflux} and below; Sec.~\ref{sec:Ye_shock_reprocessing}). This ejecta material is, in general, faster and more proton-rich than tidal ejecta. A large fraction of the material released in such waves has been heated considerably due to hydrodynamical shocks at the collision interface during the merger process and is further heated as it shocks into slower surrounding merger debris. Associated neutrino emission in such a neutron-rich environment favors electron captures onto neutrons over positron captures on protons, raising the electron fraction and causing the electron fraction to widen from the initially cold and neutron-rich conditions of the individual NSs ($Y_\el\lesssim\!0.1$) into broad distributions with prominent high-$Y_\el$ tails (Fig.~\ref{fig-histogram}). The stronger the heating, as indicated by the specific entropy distributions (Fig.~\ref{fig-histogram}), the higher the resulting final $Y_\el$ of the ejecta. Neutrino reabsorption, which also contributes to increasing $Y_\el$ in a neutron-rich environment, only plays a minor role in setting $Y_\el$ at this early dynamical stage. Overall, the more compact NS binaries (\apr{} and \sfho{}) eject a factor of $\simeq2-3$ more ejecta than the less compact \lattimer{} binary (cf.~Tab.~\ref{tab:sim_properties}), since tidal ejecta is subdominant and post-merger bounces are more energetic and lead to more violent shocks unbinding more material for more compact NSs.

\begin{figure}[tb!]
  \centering
  \includegraphics[width=1\columnwidth]{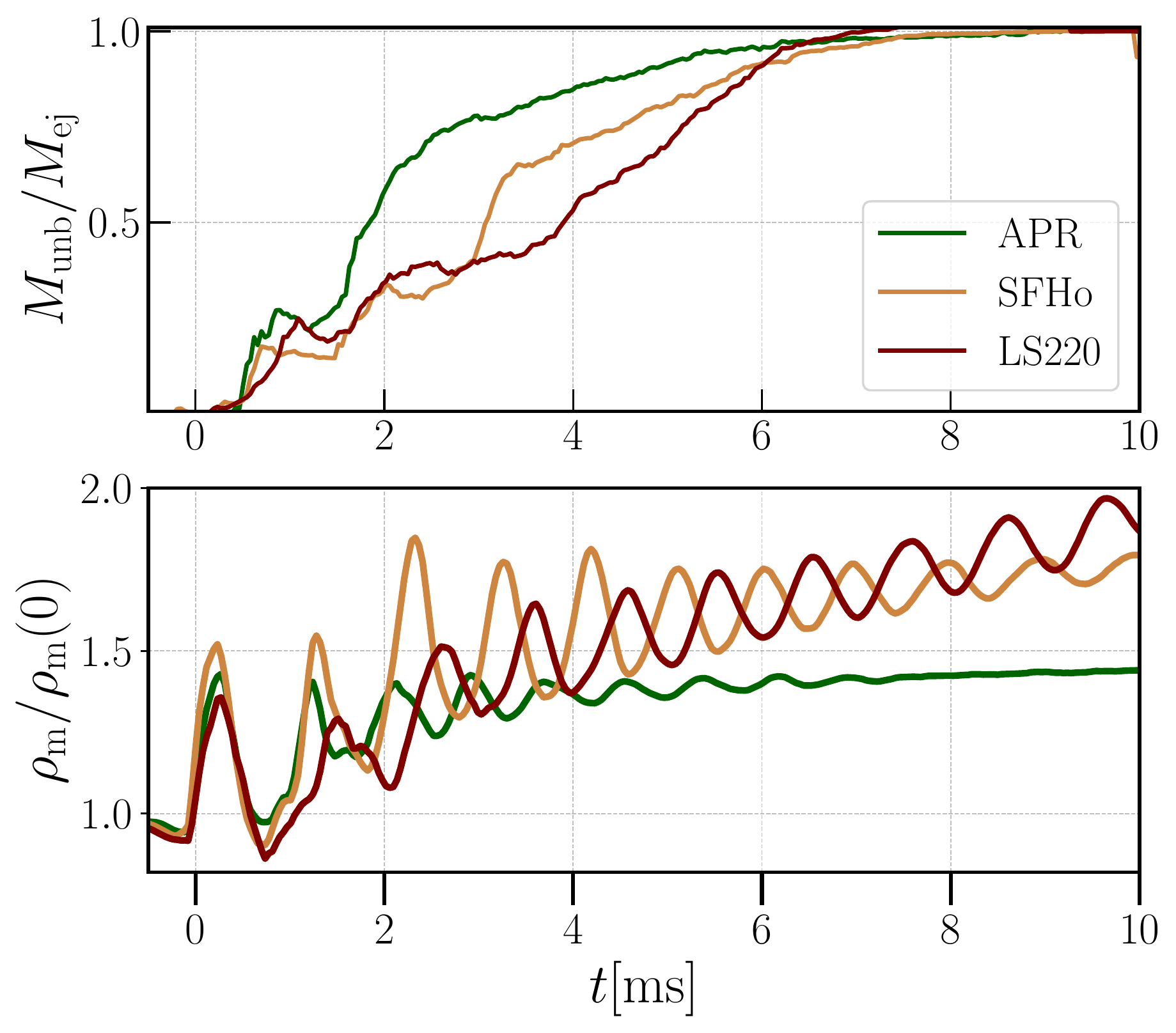}
  \caption{Upper panel: Fraction of unbound ejected mass sampled by passive tracer particles according to the Bernoulli criterion as a function of time since the onset of the merger. Lower panel: evolution of maximum density of the merger remnant relative to its initial value at merger, showing perturbations that result from radial oscillations of a double-core structure within the remnant.}
  \label{fig-unbmass}
\end{figure}

Figure \ref{fig-unbmass} summarizes a more detailed analysis of the matter ejection mechanism post-merger, correlating the generation of unbound material with radial oscillations of the remnant. Unbound mass, as sampled by passive tracers according to the Bernoulli criterion, initially increases in `steps' when the maximum density reaches a minimum during oscillations.

At this point of maximum decompression and least compactness, some amount of material becomes unbound and is ejected, the remnant undergoes a bounce, and the maximum density starts to increase again. In the softest EOS simulations, nearly $75\%$ of the outflow becomes unbound after the first three bounces, followed by a secular growth that lasts up to 10\,ms after merger, reflecting the integrated action of several weaker bounces. In the stiffer \lattimer{} case, only two individual mass ejection episodes can be discerned by this tracer method ($\simeq 30 \%$ of the total ejected mass). 

The Bernoulli criterion used here is valid for stationary fluid flows only and might overestimate the total ejecta mass near shocks (see \citealt{kastaun2015Properties} for a discussion). However, confidence in the present interpretation comes from the fact that all ejecta curves are monotonically increasing over time, together with the fact that the amount of unbound material for both the geodesic and the Bernoulli criteria at large distances agree up to saturation of the dynamical ejecta as measured by the geodesic criterion (cf.~Fig.~\ref{fig-conv-res}). The importance of post-merger bounces for mass ejection has been noted previously in smoothed-particle hydrodynamics (SPH) simulations \citep{bauswein2013Systematics} and more recently in grid-based GRHD simulations \citep{radice2018Binary,nedora2021Dynamical}.

The three (two) individual mass ejection episodes discussed above for \apr{} and \sfho{} (\lattimer{}) drive shock waves into the surrounding medium (see Fig.~\ref{fig-tri-yes}) and  manifest themselves in step-wise increases of high specific entropy mass flux ($s>20k_{\rm B}$\,baryon$^{-1}$) at large distances (see Fig.~\ref{fig-massflux}).

Figure~\ref{fig-massflux} also indicates that the first high-entropy mass flux events for \lattimer{} and \apr{} are associated with predominantly equatorial outflows (within $45^\circ$ from the plane), while the first event for \sfho{} also contains a significant fraction of high-entropy material in the polar direction (within $45^\circ$ from the polar axis). In general, mass ejection in equatorial directions dominates over polar directions, as shown by Figs.~\ref{fig-massflux} and \ref{fig-angularYe}. However, the relative ratio of equatorial to polar ejecta is larger for the softer EOSs (\sfho{} and \apr{}), owing to stronger bounces and associated shock waves, cf.~Figs.~\ref{fig-massflux} and \ref{fig-angularYe}).

Figure~\ref{fig-massflux} also illustrates that the first two (one) mass ejection episodes discussed above are (is) associated with the ejection of fast material with asymptotic speed $v_\infty>0.6 c$. We find a fast-moving ejecta component with $v_\infty>0.6c$ of mass $\approx\!(4-5)\times 10^{-6}\,\msun$ for the \apr{} and \sfho{} runs (cf.~Tab.~\ref{tab:sim_properties}), which is more than an order of magnitude larger than the corresponding swept-up atmosphere mass and is thus well captured; for the stiff \lattimer{} case, we also find a fast component of mass $\approx \!6 \times 10^{-7}\,\msun$, which is still larger than the swept-up atmosphere mass by a factor $\gtrsim 7$. Measuring the amount of fast ejecta at a closer distance of $R_{D}=300$ km, we find that the total mass of this fast tail is larger by about $5 \%$ for both \apr{} and \sfho{} and by approximately $30 \%$ for \lattimer{}. The extracted values for the maximum velocity remain essentially insensitive to the detector radius (see also Fig.~\ref{fig-conv-res} for a convergence test).

This fast outflow component is ejected by the first two bounces after merger, consistent with other grid-based GRHD simulations \citep{nedora2021Dynamical}. For \sfho{}, we observe in Fig.~\ref{fig-massflux} that these fast outflows are ejected in two impulsive trains of similar strength, while for the even softer \apr{} EOS, the first ejection event is much stronger. For the stiffer \lattimer{} EOS, there is only a single fast ejection event because the bounces are generally weaker.

\begin{figure}[tb!]
  \centering
  \includegraphics[width=1\columnwidth]{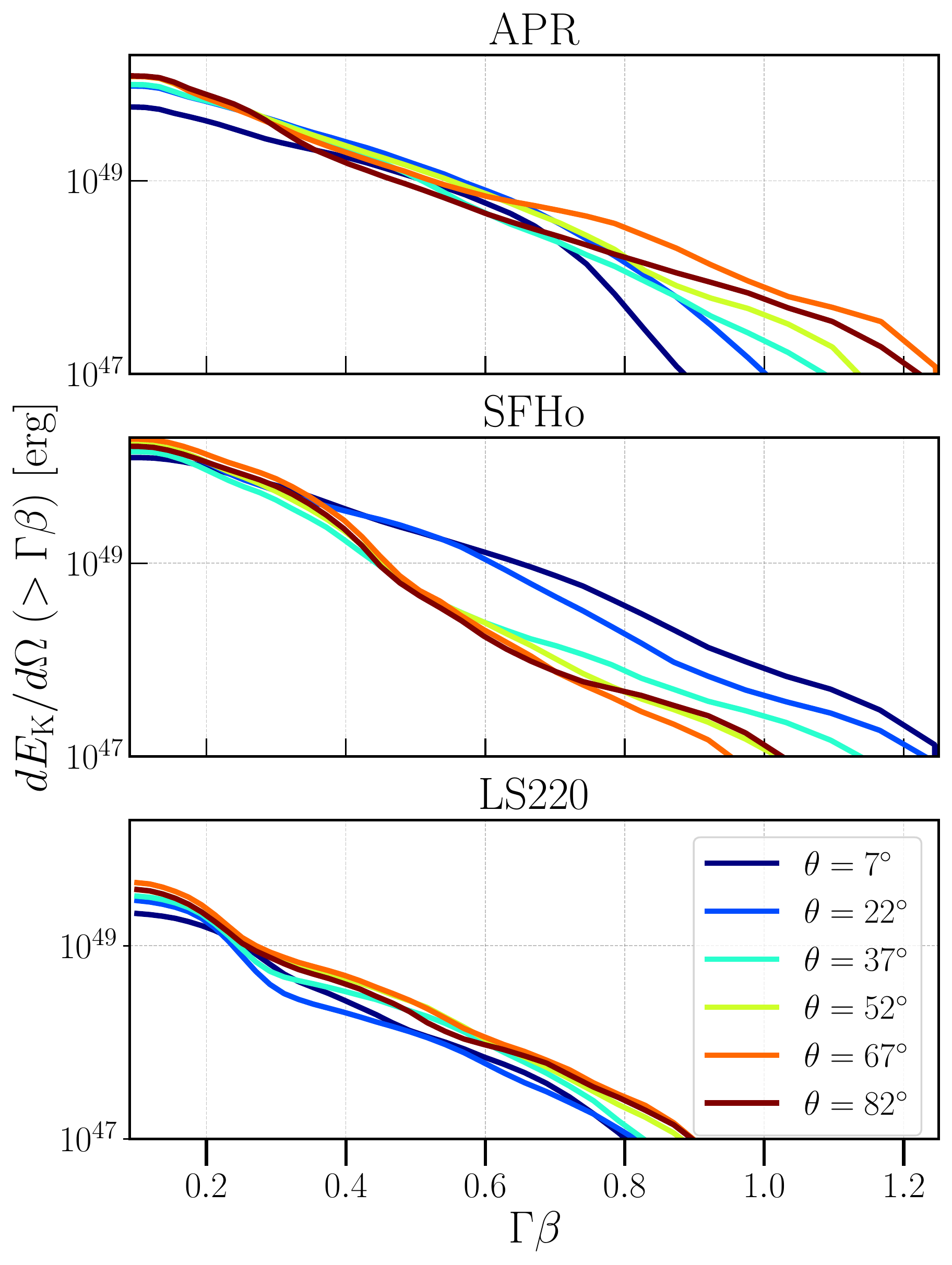}
  \caption{Isotropic equivalent cumulative kinetic energy distribution of the ejecta as a function of ejecta velocity $\beta$ in units of the speed of light times Lorentz factor $\Gamma$ for six different polar angle bins equidistant in $\cos\theta$ and different EOS.}
  \label{fig-kin-ang}
\end{figure}

The angular distribution of cumulative kinetic energy $E_{\rm K} = (\Gamma-1) M_{\rm ej}$ of the ejecta as a function of $\beta\Gamma$ is shown in Fig.~\ref{fig-kin-ang}. Here, $\beta$ denotes the ejecta speed in units of the speed of light and $\Gamma$ the corresponding Lorentz factor. We observe that the most energetic outflows for \sfho{} are ejected at latitudes close to the pole (polar angles $\approx\!0 ^\circ-35^\circ$), consistent with previous findings \citep{radice2018Binary}. For the softest EOS \apr{}, the energetics of the ejecta are dominated by near-equatorial outflows (polar angles $\approx\!60^\circ-90^\circ$). For the stiffer EOS \lattimer{}, we obtain milder energetics (mostly $\Gamma \beta <1$), and a fairly homogenous angular distribution, with more energetic outflows near the equatorial plane.

The total amount of fast ejecta measured in simulations is sensitive to different factors such as (a) grid resolution or sampling of high-velocity ejecta tails with a small number of particles in SPH simulations, (b) the grid topology, (c) atmosphere levels and magnetosphere treatment, and (d) definition of asymptotic velocity. In our present study (a) is fixed by constraints arising from computational cost. High-resolution Newtonian convergence studies \citep{dean2021Resolving} suggest that at our resolution of 180\,m convergence of the high-velocity ejecta ($v>0.6c$) may, in principle, be achieved within measurement uncertainties of a factor of $\sim\!2$. Aspect (b) is fixed by the Cartesian grid paradigm our code is based on. We elaborate on aspects (c) and (d) below.

In this work, we use lower floor/atmosphere levels than previously reported BNS simulations (e.g., smaller by a factor of 10 compared to \citealt{radice2018Binary}) to minimize the effects of an artificial atmosphere on the properties of the fast ejecta (at relevant extraction radii of $\approx\!400$ km).\footnote{We note that our simulation set-up does not use refluxing through mesh refinement boundaries; we are thus not ensuring mass conservation at round-off level for the outflows. However, the other mentioned effects are likely more important than this one \citep{reisswig2013ThreeDimensional}.}

The asymptotic velocity $v_\infty$ of unbound outflows are computed using the Lorentz factor at infinity $W_{\infty} = -u_t$, where $W_{\infty} = (1-v^2_{\infty})^{-1/2}$. Using the Bernoulli criterion instead, according to which $W_{\infty} = -h u_t$, we find a total amount of fast ejecta $15 \%$ larger, which is expected as thermal effects can still accelerate the ejecta at this detector distance. There are other techniques used in the literature to extract outflows, e.g. volume integrations \citep{kastaun2013Black, ciolfi2017General, hotokezaka2013Mass}, and other definitions of asymptotic velocity that explicitly include the gravitational potential \citep{shibata2019Merger}. \citet{nedora2021Dynamical, nedora2021Numerical} compute asymptotic velocities using the Newtonian expression $v_{\rm N, \infty}=\sqrt{2 E_{\infty}}$, where $E_{\infty}$ is the specific energy at infinity. We find this expression overestimates the speed and amount of fast ejecta: using the latter expression, we find a larger amount (by almost an order of magnitude) of fast ejecta in our simulations, comparable to that of \citet{nedora2021Numerical}. 

Previous analyses of fast-ejecta with GRHD simulations using neutrino transport \citep{radice2018Binary,nedora2021Numerical}
and Newtonian simulations at ultra-high resolution in axisymmetry \citep{dean2021Resolving} report fast-ejecta masses within factors of $5-10$ higher than found here. SPH simulations find even higher masses than grid-based simulations, of the order of $10^{-4}-10^{-5} \msun$ \citep{metzger2015Neutronpowered,kullmann2021Dynamical,rosswog2022thinking}. 
Our default mass estimates are similar to the ones obtained in the grid-based ultra-high resolution simulations (purely hydrodynamic, without weak interactions) of \citet{kiuchi2017Subradianaccuracy}, specifically, their HB EOS case, which leads to a similar NS compactness to \apr{} (see Figure 7 in \cite{hotokezaka2018Synchrotron}). Differences to other grid-based merger simulations \citep{nedora2021Numerical} can be understood in terms of the definition of asymptotic velocity (see above), while the higher level of ejecta in grid-based convergence studies \citep{dean2021Resolving} may be attributable to differences in the overall setup (such as, e.g., Newtonian vs. general-relativistic dynamics and geometric effects due to assumed symmetries), since ejecta masses are expected to be roughly converged according to these studies at our present resolution (at least up to a factor of a few). Fast ejecta tails are resolved by $\lesssim\!\text{tens}$ of SPH particles in present SPH simulations \citep{kullmann2021Dynamical, rosswog2022thinking}, and details about the dependence of fast ejecta masses with resolution (number of particles) have been reported only recently \citep{rosswog2022thinking}. \citet{rosswog2022thinking} show that the mass of fast ejecta decreases by a factor of $\approx\!10$ when the resolution is increased from $10^6$ to $5\times 10^6$ SPH particles used in the simulation. The total fast ejecta mass in the latter paper---estimated using the relativistic asymptotic velocity similar to our work---is still a factor of $\approx\!2$ higher than the value we find for a BNS merger with a similar set-up and using the APR3 EOS, but much closer than previous SPH simulations.

Introducing a dedicated tracer family to track the shock-heated fast ejecta (cf.~Sec.~\ref{sec:tracers}) allows us to sample the fast ejecta tail with velocities $v>0.6c$ by typically $\gtrsim\!500$ tracer particles. Appendix \ref{app:tracer_sampling} provides more details on tracer sampling of the fast tail of the ejecta distribution in our simulations.

\subsection{Electron fraction distribution and shock reprocessing}
\label{sec:Ye_shock_reprocessing}

\begin{figure}[tb!]
  \centering
  \includegraphics[width=1\columnwidth]{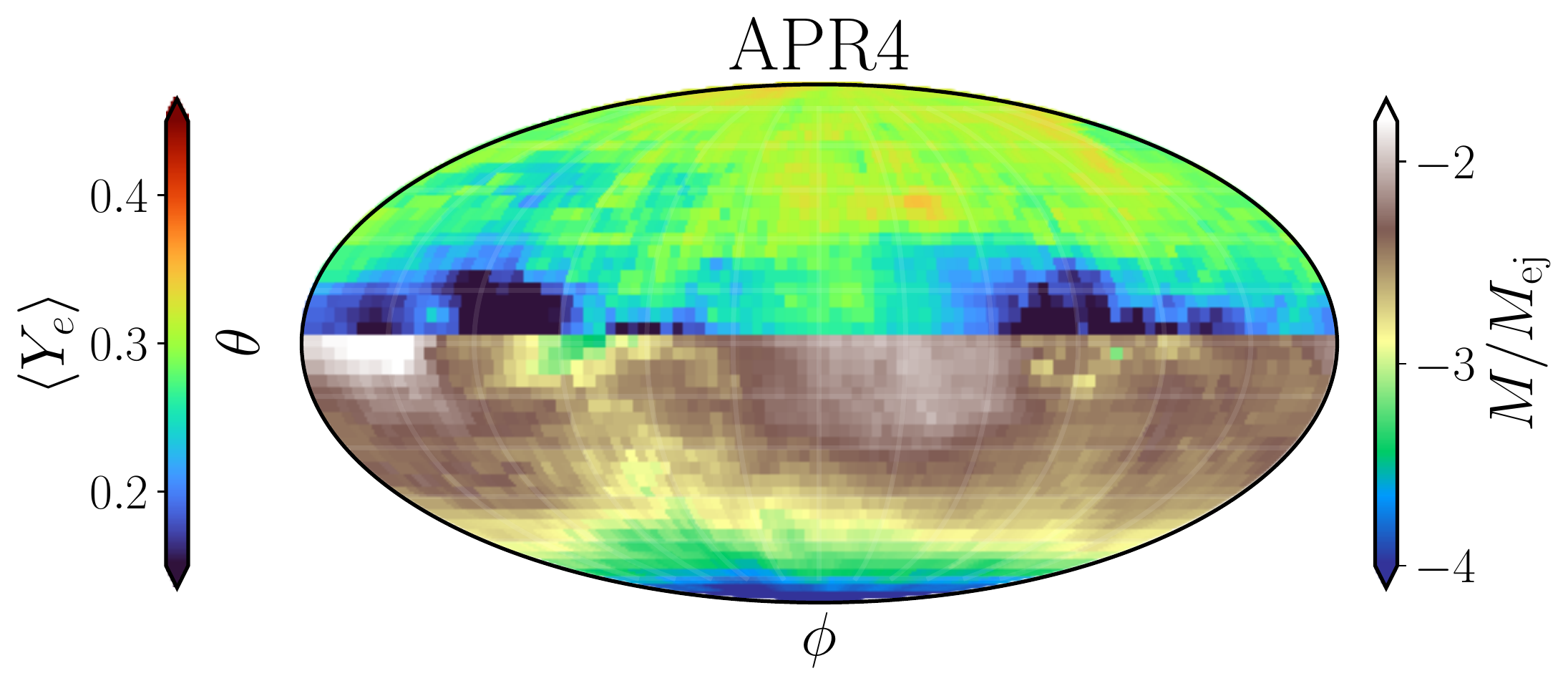}
  \includegraphics[width=1\columnwidth]{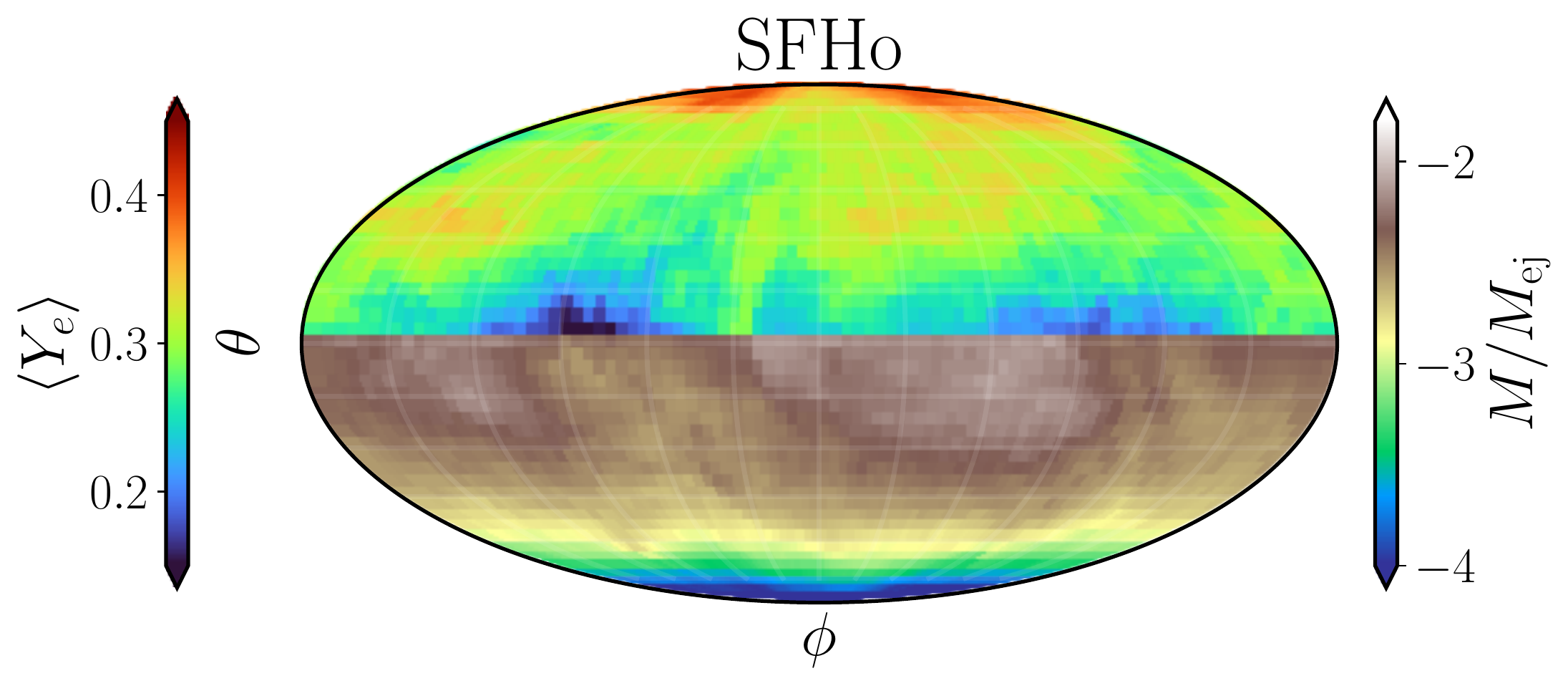}
  \includegraphics[width=1\columnwidth]{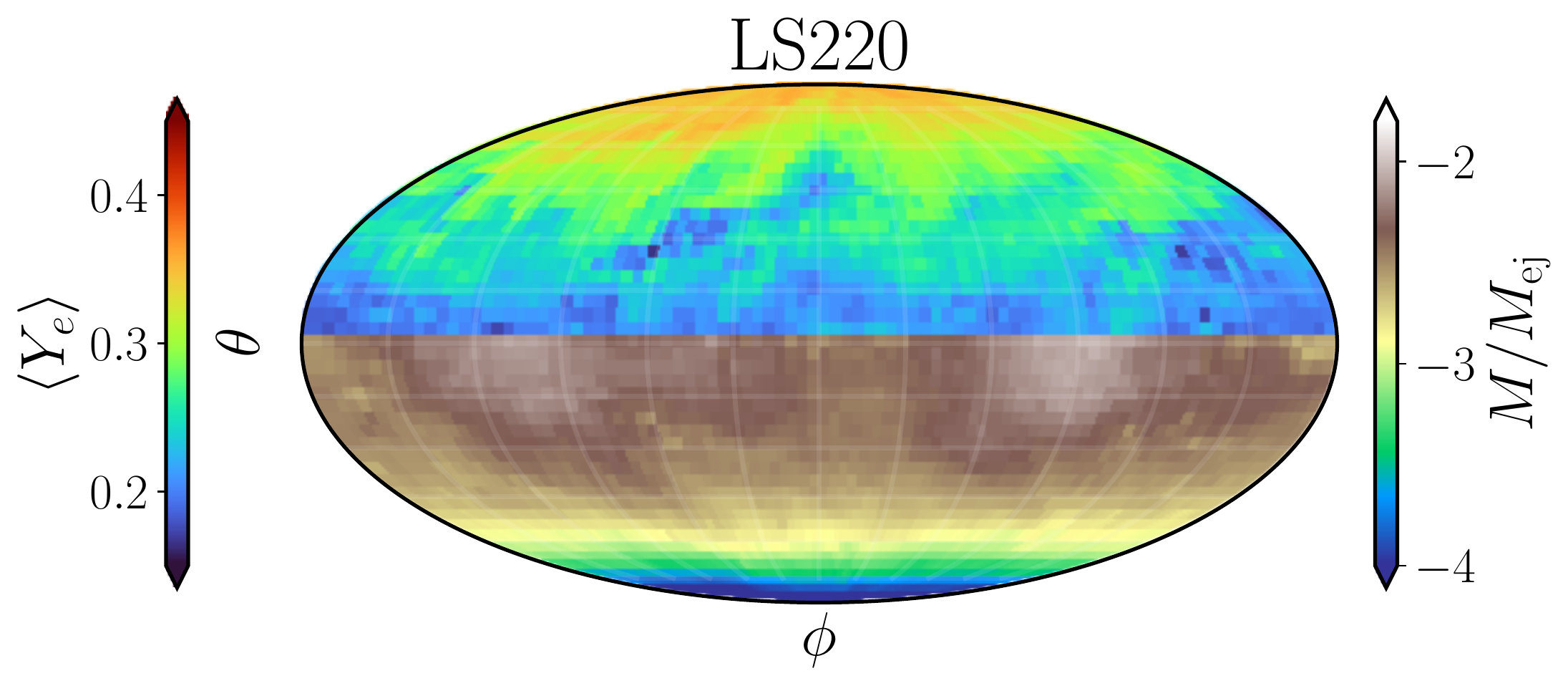}
  \caption{Sky map (Moillede) projection of the mass-weighted electron fraction (north) and of the ejected mass (south) for the total time-integrated dynamical BNS ejecta measured at $440$\,km and the three different EOS considered here. Equatorial outflows with low-to-medium electron fraction dominate the ejecta. Shock-reprocessing of tidal ejecta leads to azimuthal asymmetries in $\langle Y_{\rm e} \rangle$ for soft EOS (\textit{`lanthanide/actinide pockets'}).}
  \label{fig-angularYe}
\end{figure}

The $Y_{\el}$ distribution of the ejecta extracted at $440$\,km spans a wide range of values from $0.05$ to $0.5$ and is centered around $\approx\!0.3$ for all simulations, with a more pronounced tail to higher $Y_{\rm e}$ in the case of the softer \apr{} and \sfho{} EOS (Fig.~\ref{fig-histogram}; Tab.~\ref{tab:sim_properties}). The angular dependence of the mass-weighted $Y_{\el}$ is shown in Fig.~\ref{fig-angularYe}. Near the pole, we observe high values $>0.3$ of the electron fraction for all EOSs. This can be attributed mostly to the dominance of shock-heated ejecta in polar regions as well as to neutrino irradiation of the polar ejected material by the newly formed remnant. For \apr{} and \sfho{}, the NS remnant reaches higher temperatures due to larger compressions and oscillations after collision (owing to higher compactness of the individual stars), whereas the remnant in \lattimer{} is cooler. As a result, since the heating rate of material due to neutrino irradiation is roughly the same in all remnants, the entropy of the polar outflows in \lattimer{} is slightly higher.

Near the equatorial plane, shocks reprocess part of the neutron-rich matter to values of $Y_{\rm e} \geq 0.1$. For the soft EOSs simulations, we observe a clear azimuthal asymmetry in the $Y_\el$ angular distribution, while in \lattimer{} case the azimuthal distribution is nearly homogeneous (Fig.~\ref{fig-angularYe}). This azimuthal asymmetry for the \apr{} and \sfho{} runs occurs because the first violent bounces of the remnant produce shock waves in a preferential direction dictated by the phase of orbital rotation. 

When the two NSs merge, a rotating remnant with a bar-like ($m=2$ mode) shape is formed, which undergoes quasi-radial oscillations of its two-core structure (cf.~Sec.~\ref{sec:ejecta_dynamics}, Fig.~\ref{fig-unbmass}). The first quasi-radial bounce generates a fast shock-wave expelling material that propagates outwards without encountering significant resistance by the ambient medium. The system then develops two spiral arms due to enhanced tidal torques at maximum decompression, polluting the environment with cold neutron-rich material (see Figure \ref{fig-tri-yes}, first panel). When the remnant bounces again, it generates another violent shock wave directed along the axis formed by the double-core structure (the $x$-axis in Figure \ref{fig-tri-yes}, second panel). This fast shock wave reprocesses the previously ejected tidal tail material to high $Y_{\rm e}$ through shock-heating and associated neutrino emission in a relatively wide azimuthal wedge. The resulting angular distribution shows pockets of low-$Y_{\rm e}$ material around the equatorial plane (in directions orthogonal to the propagation of the second shock wave; Fig.~\ref{fig-angularYe}). These \textit{`lanthanide/actinide pockets'} possess much higher opacities than the surrounding shock-reprocessed material (owing to the lower electron fraction, which is conducive to forming high-opacity lanthanide and actinide elements). Observable imprints of these pockets in the kilonova emission can be explored with multi-dimensional kilonova radiation transport simulations, which are, however, beyond the scope of the present paper.

Since bounces are much less violent in the \lattimer{} run, azimuthal inhomogeneities in $Y_{\rm e}$ as described above are much less prominent. This is because a series of weaker shock waves run into previously (tidally) ejected material, which leads to an overall more homogeneous shock heating of the ejecta across all azimuthal directions and thus to a more homogeneous azimuthal $Y_{\rm e}$ profile. After the first bounces, when most of the dynamical ejecta is expelled, neutrino-driven outflows and tidal torques help launch additional material from the remnant, some of which join previously launched, circularizing debris material in forming a disk with an initial $Y_{\rm e} \approx 0.15-0.25$ and specific entropy 8-10\,$k_B$ per baryon (see cf.~Fig.~\ref{fig-tri-yes}, third panel). These entropies that arise from a complicated process of shock reprocessing as described above are almost identical to the initial entropies of $8\,k_B$ per baryon assumed in previous post-merger disk simulations (e.g., \citealt{fernandez2013Delayed,siegel2017ThreeDimensional}).

\begin{figure*}[tb!]
  \centering
  \includegraphics[width=1.0\linewidth]{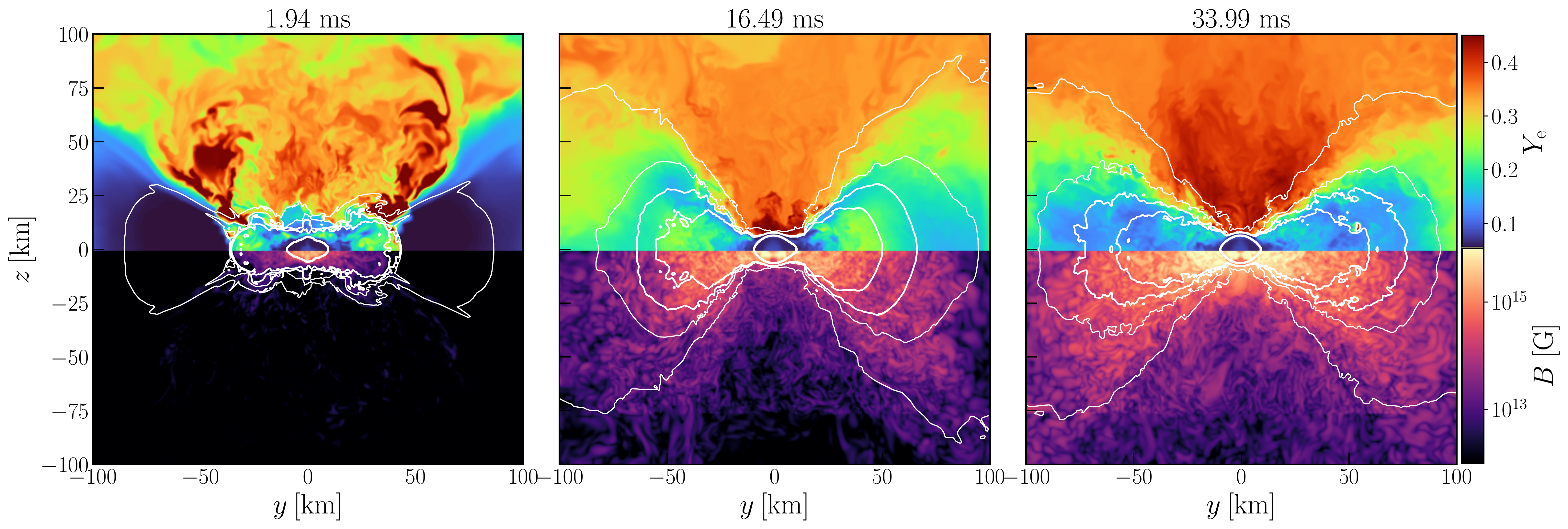}
  \caption{Snapshots of the electron fraction (upper part) and magnetic field (lower part) in the meridional plane at different times after merger for the \apr{} EOS run. White contours indicate rest-mass densities of $\rho = \lbrace 10^{9},10^{10},10^{10.5},10^{11},10^{14}\rbrace \rm ~ g\,cm^{-3}$.}
  \label{fig-tri-bye}
\end{figure*}

\subsection{Post-merger phase}

In this section, we provide a brief overview of the post-merger evolution for each simulation. We defer a more detailed discussion of the post-merger stage to a forthcoming paper.

After merger, the deformed remnant keeps shedding material to the environment through spiral arms that contribute to forming a disk. The remnant NS cores of the \lattimer{} and \sfho{} binaries keep oscillating until they merge into a single core and eventually collapse to a BH after $15$\,ms and $25$\,ms post-merger, respectively. In contrast, oscillations in the \apr{} case damp on a much shorter (few ms) timescale (Fig. \ref{fig-unbmass}).  The time of collapse of the remnant to a black hole depends critically on the angular momentum transport mechanisms present in the system. In this set of simulations, we incorporate magnetic fields and neutrino transport self-consistently, which play an important role in the evolution of the angular momentum of the remnant and its accretion flow.

Angular momentum in the disk is transported outwards by spiral waves, powered by the $m=2$ azimuthal mode of the remnant, which then develops an $m=1$ mode through hydrodynamical instabilities (see also \citealt{lehner2016Instability,radice2016Onearmed}). The disk increases in mass until $\sim\!20$ ms post-merger. At this point, the average magnetic field has increased by two orders of magnitude from its initial value and MRI-driven turbulence starts contributing to angular momentum transport in the disk, opposing outward angular momentum transport by the hydrodynamical spiral waves in the disk. As the MRI fully develops, the bulk of the disk becomes more turbulent, as can be seen qualitatively in Fig.~\ref{fig-tri-bye} (third panel). As a result of the MRI-mediated angular-momentum transport, the accretion rate increases well above the ignition threshold for neutrino cooling in the disk \citep{chen2007Neutrinocooled,metzger2008TimeDependent, metzger2008Conditions,de2021Igniting}. Thus cooling becomes energetically significant, the disk height decreases, and the disk neutronizes due to electrons becoming degenerate, keeping the disk midplane at mild electron degeneracy and $Y_{\rm e}\sim0.1-0.15$ in a self-regulated process \citep{chen2007Neutrinocooled,siegel2017ThreeDimensional,siegel2018Threedimensional}. This can be seen in Fig.~\ref{fig-tri-bye} (second and third panel), where the disk $Y_{\rm e}$ evolves from $\sim\,0.25$ to $\sim\,0.15$, the magnetic field strength increases throughout the disk, and the disk becomes less `puffy'.

Material from the disk winds and merger debris initially pollutes the polar regions right after merger, suppressing outflows driven by neutrino absorption \citep{perego2014Neutrinodriven, desai2022three}. After $\sim\!15-20$\,ms post-merger, a neutrino-driven wind emerges, ejecting only mildly neutron-rich material (Fig.~\ref{fig-tri-bye}, second and third panel). Such a wind does not develop in the \lattimer{} case, in which the remnant collapses to a black hole too early for a wind to develop. In the \sfho{} simulation, a coherent neutrino-driven wind is launched, but the remnant collapses shortly thereafter. The long-lived remnant formed in the \apr{} case does develop a steady neutrino-driven wind, which is additionally aided by magnetic fields. We defer a more detailed discussion to a forthcoming paper.

\section{Nucleosynthesis}
\label{sec-nucleosyn}

\subsection{r-process abundance pattern}
\label{sec:r-process_abundances}

Detailed nucleosynthesis analyses of the ejecta using a nuclear reaction network are performed in a post-processing step (Sec.~\ref{sec:nucleosynthesis_setup}) based on passive tracer particles that record various thermodynamic quantities of the flow along their respective Lagrangian trajectories (Sec.~\ref{sec:tracers}). In each simulation, $>\!10 \%$ of the initially placed tracers are unbound during the dynamical phase of the merger, which constitutes approximately $5 \times 10^3 - 8 \times 10^3$ tracers in total. 

We find that the injected tracer particles sample the outflow properties very well compared to those of the mass flux recorded by detector spheres on the Cartesian grid of the Eulerian observer. Figure~\ref{fig-tracers-ye} shows the ejecta mass distribution in $Y_{\rm e}$ for the \sfho{} run as recorded by a spherical outflow detector and as recorded by tracers passing through the same detector, as well as the total number of tracers per $Y_{\rm e}$ bin. The mass of each tracer is assigned as described in Sec.~\ref{sec:tracers}. Similar good agreement is obtained in all other simulations and across other quantities, such as entropy, velocity, etc.; this is of pivotal importance to ensure accurate nucleosynthesis analyses. Merely the mass distribution in velocity at high velocities $(v \gtrsim 0.6)$ is typically slightly oversampled (by factors of 2--3) with respect to grid detectors (see Appendix \ref{app:tracer_sampling}). This is a result of placing a large number of tracers in initially highly tenuous plasma (Sec.~\ref{sec:tracers}). Since only a relatively tiny fraction of mass resides at these high velocities, such oversampling does not influence the final abundance patterns (cf.~Appendix \ref{app:tracer_sampling}); it is beneficial for resolving the properties of and nucleosynthetic processes within the fast outflows.

\begin{figure}[tb!]
  \centering
  \includegraphics[width=1\columnwidth]{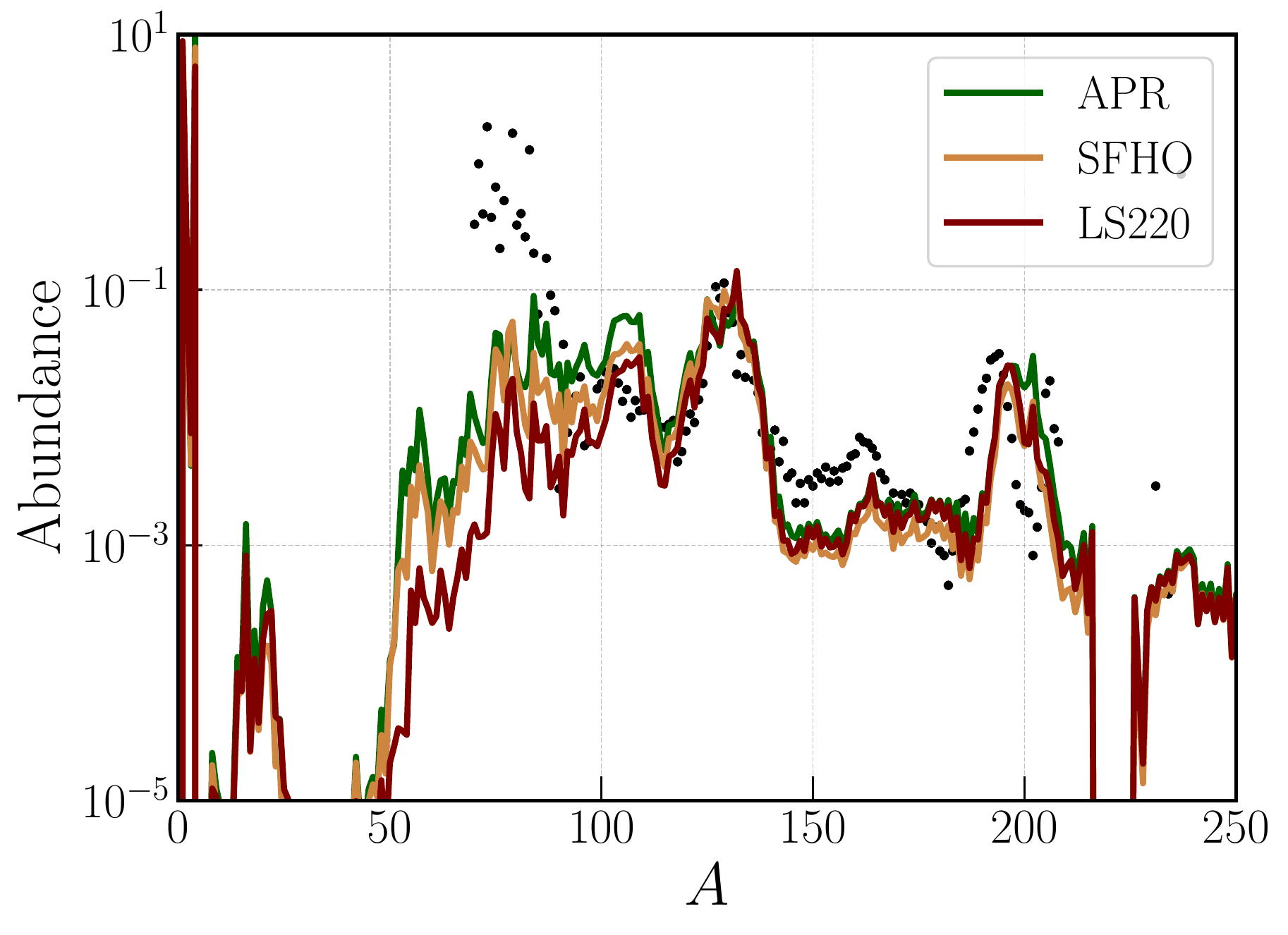}
  \includegraphics[width=1\columnwidth]{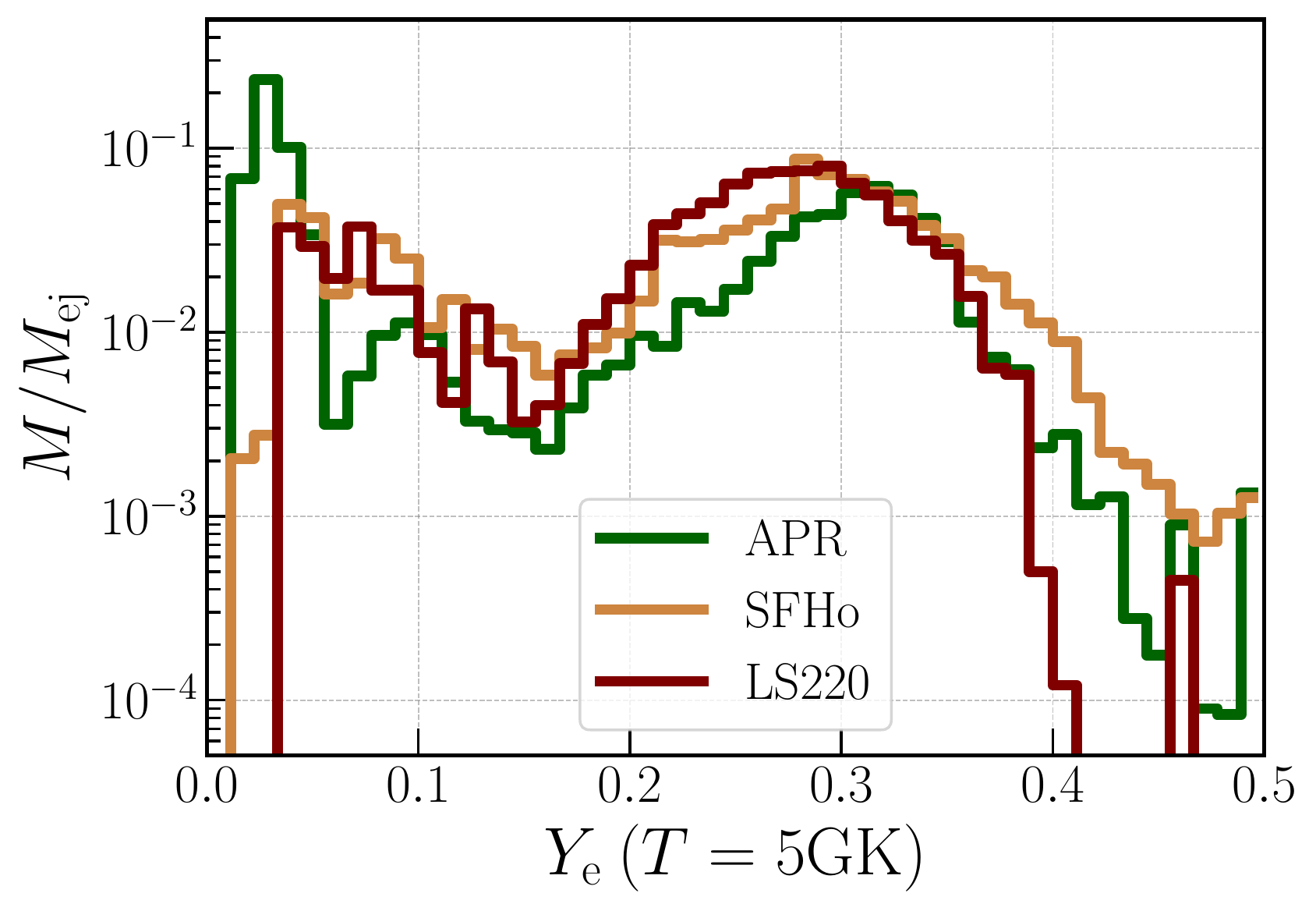}
  \caption{ Top: Mass-weighted final nuclear abundances (arbitrary units) as a function of mass number $A$ for each simulation, based on all tracers sampling the dynamical ejecta of the respective run. A robust 2nd-to-3rd peak r-process independent of the EOS is obtained. For comparison, solar r-process abundances from \citet{sneden2008NeutronCapture} are shown as black dots. Bottom: Mass distribution as a function of $Y_{\rm e}$ extracted at $T= 5 \textrm{GK}$ using tracer particles, showing the collective composition of dynamical ejecta at the onset of the r-process.}
  \label{fig-abun}
\end{figure}

The final mass-averaged abundances for the dynamical ejecta of each simulation are shown in the top panel of Fig.~\ref{fig-abun}, with solar r-process abundances added for comparison as black dots. The second (mass number $A\sim 130)$ and third ($A\sim 195)$ r-process peak are well-reproduced in all simulations---a robust 2nd-to-3rd peak r-process is obtained irrespective of the EOS. This is because, as far as dynamical ejecta is concerned, weak interactions involving both emission and absorption of neutrinos (chiefly via $e^{+} + n \leftrightarrow p + \bar{\nu}_e$ and $e^{-} + p \leftrightarrow n + \nu_e$) only have a finite ($\lesssim$ few ms; Fig.~\ref{fig-unbmass}) amount of time to increase the electron fraction of merger material from the cold, highly neutron-rich conditions $Y_{\rm e}\sim0.05-0.15$ of the colliding NSs via direct shock heating at the collision interface, via reheating of neutron-rich tidal material by shock waves from the oscillating merger remnant (Sec.~\ref{sec:ejecta_dynamics}), and via absorption of strong neutrino radiation from the hot remnant NS that is being formed. Provided a dominant fraction of the ejecta remains at $Y_{\rm e}\lesssim 0.25$ around 5\,GK when NSE breaks down, as satisfied here (cf.~Fig.~\ref{fig-abun}, bottom panel), a sufficiently high neutron-to-seed ratio can be achieved, such that a pile-up of material in the fission region at freeze-out occurs. The subsequent decay via fission then guarantees, depending on the fission model, a robust r-process pattern in the region $A\approx 120-180$ \citep{dejesusmendoza-temis2015Nuclear}. The robustness of the 2nd-to-3rd peak r-process abundance pattern largely independent of the EOS is in agreement with other recent BNS simulations including weak interactions and approximate neutrino transport \citep{radice2018Binary,kullmann2021Dynamical}. The robustness of the pattern also holds when extended to non-equal mass mergers, which suppress the amount of high-$Y_{\rm e}$ material and increase the neutron-rich tidal ejecta component (and thus the nucleon-to-seed ratio). For all runs, we find actinide abundances at the level of uranium similar to solar abundances. 

Some deviations from the solar abundance pattern visible in Fig.~\ref{fig-abun} are likely the result of nuclear input data for the nucleosynthesis calculations. Since we employ the FRDM mass model for nucleosynthesis, the third r-process peak is systematically shifted to the right (slightly higher mass numbers) for all simulations, which has been attributed to neutron captures after freeze-out combined with relatively slower $\beta$-decays of third-peak nuclei in the FRDM model \citep{eichler2015Role,dejesusmendoza-temis2015Nuclear, caballero2014local}. The trough in abundances between $A\sim140-170$ relative to solar is likely due to the fission fragment distribution employed here, as pointed out by r-process sensitivity studies (e.g., \citealt{eichler2015Role,dejesusmendoza-temis2015Nuclear}).

The first r-process peak is under-produced in all models, which is due to only partially reprocessed ejecta material with respect to the original cold, neutron-rich matter ($Y_{\rm e}\sim 0.05-0.1$) of the individual NSs (see above). Reproducing the first solar r-process abundances requires a $Y_{\rm e}$-distribution that extends well above 0.25 for a significant fraction of the ejecta (e.g., \citealt{lippuner2015RProcess}). Given that the mean of the $Y_{\rm e}$ distribution is slightly higher for \sfho{} and \apr{}, these models have a larger fraction of first peak material, which is however still under-produced with respect to solar values.
Light r-process elements in the first-to-second r-process peak region are preferentially synthesized in the post-merger phase in winds launched from a remnant neutron star and from the accretion disk around the remnant, which can give rise to broad $Y_{\rm e}$-distributions (e.g., \citealt{perego2014Neutrinodriven,lippuner2017Signatures,siegel2018Threedimensional,de2021Igniting}). We defer a more detailed discussion on nucleosynthesis including post-merger ejecta to a forthcoming paper.

\begin{figure}[tb!]
  \centering
  \includegraphics[width=1\columnwidth]{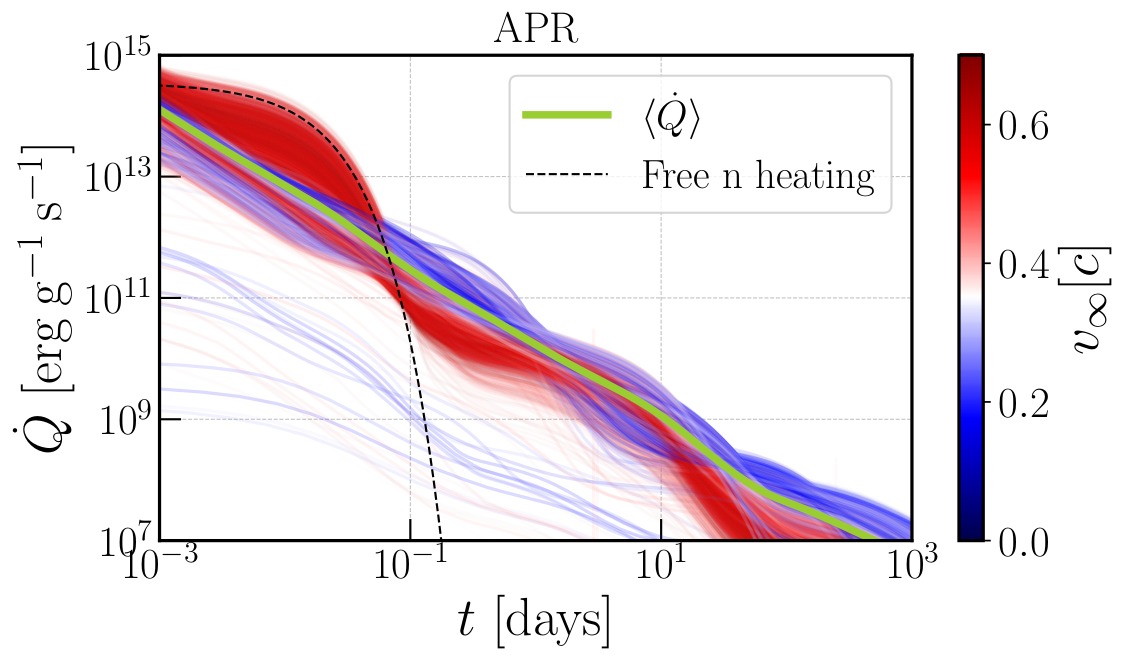}
  \includegraphics[width=1\columnwidth]{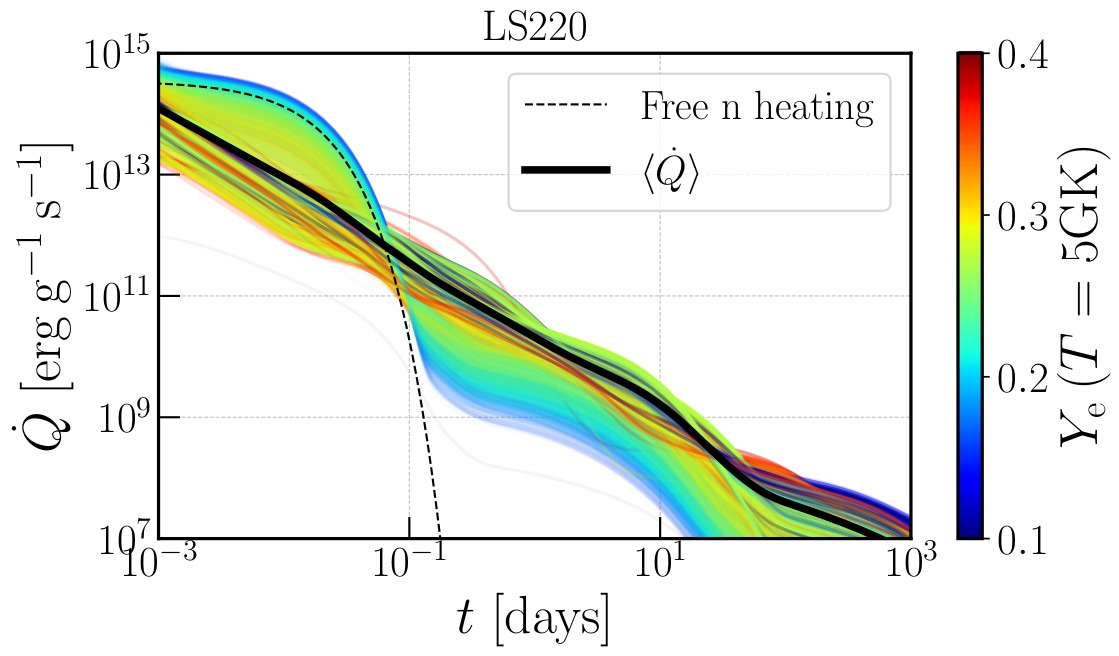}
  \caption{
  Evolution of the specific heating rate calculated with \texttt{SkyNet} for all unbound tracers of the dynamical ejecta and \apr{} (upper panel; color-coded by asymptotic velocity) as well as \lattimer{} (lower panel; color-coded by electron fraction). Thick lines represent mass averages over all tracers, which closely follow the expected $\propto t^{-1.3}$ power-law for r-process heating. 
  Black dashed-lines correspond to an analytic approximation to the heating rate of free-neutron decay \citep{kulkarni2005Modeling}. Heating due to free neutron decay is present over a wide range of EOS softness/stiffness and originates in high-velocity ($v_\infty\gtrsim 0.5c$) and low-to-moderate $Y_{\rm e}\lesssim0.2$ outflows. }
  \label{fig-qdot-vel}
\end{figure}

\subsection{Fast ejecta and free neutrons}
\label{sec:free_neutrons}

The possibility of tracing individual fluid elements allows us to evaluate the radioactive heating rate within the ejecta in more detail. In particular, we are interested in the fast portion of the ejecta that generates free neutrons, which then $\beta$-decay with a half-life of $\approx\!10$\,min and provide additional heating of the material at timescales of up to hours relative to what would be expected from pure r-process heating. Such early excess heating can power bright UV emission known as a kilonova precursor \citep{metzger2015Neutronpowered}. The amount of free neutrons generated by the ejecta sensitively depends on the expansion timescale (i.e., on the ejecta velocity) and the proton fraction $Y_{\rm e}$. 

Figure \ref{fig-qdot-vel} shows the specific heating rate as recorded by each unbound tracer that samples the dynamical ejecta. Irrespective of the stiffness/softness of the EOS, we find a fast $v_\infty\gtrsim 0.5 c$ and neutron-rich $Y_{\rm e}\lesssim 0.2$ component of tracers that generate excess heating on a $\sim\!10$\,min timescale (as expected for free-neutron decay with half-life $\approx\!10$\,min), relative to the standard $\dot{Q}\propto t^{-1.3}$ heating rate of a pure r-process. This excess self-consistently obtained from nuclear reaction network calculations is in good agreement with analytic predictions for free-neutron decay (dashed lines in Fig.~\ref{fig-qdot-vel}; \citealt{kulkarni2005Modeling}). Although excess heating from this light, fast ejecta component does not significantly alter the mass-averaged heating rate (thick solid lines in Fig.~\ref{fig-qdot-vel}) with respect to its power-law behavior, this excess heating occurs in the outermost layers of the ejecta where even a small number of free neutrons can give rise to observable emission on timescales of hours (Sec.~\ref{sec:kilonova_lightcurves}).

\begin{figure}[tb!]
  \centering
  \includegraphics[width=1\columnwidth]{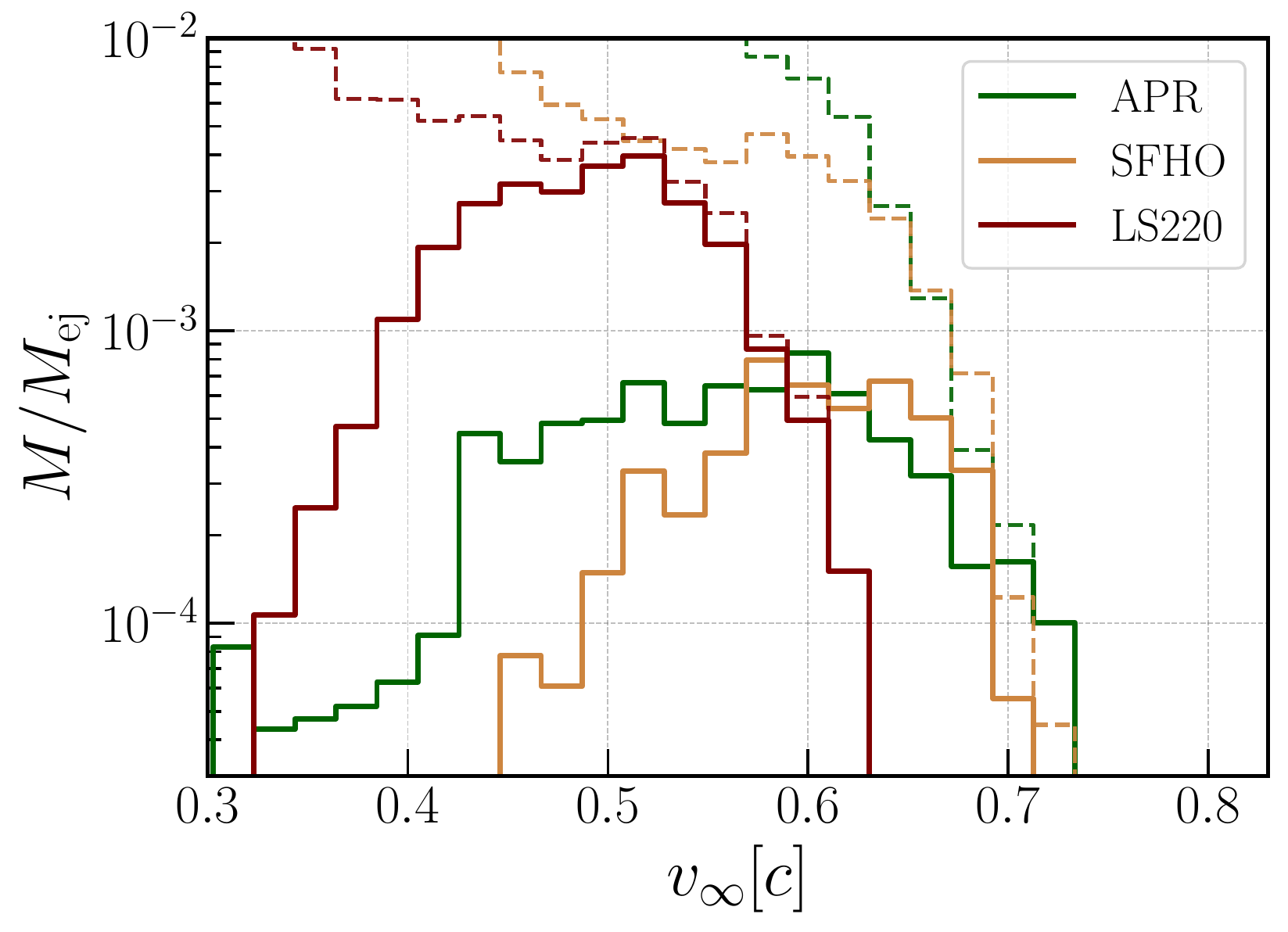}
  \caption{Distribution of dynamical ejecta mass $\mejn$ (cf.~Tab.~\ref{tab:sim_properties}) that results in free neutrons as a function of asymptotic speed and as sampled by tracer particles (solid lines). Also shown are the corresponding distributions of total dynamical ejecta mass (dashed lines).}
  \label{fig-fnhisto}
\end{figure}

Figure \ref{fig-fnhisto} shows the detailed mass distributions of free neutrons according to the asymptotic speed of ejecta. We find that almost all material with velocities faster than $\approx\!0.6c$ produces free neutrons. Slower parts of the ejecta, however, also contribute. These slower layers reside deeper within the ejecta and thus contribute to the kilonova precursor at a higher initial optical depth. Their contribution to the total luminosity is thus dimmer and peaks at slightly later times relative to injection in the outermost (fastest) layers (Sec.~\ref{sec:kilonova_lightcurves}). The total mass of free neutrons is $\approx\!2\times 10^{-5}\,\msun$ in all runs (Tab.~\ref{tab:sim_properties}).

\begin{figure}[tb!]
  \centering
  \includegraphics[width=1\columnwidth]{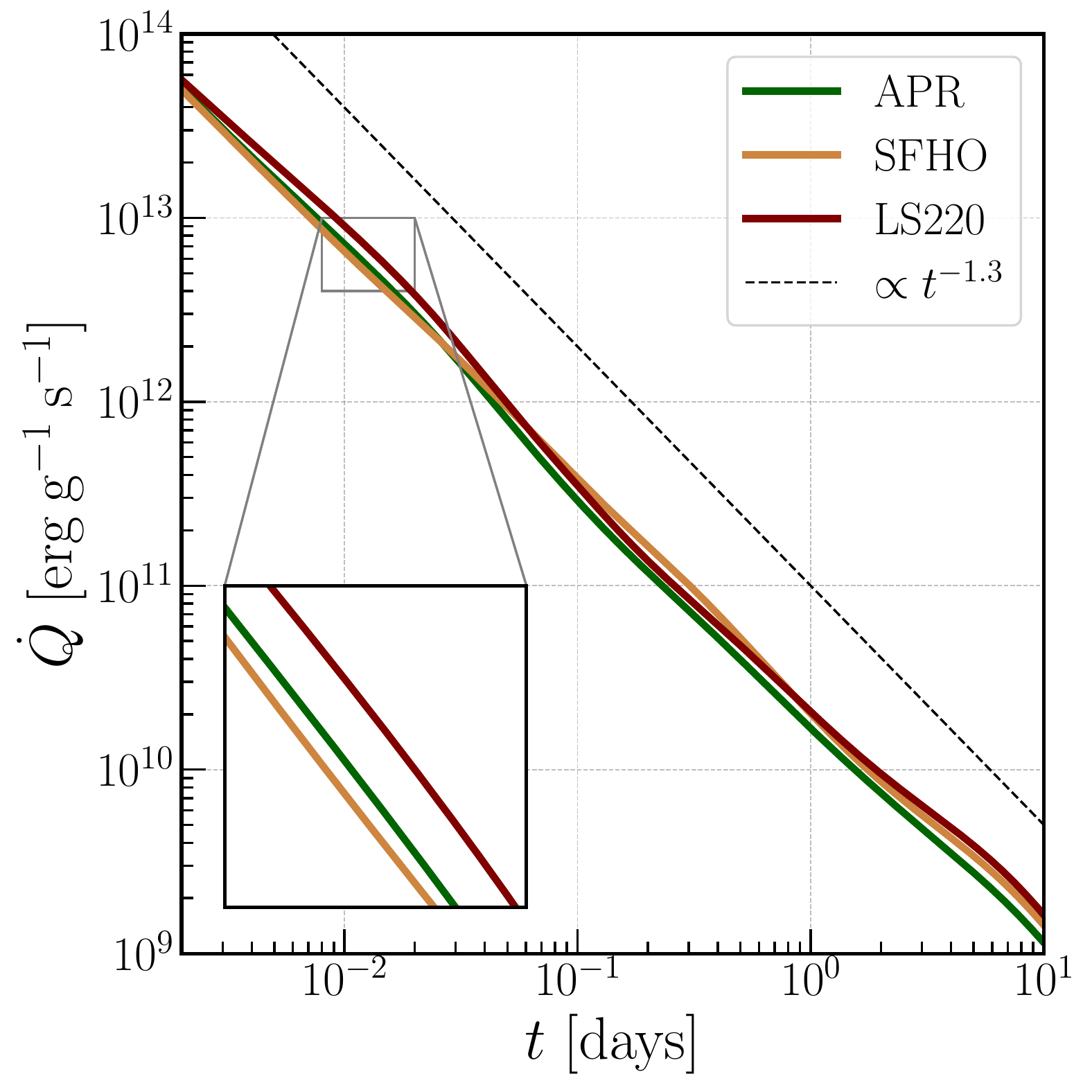}
  \caption{
  Mass-weighted specific heating rate of all tracer particles sampling the dynamical ejecta for each simulation. The inset at $t=10^{-2}$ days = $900$ s shows the differences of the mass-weighted heating rates due to free neutron decay.}
  \label{fig-qdot-all}
\end{figure}

We find that the stiffer EOS simulation \lattimer{} provides a slightly higher heating rate at $t \approx 10^{-2}$ days than the softer EOSs employed here (Fig.~\ref{fig-qdot-all}), which we mostly attribute to free neutron heating. Although the fraction of fast ejecta producing free neutrons in the former model is smaller (and slower) than those of the latter (Tab.~\ref{tab:sim_properties}, Fig.~\ref{fig-fnhisto}), it is more neutron-rich owing to less shock heating ($Y_{\rm e} \sim 0.2$ for the fast ejecta in \lattimer{}, compared to $Y_{\rm e} \sim 0.3$ for \sfho{} and \apr{}, see Fig.~\ref{fig-qdot-vel}). Overall, this results in more free neutrons nevertheless. This conclusion is expected from parametric explorations of r-process nucleosynthesis \citep{lippuner2017Signatures} (see their Fig.~3) and also supported by recent parametric studies in merger ejecta using SPH simulations (\citealt{kullmann2021Dynamical}, their Fig.~12).

\section{Electromagnetic signatures}
\label{sec-emsig}

Although dynamical ejecta are expected to be subdominant with respect to post-merger ejecta for typical BNS systems such as those considered here, but also more generally looking at the expected BNS population as a whole \citep{siegel2019GW170817, siegel2022r}, they give rise to a number of distinctive electromagnetic transients that we intend to model here in a self-consistent way based on ab-initio GRMHD simulations. Based on characteristics of dynamical ejecta extracted from our GRMHD simulations (Sec.~\ref{sec:simulation_results}) and detailed nucleosynthesis analysis (Sec.~\ref{sec-nucleosyn}), we present a model to calculate the corresponding early kilonova lightcurves taking into account the contribution of free-neutron decay (the kilonova precursor), as well as relativistic effects associated with the fast-moving outflow (Sec.~\ref{sec:kilonova_model}). We also present a model to calculate the non-thermal afterglow emission related to the mildly relativistic shock wave that the fast ejecta gives rise to as it expands into the interstellar medium (Sec.~\ref{sec:afterglows}). We build a multi-angle model that solves a 1D hydrodynamic shock-wave problem in multiple directions (Sec.~\ref{sec:afterglows_shock_dynamics}) and computes the associated radiation of a non-thermal distribution of electrons accelerated within the shock, solving the time-dependent Fokker-Planck equation (Sec.~\ref{sec:non-thermal radiation}). Results of both models based on our simulations runs and in the context of GW170817 are discussed in Secs.~\ref{sec:kilonova_lightcurves} and \ref{sec:kilonova_afterglows_GW170817}, respectively. We also investigate the potential impact of recently calculated extremely large opacities of highly ionized lanthanides on the early kilonova emission (Sec.~\ref{sec:ionization_lanthanides}) as well as signatures of high-energy gamma-ray emission due to inverse Compton and synchrotron self-Compton processes in the kilonova afterglow shock (Sec.~\ref{sec:gamma_rays}).

\subsection{Kilonova Semi-Analytical Model}
\label{sec:kilonova_model}

We calculate kilonova light curves generated by the dynamical ejecta of our simulations using a new semi-analytical model. We formulate this model based on the approach of \citet{hotokezaka2020Radioactive}, which, in turn, closely follows the work of \citet{waxman2019Latetime} and \citet{kasen2019Radioactive}. This model allows us to compute the heating rates of r-process elements using experimentally measured values for the injection energies of nuclear decay chains. Our code builds on the public version of \citet{hotokezaka2020Radioactive}\footnote{\url{https://github.com/hotokezaka/HeatingRate}}, which we optimized in terms of computational efficiency. We further developed the framework to additionally include (a) arbitrary mass distributions from numerical simulations, (b) heating rates due to free neutrons, (c) arbitrary opacities, (d) flux calculation in different wavelength bands, and (c) relativistic effects. Furthermore, we embed this 1D framework in a generalized infrastructure for computing a multi-angle kilonova signal following \citet{perego2017AT2017gfo}. For this work, however, we restrict ourselves to a spherically symmetric (angular-averaged) approximation and defer a multi-angle approach including post-merger ejecta components to forthcoming work. In the following, we briefly outline the main features of the model.

Given an angular-averaged mass distribution $M(v)$ extracted from a BNS merger simulation, we assume homologous expansion of the outflow and we discretize it into velocity shells $v_i$. Each velocity shell is characterized by a mass $M(v_i)$ and grey opacity $\kappa(v_i)$. We typically use at least 30 velocity bins, which we find is sufficient to obtain converged light curves.

The energy of each ejecta shell is determined by losses due to pdV work associated with the adiabatic expansion of the fluid ($E_{i}/t$), radiative loses due to photons escaping the flow ($L_{\rm rad, i}$), and heating from nuclear decay products that thermalize in the ejecta ($\dot{Q}_i$). From energy conservation, the equation for each shell in the comoving frame reads
\begin{equation}
\frac{dE_i}{dt} =  - \frac{E_{i}}{t} + \dot{Q}_i - L_{\rm rad, i},
\label{eq-firstlaw}
\end{equation}
which we integrate using a fourth-order Runge-Kutta algorithm. This requires the computation of the heating term $\dot{Q}_i$ and of the radiative luminosity $L_{{\rm rad}, i}$

The ejected outflow contains a variety of heavy neutron-rich isotopes synthesized by the r-process as well as free neutrons (Sec.~\ref{sec-nucleosyn}). Unstable lighter nuclei will mainly $\beta$-decay, while translead nuclei will also generate $\alpha$-decays and undergo spontaneous fission. The products of these processes, mainly fast-moving electrons, $\alpha$-particles, and gamma-rays, will in part be absorbed by the outflow. The deposition of energy from charged particles in radioactive decay occurs through collisional ionization and Coulomb collisions. It is determined by the energy-loss rate of fast particles $\beta K_{\rm st}$, where $K_{\rm st}$ is the stopping cross-section per unit mass in units of MeV\,cm$^{2}$\,g$^{-1}$ and $\beta$ is the velocity of the particle in untis of the speed of light $c$. As in \citet{hotokezaka2020Radioactive}, we take the experimentally measured values of the stopping cross-sections from ESTAR and ASTAR of the NIST database\footnote{\href{http://physics.nist.gov/Star}{http://physics.nist.gov/Star}}.

If the timescales of energy loss of the charged particles created in radioactive decays are much shorter than the dynamical time of the flow, their energy is entirely deposited in the ejecta. At later times, thermalization is not as efficient anymore, and we need to follow the energy loss of the particles. Let us consider a nuclear element $j$ that injects an electron with energy $e_0$ at time $t_0$. Taking into account adiabatic losses, the energy $e_j$ of the electron follows \citep{kasen2019Radioactive}
\begin{equation}
    \frac{de_{j}}{dt} = -3(\gamma(p)-1) \frac{e_{j}}{t} - \beta K_{\rm st}(e_j) \rho(t) c,
\end{equation}
where $\gamma(p)$ is the adiabatic index of charged particles \citep{nakar2020Electromagnetic} and $\rho$ is the density of a mass shell. Once we have solved for $e_j(t; e_0,t_0)$, the deposited energy into the mass shell is obtained by integrating the ionization losses as a function of time, $\beta K_{\rm st}(t; e_0,t_0)$. In order to calculate the total heating rate of charged decay products at time $t$, one also needs to take into account particles injected at earlier times as they may not have previously lost all of their energy. Summing over all nuclear species, the total heating rate per unit mass is
\begin{equation}
   \dot{q}_{\rm th, e}(t) = \sum_j \int^t_{t0_j} \beta K_{\rm st}(t; e_0, t') \rho(t) c  \frac{N_j(t')}{\tau_j} dt',
   \label{eq:totheating}
\end{equation}
where $N_j(t')$ is the number of a radioactive element $j$ per unit mass, $\tau_j$ the mean lifetime of that element, and $t0_j$ the injection time of the oldest non-thermal particle surviving at $t$. To compute the initial injection energies and mean lifetimes we use the Evaluated Nuclear Data File library (ENDF/B-VII.1; \citealt{chadwick2011endf}). We use abundance distributions of nuclei with $A\ge 70$ directly extracted from our nucleosynthesis calculations (Sec.~\ref{sec-nucleosyn}; Fig.~\ref{fig-abun}). We compute Eq.~\eqref{eq:totheating} separately for $\beta$-decay, $\alpha$-decay, and fission, using the corresponding stopping cross-sections for each process.

Thermalization of gamma-rays produced in radioactive decays occurs through Compton scattering, absorption, and pair creation. The fraction of energy that is deposited by gamma-rays is calculated as $f_{\gamma}= 1-\exp(-\tau_{\rm eff})$, where the effective optical depth is defined as $\tau_{\rm eff}=\kappa_{\rm eff} \Sigma(t)$, with $\kappa_{\rm eff}$ the absorptive opacity, and $\Sigma(t)$ the mass weighted column density of the mass shell (see Sec.~3.2 in \citealt{hotokezaka2020Radioactive} for more details). The specific heating rate is then $\dot{q}_{\rm th, \gamma} = f_{\gamma}(t) \dot{q}_{\gamma}$, where, again, we use data from ENDF/B-VII.1 to calculate $\dot{q}_{\gamma}$. 

Summing the contributions to the heating rates from $\alpha$ and $\beta$-decays as well and from fission by both charged particles and gamma-rays, we obtain $\dot{q}_{\rm r}$, the total net heating rate per unit mass of the synthesized r-process elements. Upon adding the contribution of free neutrons that undergo $\beta$-decay, the total heating rate of each shell $\dot{Q}_i$ in Eq.~\eqref{eq-firstlaw} is obtained as
\begin{equation}
    \dot{Q}_i = M_i(1-X_{\rm fn, i})\,\dot{q}_{\rm r}(t) + M_i X_{\rm fn, i}\, \dot{q}_{\rm fn}(t),
\end{equation}
where $X_{\rm fn, i}$ is the mass fraction of free neutrons in each velocity shell as extracted from the nuclear reaction network calculations based on the simulation runs; for the specific heating rate $\dot{q}_{\rm fn}(t)$ of free neutrons we set \citep{kulkarni2005Modeling}
\begin{equation}
    \dot{q}_{\rm fn}(t)= 3.2 \times 10^{14} {\rm erg\,g ^{-1}\,s^{-1}}\, \exp{(-t/\tau_{\rm N})}, \label{eq:free_n_heating}
\end{equation}
which provides a good approximation to the free-neutron decay heating rate in our reaction network calculations (Sec.~\ref{sec-nucleosyn}, Fig.~\ref{fig-qdot-vel}). Here, $\tau_{\rm N}\simeq 880$\,s is the mean lifetime of a free neutron.

Having calculated the effective heating rate for all mass shells, we need to estimate how photons escape from each shell. In general, the frequency-integrated (grey) opacity of r-process ejecta sensitively depends on the mass fraction of lanthanides and actinides (owing to their valence f-shells, which result in an increase in line expansion opacities by orders of magnitude with respect to light r-process elements) as well as on the temperature of the outflow. Since we focus on the early light curve evolution from the outermost (dynamical) ejecta, the photospheric temperatures satisfy $T>4000-5000$\,K, at which opacities are largely insensitive to temperature (e.g., \citealt{kasen2013Opacities,tanaka2020Systematic, banerjee2020simulations}). Recent opacity calculations by \citet{banerjee2022opacity} find a strong enhancement in the opacity of lanthanide elements when highly ionized at temperatures $\approx 8\times 10^4\,\text{K}$. We discuss in Sec.~\ref{sec:ionization_lanthanides} that this temperature regime is not of concern for the ejecta structure of our binary simulations presented here, as the photosphere never reaches this temperature regime. The opacity is largely controlled by the lanthanide content, which, in our scenario, mainly depends on the electron fraction of the ejecta. We thus ignore the temperature dependence and use a 1D-parametrization of the opacity as a function of $Y_{\rm e}$ proposed by \citet{wu2021radiation},
\begin{equation}
    \kappa(Y_{\rm e})  = \frac{9}{1+(4 Y_{\rm e})^{12}} \quad \rm cm^{2} \rm g^{-1}, \label{eq:kappa_Ye}
\end{equation}
which models the expected range of grey opacity, with a characteristic rapid drop around $Y_{\rm e}\approx 0.25$ (above which lanthanide production is suppressed; \citealt{lippuner2015RProcess}). For each velocity shell $v_i$, we calculate a mass-averaged electron fraction $Y_{\rm e}$ to obtain $\kappa(v_i)$ according to Eq.~\eqref{eq:kappa_Ye}. Although there is some (residual) mixing of material (and thus of different $Y_{\rm e}$ components) within and in-between neighboring velocity shells, we find this mapping is robust when applied at different extraction radii.

The radiative luminosity of each shell, $L_{\rm rad, i}$, is calculated by taking into account, approximately, the trapping, diffusive, and free-streaming radiative regimes. Generalizing the single-zone model of \cite{piro2013WHAT}, each velocity shell is assigned an energy escape fraction as
\begin{equation}
    L_{\rm rad,i} = \frac{f_{\rm Esc, i} E_i}{t_{\rm Esc, i}},
\end{equation}
where $f_{\textrm{Esc},i}=\textrm{erfc}(\sqrt{t_{\rm diff,i}/t})$ is the energy escape fraction from each shell. Here, erfc denotes the complementary error function, $t_{\textrm{diff}, i} = \tau_i v_i t /c$ is the diffusion time, and the escape time $t_{\textrm{Esc}, i}$ of the shell is defined by $t_{\textrm{Esc}, i}= \text{min}(t_{\rm diff},t)+ v_i t /c$ (see also Eqs.~(30)--(32) in \citealt{hotokezaka2020Radioactive} for more details). The optical depth is determined by $\tau_i (t) = \int^{\infty}_{v_i t} \kappa(r) \rho(r) dr$, where $\rho(r)$ is the density of the shell. The radiative luminosity is assigned a blackbody spectrum $L_{{\rm rad},i,\nu}$, defined by the effective temperature that is determined by the shell's radius, $T_{{\rm eff},i} = [L_{{\rm rad}, i}/(4\pi\sigma_{\rm SB}v_{i}^2t^2)]^{1/4}$, where $\sigma_{\rm SB}$ is the Stefan-Boltzmann constant. The total luminosity in the comoving frame is obtained by adding the radiative luminosity of all shells, $L_{\rm bol}=\sum_i \int\mathrm{d}\nu\,L_{{\rm rad},i,\nu}$. For reference, we also define the location of the photosphere as the shell $i$ with radius $R_{\rm ph}$ where $\tau_i = 1$. The corresponding photospheric temperature is then defined as $T_{\textrm{ph}}\equiv [ L_{\rm bol}/(4 \pi \sigma_{\rm SB} R_{\rm ph}^2))]^{1/4}$.

Finally, since we are dealing with mildly relativistic outflows, we need to incorporate relativistic effects. Since the observed luminosity strongly scales with the Doppler factor, even mildly relativistic ejecta shell velocities can have an appreciable impact on the observed lightcurves. Luminosity and fluxes in the observer frame are computed taking into account all relevant relativistic effects (the relativistic Doppler effect, the time-of-flight effect, and relativistic beaming) following the reconstruction method based on an `energy package' approach presented by \citet{siegel2016Electromagnetic}. The photosphere at a given time $t$ is discretized in angular bins and the emergent radiation from each patch of the surface is further discretized into frequency bins. The resulting energy packages $\Delta E(\theta_k, \nu_l)$ are emitted from the instantaneous photosphere in the comoving frame and are received by the observer at their respective arrival times. The observer light curve is reconstructed by binning these packages into observer time and frequency bins accounting for relativistic effects (see Sec.~5.7 in \citealt{siegel2016Electromagnetic} for details).

\subsection{Kilonova Light Curves}
\label{sec:kilonova_lightcurves}

\subsubsection{Bolometric light curves}
\label{sec:bolometric_kilonova_lightcurves}

\begin{figure}[tb!]
  \centering
  \includegraphics[width=1\columnwidth]{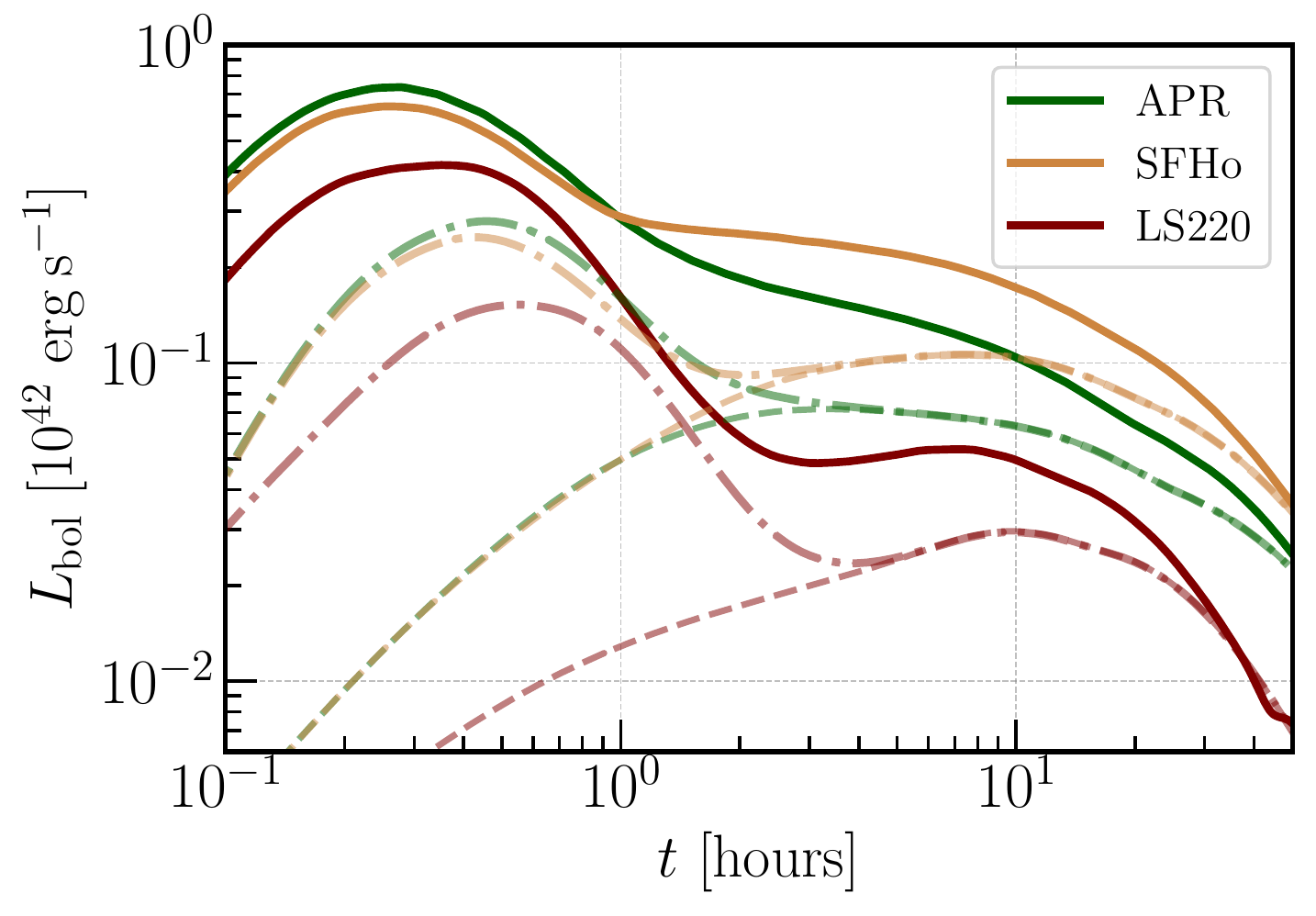}
  \caption{Bolometric light curves of kilonovae and neutron precursor emission produced by the dynamical ejecta only for each simulation, taking into account free-neutron decay and relativistic effects (thick lines), free-neutron decay without relativistic effects (dot-dashed lines), and r-process heating only (dashed lines).}
  \label{fig-knlc-eos}
\end{figure}

Figure \ref{fig-knlc-eos} presents the bolometric luminosity of the kilonova emission and neutron precursor emission resulting from the dynamical ejecta of all our runs as computed with the model discussed in Sec.~\ref{sec:kilonova_model}. Shown are separate calculations that distinguish the effects of heating due to free neutron decay and relativistic effects from pure r-process heating only.

Heating from free-neutron decay enhances the total luminosity at early times ($\sim\!0.5$\,h after merger) by a factor of 8--10. Differences to \citet{metzger2015Neutronpowered}, who find an enhancement of the order of 15--20, are likely due to their mass of free neutron material being an order of magnitude larger and distributed at larger velocities ($v\sim 0.8c$) than what we find in our present simulations (see the discussion in Sec.~\ref{sec:ejecta_dynamics}). We expect relativistic effects (chiefly Doppler boosting) to enhance the observed luminosity roughly by the Doppler factor to the third power (cf.~\citealt{siegel2016Electromagnetic}).\footnote{This follows from the fact that $I/\nu^3$ is a Lorentz invariant, where $I$ is the specific intensity and $\nu$ the frequency of emitted radiation. Since, $\nu'=D\nu$ in another inertial frame, where $D$ is the Doppler factor, one has $I'=I D^3$.} We crudely estimate the angular averaged enhancement factor to
\begin{equation}
    D^3\approx \left(\frac{\sqrt{1-\beta^2}}{1-\beta\langle\cos\theta\rangle}\right)^4 \approx 3.5,
\end{equation}
where we used a typical (solid angle averaged) value of $\langle\cos\theta\rangle = \pi/4$. This estimate is consistent with Fig.~\ref{fig-knlc-eos} which shows an additional enhancement by a factor of 3--4. In total, at early times ($
\lesssim 1\,\text{h}$), we find the bolometric luminosity of this kilonova with an amount of $\approx\!2\times\!10^{-5} \msun$ of free-neutron producing material and typical velocity of $\langle v \rangle \sim 0.6c$ (cf.~Sec.~\ref{sec:free_neutrons}; Tab.~\ref{tab:sim_properties}) is a factor $\lesssim\!10-20$ higher than a non-relativistic kilonova powered by freshly synthesized r-process elements only. Relativistic effects subside as the photosphere recedes to lower expansion speeds further within the ejecta; we find noticeable enhancements in observed luminosity up to $\lesssim 10$\,h.

The neutron precursor signal is expected to peak in the source frame when the free-neutron producing ejecta becomes transparent to photons, $t_{\rm diff} \approx t$, which, for typical values in our simulations, can be roughly estimated to
\begin{eqnarray}
    t_{\rm p} &\approx& \left(\frac{3 \mejn \kappa}{4\pi\alpha vc}\right)^{1/2} \label{eq:t_peak_n}\\
    &\approx& 0.4\,\text{h}\left(\frac{\mejn}{2\times 10^{-5}M_\odot}\right)^{\frac{1}{2}} \mskip-5mu\left(\frac{\kappa}{1\,\text{cm}^2\text{g}^{-1}}\right)^{\frac{1}{2}}\mskip-5mu\left(\frac{v}{0.6c}\right)^{-\frac{1}{2}},\nonumber
\end{eqnarray}
in good agreement with the peak timescales of the dashed-dotted curves in Fig.~\ref{fig-knlc-eos}. Here, we have assumed a steep outermost ejecta profile with $M(v)\propto v^{-\alpha}$, where $\alpha\approx 10$ (cf.~Figs.~\ref{fig-ek-ag} and \ref{fig-fnhisto}). Relativistic effects then decrease that timescale roughly to $t_{\rm p, obs}=\sqrt{1-\beta^2}t_{\rm p}\approx 0.8 t_{\rm p}\approx 0.3\,h$ for $v = \beta c = 0.6c$, consistent with the shift evident from Fig.~\ref{fig-ek-ag}. The observed peak timescale is thus closer to the timescale of free neutron heating of about $0.25\,\text{h}\approx 900$\,s (cf.~Eq.~\eqref{eq:free_n_heating}).

Softer EOSs show a brighter transient at peak light with a slightly earlier peak time, while \lattimer{} shows a somewhat broader peak and slightly delayed peak time. While this behavior is expected from the fact that softer EOSs lead to larger velocities of the outermost layers of the dynamical ejecta (cf.~the discussion in Sec.~\ref{sec:ejecta_dynamics}, Figs.~\ref{fig-histogram}, \ref{fig-kin-ang}), the electron fraction of the outflows complicates the picture. Part of the fastest ejecta is shock-heated and thus turns less neutron-rich, with softer EOS giving rise to more shock heating. As free neutrons become increasingly buried within slower ejecta shells, their heating acts at higher initial optical depths, which implies longer peak timescales, hence a broader peak, and a dimmer light curve at peak light (cf.~\lattimer{} in Fig.~\ref{fig-knlc-eos}). We note that despite the smaller ejecta velocities, heating due to free neutron decay can still be significant for stiffer EOS due to a higher neutron-richness in the fast outflows.

\subsubsection{Application to GW170817}
\label{sec:kilonova_GW170817}

\begin{figure}[tb!]
  \centering
  \includegraphics[width=1\columnwidth]{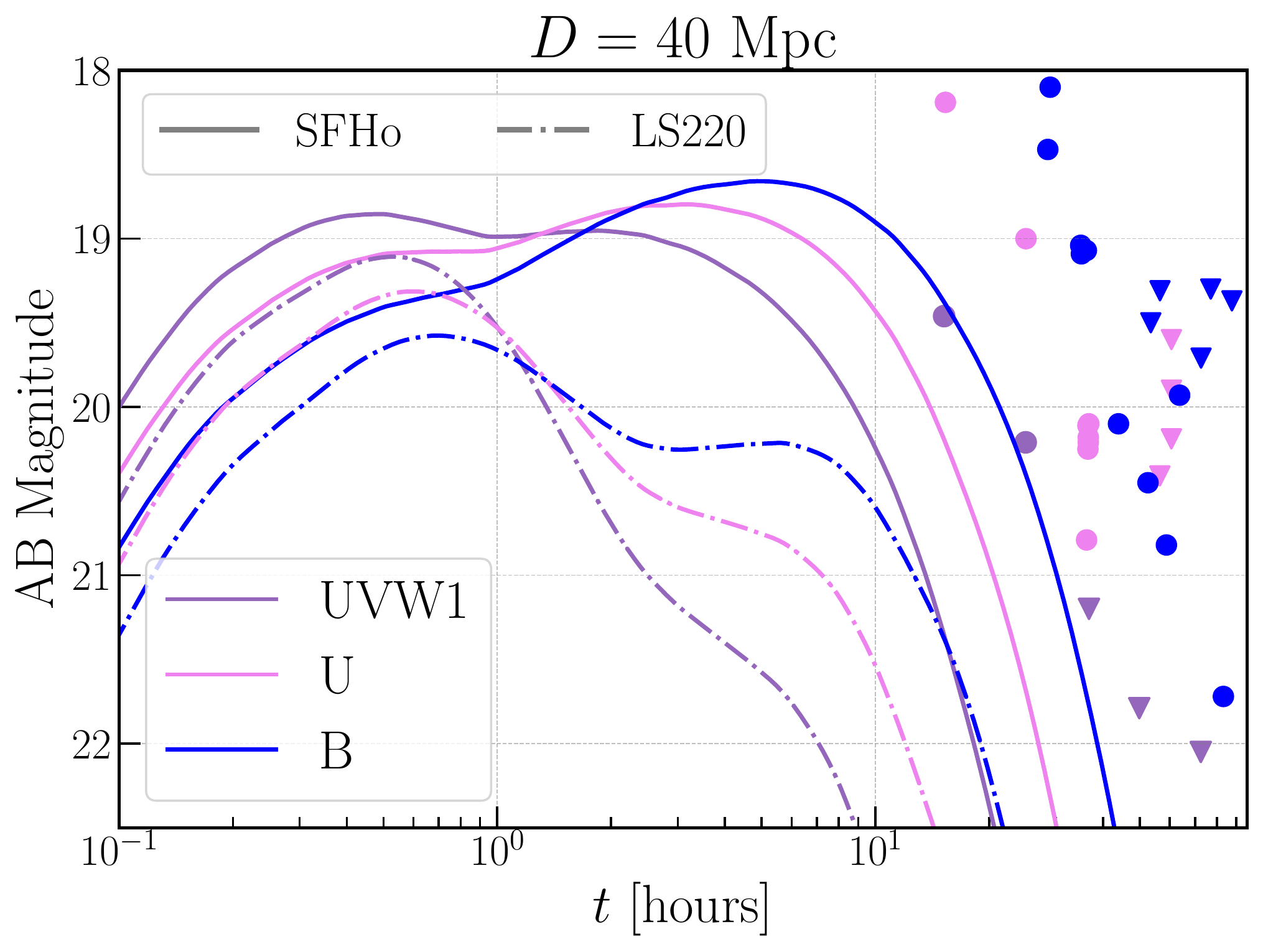}
  \caption{AB magnitude of blue wavelength bands for the \sfho{} (solid lines) and the \lattimer{} model (dashed lines). Also shown as dots with corresponding colors are early kilonova observations of GW170817 as compiled in \citet{villar2017Combined}.}
  \label{fig-knlc-data}
\end{figure}

Given the high photospheric temperatures, the kilonova precursor emission is most prominent in the U-V/U band. The binary configurations explored here are compatible with the binary parameters inferred for GW170817 (Sec.~\ref{sec-simusetup}). Figure \ref{fig-knlc-data} shows the predictions of our kilonova model (Sec.~\ref{sec:kilonova_model}) for the UV/U/B bands based on the \sfho{} and \lattimer{} runs when applied at the distance of GW170817 of $\approx\!41$\,Mpc, together with kilonova observations in these bands starting at 11\,h after merger as compiled in \citet{villar2017Combined}. Our results suggest that if earlier observations could have been conducted, the neutron precursor signal would likely have been detected. Based on our modeling, we find that the $\approx 2\times 10^{-5}\,M_\odot$ neutron precursor generated by the binary systems typical of Galactic double neutron-star systems considered here can be detected with planned missions such as the wide-field ULTRASAT satellite \citep{sagiv2014science}\footnote{\url{https://www.weizmann.ac.il/ultrasat/}} out to $\lesssim\!250$\,Mpc, assuming a $5\sigma$ limiting magnitude of $\approx\!23$ for a 1\,h target-of-opportunity integration at a wavelength of 220--280 nm.

The double-peaked structure in the UV/U/B bands ($\lambda=268,365,445$ nm, respectively) evident here is due to the neutron precursor and conventional r-process heating, respectively. Figure \ref{fig-knlc-data} illustrates that the relative strength of the first and second peak (i.e. neutron precursor to dynamical r-process ejecta) encodes information about the softness of the EOS, which might prove useful to place additional EOS constraints in future merger events. As also evident from Fig.~\ref{fig-knlc-data} (cf.~the short peak timescales), the dynamical ejecta in our runs is not massive enough to explain the blue kilonova data in GW170817 \citep{just2022dynamical}. This is consistent with previous conclusions that the blue kilonova emission in GW170817 likely requires a substantial contribution from post-merger ejecta \citep{siegel2019GW170817,metzger2020Kilonovae}, such as a magnetar-accelerated neutrino-driven wind \citep{metzger2018Magnetar,mosta2020Magnetar, curtis2021Process}.

\subsection{Impact of highly ionized lanthanides on early kilonova emission}
\label{sec:ionization_lanthanides}

Recent opacity calculations by \citet{banerjee2022opacity} point out that high ionization states of lanthanide elements reached at temperatures $\approx\!80000$\,K can boost the opacity by a factor of 10--100, up to $\approx 10^3\,\text{cm}^2\,\text{g}^{-1}$. If this temperature regime is realized in the early evolution of merger ejecta $\sim\!1$\,h that contains only a small fraction of lanthanides, \citet{banerjee2022opacity}  show that the boosted opacity can significantly dim the early $\sim\!1\,\text{h}$ kilonova lightcurve. This affecting, in particular, the early UV brightness is a potential concern for detection prospects of the neutron precursor emission as discussed in the previous sections.

\begin{figure}[tb!]
  \centering
  \includegraphics[width=1\columnwidth]{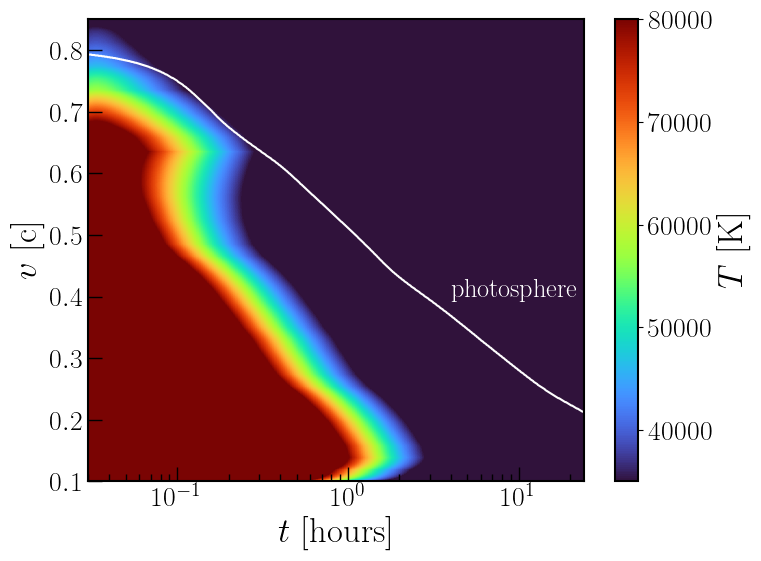}
  \caption{Gas temperature of our ejecta structure as a function of velocity coordinate and time for the \sfho{} model. The white line represents the location of the photosphere, indicating that the local gas temperature stays below the temperature regime of $\approx 80000$\,K for an opacity boost due to highly ionized states of lanthanides \citep{banerjee2022opacity}.}
  \label{fig:temp_ph}
\end{figure}

Figure \ref{fig:temp_ph} shows the gas temperature of our ejecta structure in the \sfho{} case as a function of velocity coordinate as well as time. Similar behavior applies to our other merger simulation runs. We reconstruct the gas temperature in each ejecta velocity shell $i$ with mass $dM_i$ and volume $V_i$ from the kilonova model (Sec.~\ref{sec:kilonova_model}) assuming contributions of radiation and ions to the internal energy $E_i$,
\begin{equation}
    E_i = a V_i T^4 + \frac{3}{2} \frac{dM_i}{\langle A \rangle m_u} k_{\rm B} T,
\end{equation}
where $a$ is the radiation constant, $\langle A \rangle$ is the average mass number as obtained from the nucleosynthesis calculations (Sec.~\ref{sec-nucleosyn}), and $m_u$ is the atomic mass unit (see also \cite{just2022dynamical}). In agreement with the model calculations of \citet{banerjee2022opacity}, we find a 70000\,K temperature regime at low velocity coordinates $v\sim0.2$\,c and early times $\lesssim\!1$\,h, in which high ionization states of lanthanides can be realized. However, the photosphere in our merger ejecta resides at much lower gas temperatures at all times, leading us to conclude that the opacity boost from highly ionized lanthanides is unlikely to significantly affect the early kilonova emission in practice. 

\subsection{Impact of ``lanthanides pockets'' on the kilonova and its nebular phase}
\label{sec:lanthanide_pockets}

\review1{At late times, the kilonova reaches a \textit{nebular phase}, in which the ejecta material becomes optically thin and the electromagnetic emission driven by radioactivity is dominated by emission lines \citep{hotokezaka2021nebular}. The spectral signatures of this late-time phase can contain unique information about atomic species synthesized by neutron capture.}

\begin{figure}[tb!]
  \centering
  \includegraphics[width=1\columnwidth]{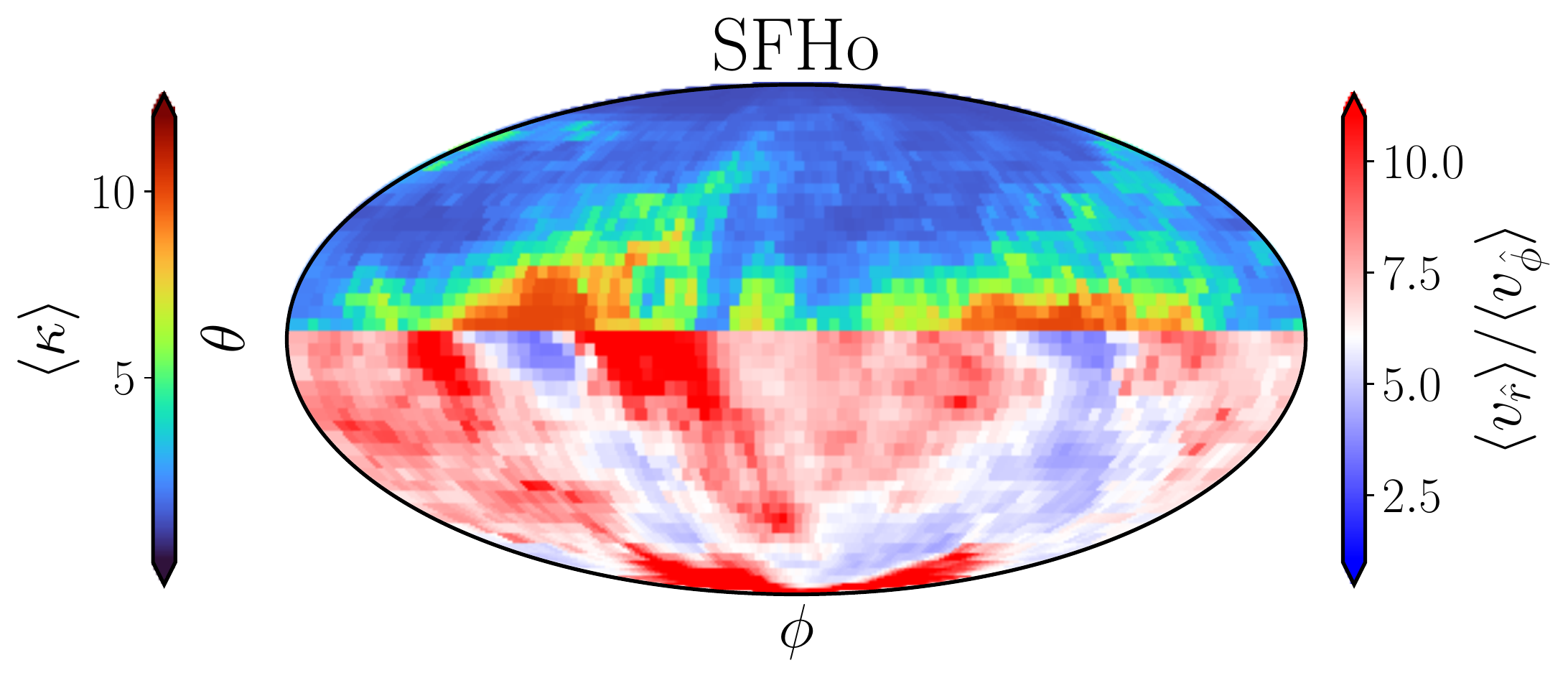}
  \caption{Upper part: Skymap of mass-weighted opacity for unbound dynamical ejecta in soft EOS runs (here: \sfho{}), computed using the $Y_{\rm e}$ distribution extracted from the simulation at a distance of ${\approx}440$\,km from the merger site and using Eq.~\eqref{eq:kappa_Ye}. For soft EOS a latitudinally and azimuthally dependent opacity profile emerges due to inhomogeneous shock heating/reprocessing of tidal ejecta (Sec.~\ref{sec:Ye_shock_reprocessing}), giving rise to `lanthanide/actinide pockets', which might lead to interesting observational consequences both in the early kilonova lightcurve and the late-time nebular phase. \review1{Lower part: Corresponding skymap of mass-weighted quotient of radial and azimuthal velocities.}}
  \label{fig-skymap-kappa}
\end{figure}

\begin{figure}[tb!]
  \centering
  \includegraphics[width=1\columnwidth]{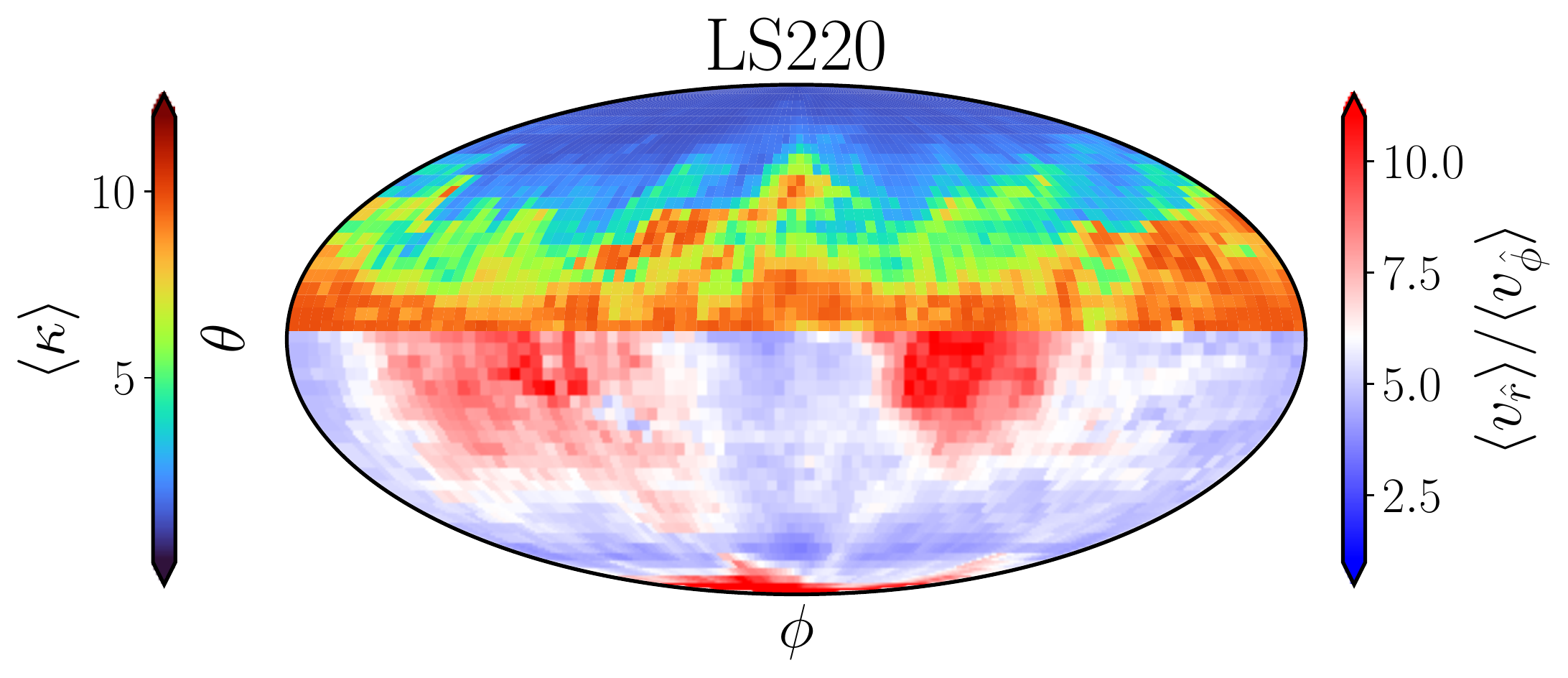}
  \caption{\review1{Same as Figure \ref{fig-skymap-kappa}, but for the stiff \lattimer{} EOS simulation. The opacity distribution is more uniform in the azimuthal direction and shows the expected latitudinal dependence.}}
  \label{fig-skymap-kappa-ls220}
\end{figure}

\review1{Our simulations show that the azimuthal distribution of the electron fraction for soft EOSs is non-uniform because of the preferred direction of core bounces (see Sec.~\ref{sec:Ye_shock_reprocessing} and Fig.~\ref{fig-angularYe}). The ejected material contains ``pockets'' of neutron-rich material that synthesizes lanthanides and actinides. The opacity of these pockets is thus significantly higher than the surrounding ejecta, which has undergone stronger heating due to shocks during the dynamical merger phase. The resulting ``opacity pockets'' are illustrated in the upper panel of Figure \ref{fig-skymap-kappa}, extracted at a distance of ${\approx}440$\,km from the merger site. For the stiff EOS, these lanthanide pockets are absent and the opacity distribution is more uniform, see Figure \ref{fig-skymap-kappa-ls220}.}

\review1{
If the structure of these pockets is preserved to larger distances (late times), this will have an impact on the observable properties of the nebular emission since the dominant nebular lines are determined by elements that dominate the cooling of the plasma. We estimate angular spreading of the ejecta based on the mass-weighted radial ($\langle v_{r}\rangle$) and azimuthal ($\langle v_{\phi}\rangle$) velocities extracted at ${\approx}440$\,km, as shown in the lower panel of Fig.~\ref{fig-skymap-kappa} in the case of \sfho{}. In general, we observe that radial velocities in these pockets dominate by at least a factor of ${\approx}3-5$ over azimuthal velocities, even though this low-$Y_{\rm e}$ component arises mostly from the tidal tails. For surrounding material this ratio is larger, typically ${\gtrsim}6-10$. As a result of the ratio $\langle v_{r}\rangle$/$\langle v_{\phi}\rangle$, material from within the pockets and from the surroundings will azimuthally spread and tend to disperse over time. A rough estimate for the angular dispersion of opacity structures in Fig.~\ref{fig-skymap-kappa} is given by
\begin{equation}
    \Delta \phi \approx \frac{d_\phi (t)}{r(t)} \sim \frac{\langle v_{\phi}\rangle}{\langle v_{r}\rangle} \lesssim \left(\frac{1}{5}-\frac{1}{3}\right) \approx (10-20)^\circ, \label{eq:pocket_dispersion}
\end{equation}
where $d_\phi(t)$ and $r(t)$ denote azimuthal and radial distance traveled by the ejecta, respectively. It is therefore plausible that such structures may persist at timescales of interest for both early photospheric ($\sim$days) and late-time ($\sim$weeks) nebular emission.
}

\subsection{Kilonova afterglows}
\label{sec:afterglows}

As the mildly relativistic dynamical ejecta expands into the interstellar medium (ISM), it sweeps up ISM material and generates a long-lived blast-wave \citep{nakar2011Radio, piran2013Electromagnetic, margalit2015Radio, hajela2022Evidence, hotokezaka2018Synchrotron}. Within the shock, randomly oriented magnetic fields are generated, amplified, and particles, mainly electrons, are accelerated to non-thermal distributions and generate synchrotron emission \citep{sari1998Spectra}. In this section, we present a model to calculate the dynamics of the shock propagation and the generation of non-thermal radiation directly based on the results of numerical relativity simulations.

\subsubsection{Shock dynamics} 
\label{sec:afterglows_shock_dynamics}

To describe the hydrodynamical propagation of the blast wave the ejecta runs into the ISM, we assume the ejecta has entered homologous, quasi-spherical expansion with a mass profile $M(v)$ as, e.g., in Fig.~\ref{fig-histogram}. We assume this shock sweeps up ISM material, which remains concentrated in a thin slab close to the shock front, where most of the electrons are accelerated. If we assume that the shock is adiabatic, and the EOS of the fluid is trans-relativistic \citep{mignone2007Equation, nava2013Afterglow, vaneerten2013Gammaray}, the evolution of the outermost ejecta with velocity $\beta$ (in units of $c$) and associated Lorentz factor $\Gamma$ is determined by energy conservation (see, e.g., \citealt{vaneerten2013Gammaray, ryan2020GammaRay} for more detailed discussion), and the velocity $\dot{R}$ of the shock at radial position $R$ from the merger site is determined by the Rankine-Hugoniot jump conditions,
\begin{equation}
\frac{\dot{R}}{c} = \frac{4 \Gamma u}{4u^2+3},
\label{eq-shockvel}
\end{equation}
where $u = \Gamma \beta$ is the four-velocity. 

\begin{figure}[tb!]
  \centering
  \includegraphics[width=1\columnwidth]{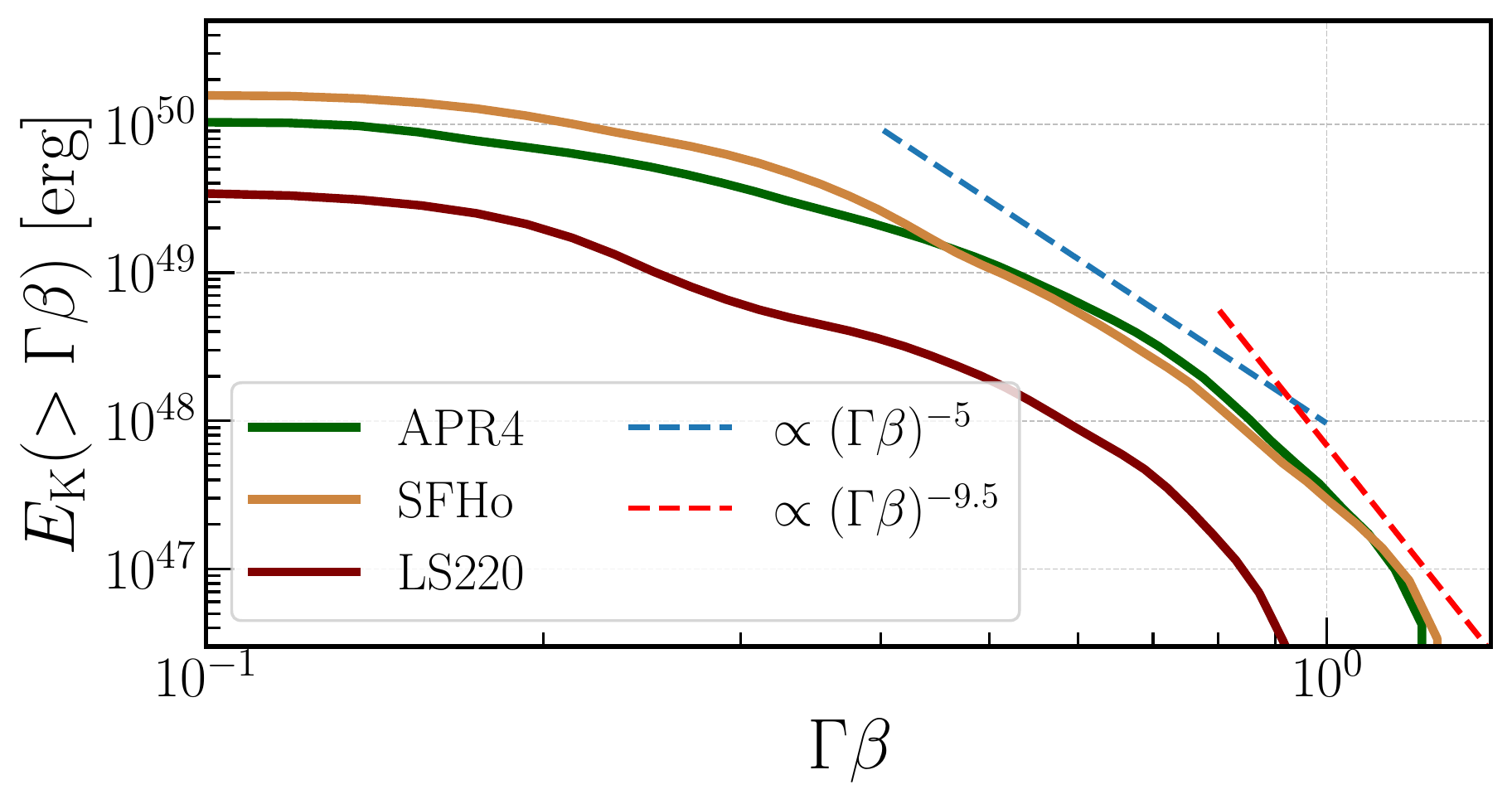}
  \caption{Cumulative kinetic energy of the dynamical ejecta as a function of four-velocity $u=\Gamma \beta$ for the three BNS merger simulation runs considered here.}
  \label{fig-ek-ag}
\end{figure}

In a merger event, the outflow has a radial velocity structure (Fig.~\ref{fig-histogram}). As the fastest part of the outflow starts to decelerate because of its interaction with the ISM, velocity shells deeper within the ejecta inject kinetic energy into the shock region, leading to a so-called refreshed shock \citep{panaitescu1998Multiwavelength, rees1998Refreshed}. In this case, conservation of energy is expressed by
\begin{equation}
E_{\rm K}(>\!u) = (\Gamma-1) M_0 + \frac{4}{9} R^3 \rho_{\rm ISM} \frac{u^2(4u^2+3)}{1+u^2},
\label{eq-shocken}
\end{equation}
where $E_{\textrm{K}}(>\!u)$ is the kinetic energy of the flow extracted from simulations (see Fig.~\ref{fig-ek-ag}), $M_0 = M(u_{\rm max})$ is the mass of the outer-most part of the fluid (we impose a mass cut-off of $M_0=10^{-8} \msun$), and the second term is the kinetic plus thermal energy of the shocked ISM material that has been swept up by the blast wave. Here, $\rho_{\rm ISM}$ is the constant density of the ISM. The fluid velocity as a function of shock position $R$ can be obtained from the non-linear algebraic Eq.~\eqref{eq-shocken}, and then the shock Lorentz factor $\Gamma_{\rm sh}(R)$ can be found using Eq.~\eqref{eq-shockvel}. We note that in the non-relativistic limit, Eq.~\eqref{eq-shocken} reduces to the solution used by \citet{piran2013Electromagnetic} and \citet{chevalier1982Selfsimilar}.

\begin{figure}[tb!]
  \centering
  \includegraphics[width=1\columnwidth]{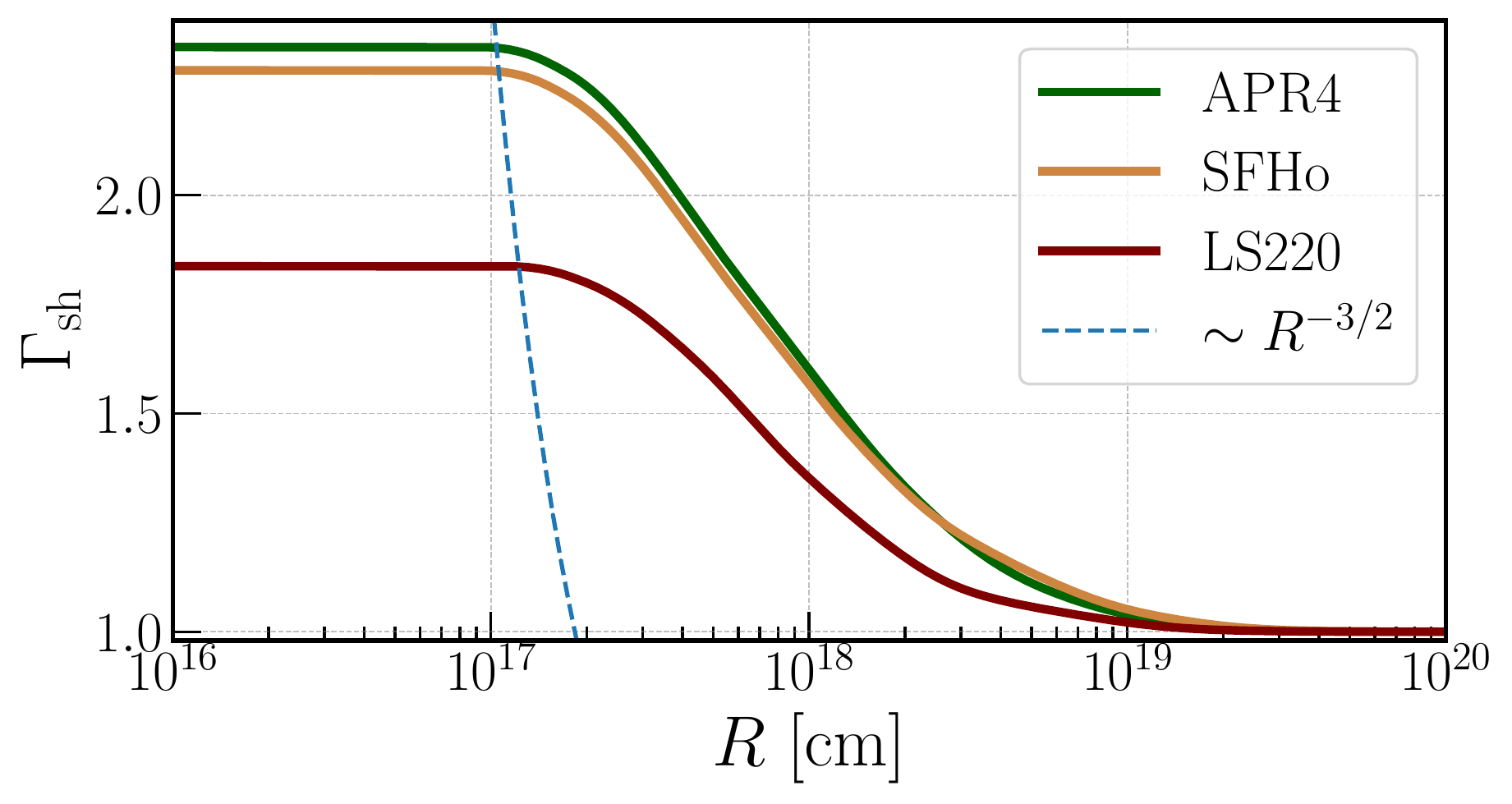}
  \caption{Evolution of the Lorentz factor $\Gamma_{\rm sh}$ of the shock driven into the ISM (with density $n=0.001\,\text{cm}^{-3}$) by the dynamical ejecta as a function of radial distance to the merger site and for the three different BNS mergers simulated here. The shock waves decelerate more slowly than a single-shell model (dashed line; $\Gamma_{\rm sh}\propto R^{-3/2}$) due to the radially structured ejecta refreshing the shock wave at late times.}
  \label{fig-gammash}
\end{figure}

Employing an ISM density of $n = \rho_{\rm ISM}/m_p = 0.001\,\rm cm^{-3}$, where $m_p$ is the proton mass, we compare the evolution of the shock Lorentz factor $\Gamma_{\rm sh} = \left[1-(\dot{R}/c)^2\right]^{-1/2}$ between the different simulation runs as a function of $R$ in Fig.~\ref{fig-gammash}. The (forward) shock propagates at mildly relativistic velocities at early times, with differences of tens of percent between different EOSs and larger Lorentz factors for softer EOS. The shock waves become Newtonian ($\Gamma_{\rm sh}\approx 1$) once reaching their respective Sedov radius, defined as the radius $R_{\rm d}$ within which the rest mass energy of the material within the blast wave equals the energy $E_{\rm K}$ of the explosion,
\begin{eqnarray}
    R_{\rm d} &\approx& \left(\frac{E_k}{(4\pi/3)n m_p c^2}\right)^{1/3} \label{eq:R_d}\\ 
    &\approx& 2.5\times 10^{18}\,\text{cm}\, \left(\frac{E_{\rm K}}{10^{50}\,\text{erg}}\right)^{1/3}\left(\frac{n}{10^{-3}\,\text{g\,cm}^{-3}}\right)^{-1/3}, \nonumber
\end{eqnarray}
in good agreement with the behavior shown in Fig.~\ref{fig-gammash}. Here, we have evaluated the expression for typical isotropic equivalent energies of the dynamical ejecta of our simulations (Fig.~\ref{fig-ek-ag}). The shock waves decelerate more gradually than expected from a single-shell scenario $(\Gamma_{\rm sh}\propto R^{-3/2})$ due to the fact that the shock is continuously being refreshed by kinetic energy of shells deeper within the ejecta that catch up with the decelerating shock front.

\subsubsection{Non-thermal radiation} 
\label{sec:non-thermal radiation}

We assume that a fraction $\epsilon_e$ of the thermal energy $e=4(\Gamma-1)\Gamma n m_p c^2$ generated in the forward shock at time $t$ accelerates electrons into a power-law distribution in Lorentz factor $\gamma_e$ in the frame comoving with the shock \citep{sari1998Spectra}, 
\begin{equation}
    Q(\gamma_e,t)=Q_0 \gamma^{-p}_e \Theta(\gamma_e; \gamma^{\rm min}_e, \gamma^{\rm max}_e).
\end{equation}
Here, $Q_0$ is a time-dependent normalization factor, $p$ is the power-law index, and $\Theta$ denotes the Heaviside function that restricts the injection to an interval between the minimum energy given by normalization constraints, 
\begin{equation}
    \gamma^{\rm min}_e= \epsilon_e \frac{(p-2)}{(p-1)} \frac{m_p}{m_e} (\Gamma-1),
\end{equation}
and a maximum energy given by balancing acceleration and synchrotron losses \citep{de1996gamma}
\begin{equation}
    \gamma^{\rm max}_e= \sqrt{\frac{6 \pi e \epsilon_{\rm acc}}{\sigma_T B}}.
\end{equation}
Here, $e$ is the electron charge, $\sigma_T$ is the Thomson cross-section, $B$ is the co-moving magnetic field strength, and $\epsilon_{\rm acc}$ is the ratio of acceleration rate to the maximum possible particle energy-gain rate that we fix to $0.35$ following \citet{zhang2020Inverse}. Furthermore, we assume a fraction $\epsilon_B$ of the instantaneously generated post-shock thermal energy is transformed into magnetic energy \citep{sari1998Spectra}, which results in a co-moving field field given by 
\begin{equation}
    B  = \sqrt{8 \pi (\Gamma -1) (4 n m_p \Gamma c^2) \epsilon_{\rm B}}.
\end{equation}

The evolution of the electron distribution $n_e(\gamma_e,t)$ in the shock front is determined by the continuity equation for particle number. We numerically solve the full resulting Fokker-Planck equation for the distribution $n_e(\gamma_e,t)$ in the comoving frame of the shock, 
\begin{eqnarray}
    \frac{\partial n_e(\gamma_e,t)}{\partial t} &=& 
    -\frac{\partial}{\partial \gamma_e} [\dot{\gamma_e}(\gamma_e,t) n_e(\gamma_e, t)] \nonumber\\
    &&+ Q(\gamma_e,t) - \frac{n_e(\gamma_e,t)}{t_{\rm esc}}.
\end{eqnarray}
Here, $t_{\rm esc}$ is the escape time of particles from the shock layer and $\dot{\gamma_e}$ represents the energy gains/losses determined by various radiative processes. The comoving time $t$ here is related to the evolution of the shock as seen in the inertial frame of the merger site with time $t'$ by $t = \int_0^{\rm t'} dt' \Gamma^{-1}_{\rm sh}(t')=\int_0^R dR/(\Gamma^{-1}_{\rm sh} v_{\rm sh})$. To solve this equation numerically and to compute the radiation fluxes we use the code \texttt{Paramo} \citep{rueda-becerril2021Numerical}, taking into account synchrotron and inverse-Compton cooling, together with synchrotron self-Compton/self-absorption, as well as all relativistic effects required for computing fluxes in the observer frame.

Most studies of non-thermal afterglows from BNS mergers (e.g., \citealt{piran2013Electromagnetic,hotokezaka2015Mass,hotokezaka2018Synchrotron}) employ analytic approximations for the synchrotron spectrum, e.g., a simply connected power law as a function of frequency \citep{sari1998Spectra}. Here, we instead numerically obtain the SED solving for the particle distribution, taking into account self-consistently radiative cooling due to inverse Compton scattering, and the full synchrotron emissivities (for details on the numerical methods, see \citealt{rueda-becerril2021Blazar,rueda-becerril2021Numerical,davis2022Balancing}). While providing more accurate predictions for non-thermal emission in general, this approach, in particular, replaces the need to rely on order-of-magnitude estimates for the cooling frequencies, which critically shape the synchrotron light curves \citep{hotokezaka2018Synchrotron}.

In order to take the angular dependence of the ejecta structure and viewing angle effects into account, we follow the approach of \citet{margalit2015Radio}. We assume an axisymmetric outflow and discretize the outflow into angular bins according to polar angle $\Theta$, equidistantly spaced in $\cos\Theta$ 
, each bin $i$ having its kinetic energy distribution $E_{{\rm K},i} (u, \Theta_i)$ (cf.~Fig.~\ref{fig-kin-ang}). We solve the shock and Fokker-Planck equations for each angular bin separately, assuming no lateral spreading, and subsequently add the emergent radiation from each annulus in solid angle in the observer frame. Angular variation in the ejecta profiles can shift the peak time of the light curves, although in our equal-mass models the ejecta is sufficiently isotropic for this effect not to be significant \citep{margalit2015Radio}.

The free parameters of our model are 
\begin{equation}
    \{ p, \epsilon_e, \epsilon_B, n, D_L, z, \theta_{\rm obs}, u_{\rm rad} \},
\end{equation}
where $D_L$ is the luminosity distance to the merger site, $z$ the corresponding cosmological redshift, $\theta_{\rm obs}$ the observer angle relative to the orbital plane, and $u_{\rm rad}$ the external photon field density relevant for inverse Compton scattering. As a fiducial photon field, we use the thermal photons of the Cosmic Microwave Background, with $u_{\rm rad}=0.26$\,eV\,cm$^{-3}$. 

As a baseline test of our afterglow code, we consider a spherically symmetric shock powered by ejecta following a power-law distribution in velocity focusing on synchrotron emission only. Comparing to the public code \texttt{afterglowpy} \citep{ryan2020GammaRay}\footnote{https://github.com/geoffryan/afterglowpy} we find very good agreement regarding the synchrotron emission.

\subsubsection{Kilonova Afterglow Light Curves and GW170817}
\label{sec:kilonova_afterglows_GW170817}

\begin{figure}[tb!]
  \centering
  \includegraphics[width=1\columnwidth]{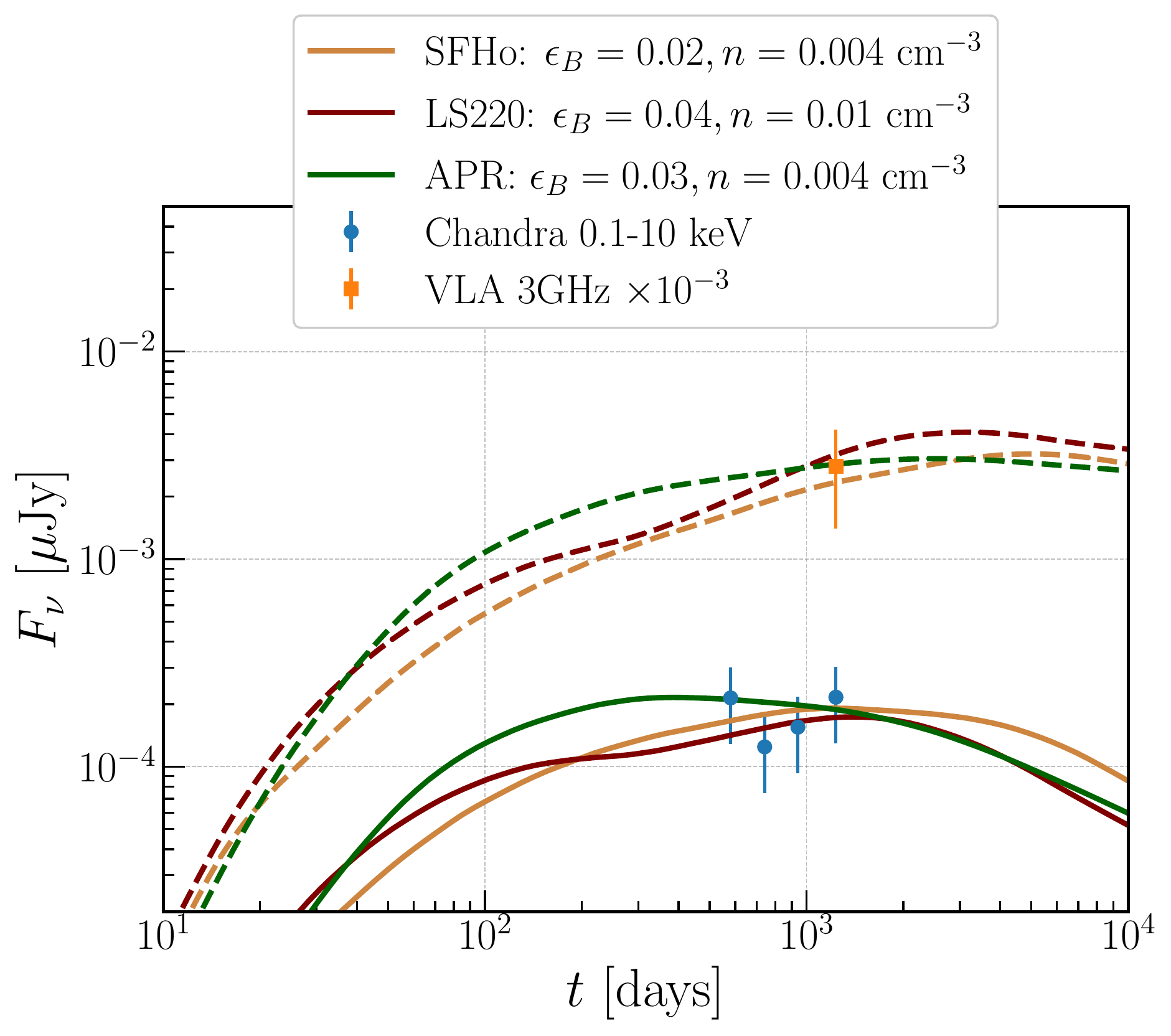}
  \caption{Representative kilonova non-thermal afterglow light curves generated by the dynamical ejecta of our BNS simulation runs, labeled by the corresponding EOSs employed. Also shown are X-ray (blue dots) and radio (orange dot) observations reported by \citet{hajela2022Evidence}, who present tentative evidence for X-ray and radio rebrightening following the GW170817 event. Our models are consistent with the observations for typical microphysical parameters (see the text for details).}
  \label{fig-ag-lc}
\end{figure}

We apply our non-thermal afterglow model described above to the observations of \citet{hajela2022Evidence}, who present evidence for a late-time ($\approx\!1000$\,days post-merger) excess of X-ray and radio emission above the gamma-ray burst broad-band afterglow of GW170817. The existence of this late-time rebrightening is still a matter of debate \citep{balasubramanian2022, connor2022continued} and, if present, could have other possible origins as well (e.g., \citealt{metzger2021neutrino}).

We fix $\epsilon_e$ to a standard value of $\epsilon_e = 0.1$ and, as usual, we vary the microphysical parameters $(p, n, \epsilon_B)$. We find that our models directly based on numerical relativity simulation data are compatible with the observational data obtained so far over a range of typical parameter values; Fig.~\ref{fig-ag-lc} shows representative models for each EOS model. The stiffer \lattimer{} simulation generates less energetic outflows and thus tends to prefer a somewhat higher density of the ISM as well as higher magnetic energy compared to the softer EOSs in order to reach the same observed fluxes at $\approx\!1230$\,days. The properties of peak flux (timescale and flux level) sensitively depend on the ISM density and the velocity of the fastest (outermost) ejecta shell. We observe that the \apr{} run with the fastest ejecta profile tends to peak earliest, although all simulations considered here produce overall similar results that are consistent with the data. 

Due to the mildly relativistic velocity structure of the outflow, the light curves show long peak timescales of thousands of days and associated broad peaks. The X-ray fluxes almost plateau around the peak timescale between $\sim\!200$ to $\sim\!2000$ days, before decreasing again. As also evident from Fig.~\ref{fig-ag-lc}, the X-ray light curves tend to have a faster rise and decay timescale than the radio signal, which indicates that the cooling frequency has crossed the X-ray band by $\sim\!2000$\,days. The radio emission plateaus for several years.

Compared to \citet{nedora2021Dynamical} who use a large set of numerical simulations and a semi-analytical afterglow model, we find that our model is compatible with similar ISM densities but tends to require higher magnetic energies. This is because the extraction of asymptotic ejecta velocities in our simulations leads to overall slower speeds (see Sec.~\ref{sec:ejecta_dynamics}). We also find that our model favors a low spectral index, $p \leq 2.1$, in agreement with \citet{hajela2022Evidence}. Finally, we note that at this early stage of what may represent excess emission, the gamma-ray burst afterglow may still be contributing in part to the observed light curve (not included in the emission model here), and thus the derived microphysical parameters would likely need further adjustment if the gamma-ray burst afterglow component was included self-consistently in this analysis.

\subsubsection{High-energy $\gamma$-ray emission}
\label{sec:gamma_rays}

Given the mildly relativistic nature of the shock that the ejecta drives into the ISM with Lorentz factor $\Gamma_{\rm sh} \approx 2$ (cf.~Fig.~\ref{fig-gammash}), and the low-density environment of photons and baryonic matter, gamma-ray emission from inverse Compton and synchrotron self-Compton processes in these types of afterglows is not expected to be detectable unless the merger event occurs sufficiently nearby and/or the efficiency $\epsilon_e$ of converting kinetic energy of the shock into accelerating electrons is very high \citep{takami2014high}.

However, if the merger occurs within a medium of high radiation density of soft photons $u_{\rm rad} \sim 10-10^3$\,eV\,cm$^{-3}$, such as in globular clusters \citep{venter2009predictions}, for example, there may be noticeable emission of gamma-rays generated by inverse Compton scattering. For $u_{\rm rad}\sim\! 10^3$\,eV\,cm$^{-3}$, we find fluxes of $\sim$\,TeV gamma-rays of $\nu F_{\nu} < 10^{-14}$\,erg\, s$^{-1}$\,cm$^{-2}$ for $D_L > 40$ Mpc. This still is too faint to be detected with the Cherenkov Telescope Array (CTA) or $10$\,years of \textit{Fermi} Large Area Telescope (LAT) data \citep{atwood2009large}. However, even in the absence of direct detection of gamma-rays, inverse Compton processes may shape the observed non-thermal afterglow emission by helping cool electrons at later times. We find that the resulting decrease in X-ray emission is significant if $u_{\rm rad} \gtrsim 10^2$\,eV\,cm$^{-3}$ at timescales of $t\gtrsim 1$ year.

Furthermore, at early times ($<\!10-15$\,d post-merger) the ejecta shock is accompanied by thermal radiation from the kilonova with radiation density
\begin{eqnarray}
    u_{\rm rad} &=& \frac{L_{\rm bol}}{4\pi R^2c}\frac{1}{\Gamma_{\rm sh}^2}\\
    &\gtrsim& 8.7\times 10^{5} \frac{\text{eV}}{\text{cm}^{3}}\,\left(\frac{L_{\rm bol}}{6\times 10^{40}\text{erg\,s}^{-1}}\right)\left(\frac{t}{10\,{\rm d}}\right)^{-3.33}\mskip-40mu.\nonumber
\end{eqnarray}
Here, we have scaled to the bolometric luminosity of the GW170817 kilonova of $\approx\!10^{42}$\,erg\,s$^{-1}$ at $0.5$\,d \citep{drout2017Light}, rescaled according to $\propto t^{-0.85}$ up to 5.5\,d and $\propto t^{-1.33}$ thereafter \citep{drout2017Light}, and have used a typical shock Lorentz factor of $\Gamma_{\rm sh}\lesssim 2$ (cf.~Fig.~\ref{fig-gammash}). In a similar manner as in the case of the gamma-ray burst afterglow shock \citep{linial2019cooling}, electrons in the kilonova ejecta shock interact with the low-energy kilonova photon field and cool by inverse Compton scattering. We find that the X-ray radiation of the ejecta shock may decrease by tens of percent at very early times ($<10-15$\,d), provided strong kilonova afterglow emission due to sufficiently fast ejecta.

\section{Conclusions}
\label{sec-conclusions}

We have presented a comprehensive model of electromagnetic signatures arising from the dynamical ejecta in BNS mergers. For this purpose, we performed GRMHD simulations that incorporated, for the first time, tabulated three-parameter nuclear EOS, magnetic fields, and approximate neutrino transport via a ray-by-ray transport scheme. We provide a detailed analysis of the characteristics of dynamical ejecta and its ejection mechanisms. We employ the dynamical ejecta profiles extracted from these simulations, characteristics of the dynamical ejecta as recorded by several families of tracer particles, and results of detailed nucleosynthesis calculations with the nuclear network \texttt{SkyNet} to compute the neutron-precursor thermal transient, kilonova emission, as well as the broad-band non-thermal kilonova afterglow. The modeling of electromagnetic counterparts presented here is directly based on ab-initio numerical simulations of the BNS merger process and provides a comprehensive picture of observable phenomena related to dynamical ejecta in a self-consistent framework.

Our main results and conclusions can be summarized as follows:

\begin{itemize}

\item We performed a detailed analysis of the dynamical ejecta properties and ejection mechanisms of 1.35--1.35\,$M_\odot$ binaries typical of the distribution of Galactic BNS systems and consistent with the inferred system parameters of GW170817 using three nuclear EOSs that span the range of NS radii as allowed by current observational constraints (Fig.~\ref{fig-tov}; Tab.~\ref{tab:sim_properties}). We showed that matter is unbound during the decompression phase in quasi-radial bounces of the remnant NS double-core structure. We find that for the soft EOS cases, $\sim\!75\%$ of dynamical ejecta is unbound in the first three bounces, while for the stiffer \lattimer{} case, material is unbound more secularly in a combination of torques and shock waves sourced by somewhat weaker bounces. 

\item All binaries generate a fast tail of material ($v>0.6c$) that can be traced back to ejection during the first two bounces of the double-core structure (Fig.~\ref{fig-massflux}). While the amount of such fast material ($\text{few}$ $10^{-6}\,\msun$, Tab.~\ref{tab:sim_properties}) is in good agreement with some grid-based GRHD simulations of ultra-high resolution \citep{kiuchi2017Subradianaccuracy,hotokezaka2018Synchrotron}, differences to other grid-based GRHD simulations \citep{nedora2021Dynamical} can be understood in terms of the criteria to extract unbounded mass and to compute the asymptotic velocities (see Sec.~\ref{sec:ejecta_dynamics}). We conjecture that based on these results and convergence studies \citep{dean2021Resolving}, the amount of fast ejecta should be roughly converged at our grid resolution and atmosphere level. Remaining differences to Newtonian setups and SPH simulations \citep{dean2021Resolving,kullmann2021Dynamical,rosswog2022thinking} remain an open question, but may likely be attributable to differences in the overall setup, such as Newtonian dynamics, velocity extraction, geometric effects due to assumed symmetries, and not sufficiently converged ejecta tails in SPH simulations with $\lesssim\!\text{tens}$ of particles (see the discussion in Sec.~\ref{sec:ejecta_dynamics}).

Although our present GRMHD setup includes neutrino absorption and employs a lower atmosphere floor than previous grid-based GRHD simulations with tabulated EOS, we are still not sensitive to a $\sim\!10^{-8}-10^{-7}M_\odot$ ultra-relativistic ejecta envelope as suggested by \citet{beloborodov2020relativistic}, who employ this envelope as a breakout medium for a relativistic jet to explain the prompt gamma-ray emission of GRB 170817A associated with the BNS merger event GW170817.

\item We showed that for soft EOSs, the second quasi-radial bounce after merger produces a shock wave that reprocesses the cold, previously ejected neutron-rich tidal tail material, resulting in an asymmetric, azimuthal pattern in $Y_{\rm e}$, giving rise to `pockets' of neutron-rich material while other angular parts are shock-heated to higher $Y_{\rm e}$ values (Sec.~\ref{sec:Ye_shock_reprocessing}, Figs.~\ref{fig-tri-yes} and \ref{fig-angularYe}). For stiffer EOS, however, the action of multiple weaker shock waves generates a more homogeneous $Y_{\rm e}$ pattern as a function of azimuthal angle. Once r-process nucleosynthesis proceeds the low-$Y_{\rm e}$ regions in the soft EOS cases translate into high-opacity `lanthanide/actinide' pockets. This is illustrated in Fig.~\ref{fig-skymap-kappa}. 

\item Future work will need to explore the observational consequences of the azimuthal opacity effect (Fig.~\ref{fig-skymap-kappa}) with radiative transfer calculations of the early kilonova emission. If preserved to large radii, these ``pockets'' could also significantly impact the nebular phase of the kilonova once the ejecta becomes transparent at late times. \review1{Indeed, the radial velocities of these pockets are larger than their azimuthal velocities by a factor of at least ${\sim}3-5$, translating into angular dispersion of $\lesssim(10-20)^\circ$ (see Eq.~\eqref{eq:pocket_dispersion}). This indicates that these structures could be preserved to large distances and thus both to the early photospheric ($\sim$days) and late-time nebular ($\sim$weeks) emission phases.}. The dominant nebular lines are determined by the elements that dominate the cooling \citep{hotokezaka2021nebular}, i.e. lanthanide/actinide-rich pockets may produce different nebular emission relative to other regions with equal mass but less lanthanide content. Such nebular composition diagnostics may be performed with the James Webb Space Telescope. \review1{This may provide an additional avenue to infer the softness/stiffness of the EOS}.


\item Our setup of several families of passive tracer particles injected into the simulation (Sec.~\ref{sec:tracers}) allows us to track and characterize different ejecta components and to perform detailed nucleosynthesis calculations of the dynamical ejecta with the nuclear reaction network \texttt{SkyNet} (Sec.~\ref{sec-nucleosyn}). We obtain a mass-averaged abundance pattern in good agreement with solar abundances for $A>80$, including, in particular, a robust pattern between and including the 2nd and 3rd r-process peak irrespective of the EOS (cf.~Fig.~\ref{fig-abun}). An underproduction of first-peak elements relative to solar for $A <80$ evident here is expected due to limited possibilities to increase $Y_{\rm e}$ in dynamical ejecta via neutrino absorption and emission. Softer EOSs lead to slightly higher abundance in the $A = 50 -80$ region due to somewhat stronger shock heating and the resulting increase in $Y_{\rm e}$.

Directly post-processing with the nuclear reaction network, we find that in all simulations a typical amount of $\approx\!2\times 10^{-5}\,M_\odot$ of the fast-ejecta tail leads to free neutrons (cf.~Fig.~\ref{fig-fnhisto}; Tab.~\ref{tab:sim_properties}) whose decay provides early heating on the free-neutron decay timescale of $\approx\!900$\,s in excess to that of the r-process (cf.~Fig.~\ref{fig-qdot-vel}). We find that although stiffer EOS tend to generate less fast dynamical ejecta, a similar amount of free neutrons can be generated due to relatively more neutron-rich outflows.

\item We present a novel kilonova model that incorporates a detailed method for thermalization of radioactive decay products according to \citet{hotokezaka2020Radioactive}, heating due to free-neutron decay, and relativistic effects to reconstruct the observer lightcurve (Sec.~\ref{sec:kilonova_model}). 

\item We find that the combination of free-neutron heating and relativistic effects results in an enhancement of the kilonova emission by a factor $10-20$ at $\lesssim\!1$\,h for all EOS cases (cf.~Fig.~\ref{fig-knlc-eos}), with relativistic effects leading to noticeable enhancement up to $\lesssim\!10$\,h. A moderate shift in the peak timescale of the precursor emission due to relativistic effects to values closer to the neutron decay timescale of $\approx\!0.3$\,h complicates confusion with possible early cocoon heating \citep{gottlieb2018cocoon, kasliwal2017Illuminating}. We find that this modeling of early kilonova and neutron-precursor emission based on ab-initio numerical simulations is compatible with observations of GW170817 in that i) the main blue emission component of the GW170817 kilonova cannot be explained with dynamical ejecta only as found here and likely requires substantially more ejecta from the post-merger phase and ii) earlier observations in UV and optical bands would likely have detected the precursor emission (cf.~Fig.~\ref{fig-knlc-data}). The boost in UV/optical brightness by a factor of a few due to relativistic effects found here provides promising prospects for detection of the neutron precursor following future merger events with UV telescopes such as \emph{Swift} or with planned missions such as the wide-field ULTRASAT satellite \citep{sagiv2014science}. Based on our precursor modeling, we estimate a detection horizon of $\lesssim\!250$\,Mpc for ULTRASAT.

\item The ejecta structure in our simulations suggests that in typical merger systems such as those investigated here an opacity boost from highly ionized lanthanide nuclei arising at temperatures $\gtrsim\!70000$\,K as recently reported by \cite{banerjee2022opacity} is unlikely to affect the early UV/optical emission on timescales $\approx\!1$\,h (Sec.~\ref{sec:ionization_lanthanides}, Fig.~\ref{fig:temp_ph}).

\item Finally, we present a novel model to compute the non-thermal kilonova afterglow emission generated by a shock wave that the fast dynamical ejecta runs into the interstellar medium (Sec.~\ref{sec:afterglows}). We numerically solve the 1D hydrodynamics of a refreshed shock propagating into the interstellar medium and compute the synchrotron and inverse-Compton radiation by numerically solving the corresponding Fokker-Planck equation. We find that for typical shock microphysics parameters and typical densities of the interstellar medium inferred for GW170817, our model based on ab-initio numerical simulations is compatible with the recent data by \citet{hajela2022Evidence}, who report tentative evidence for a rebrightening of the broad-band GW170817 GRB afterglow emission emerging at a timescale of $10^3$\,days post-merger (Fig.~\ref{fig-ag-lc}). The presence of this rebrightening, however, is still being debated \citep{balasubramanian2021Continued,connor2022continued}.

\item High-energy gamma-ray emission arising from inverse-Compton and synchrotron self-Compton processes associated with the kilonova afterglow is unlikely directly observable in future merger events. However, we find that in the presence of intense photon fields (e.g., the kilonova at early times and/or in environments such as globular clusters) indirect consequences such as additional cooling of electrons leave observable imprints in the X-ray radiation (leading to a decrease by $\sim\!\text{tens}\,\%$ relative to the radio band).

\end{itemize}

\begin{acknowledgments}
We thank Vassilios Mewes, Federico Lopez Armengol, Michael M\"uller, Aman Agarwal, Dhruv Desai, Xinyu Li, Rodrigo Fern\'andez, Brian Metzger, and Santiago del Palacio for fruitful discussions and comments. We thank Jes\'us Rueda-Becerril and Geoffrey Ryan for discussions and for providing help with \texttt{Paramo} and \texttt{afterglowpy}. This research was enabled in part by support provided by SciNet (www.scinethpc.ca) and Compute Canada (www.computecanada.ca). DMS acknowledges the support of the Natural Sciences and Engineering Research Council of Canada (NSERC), funding reference number RGPIN-2019-04684. LC is a CITA National fellow and  acknowledges the support by the Natural Sciences and Engineering Research Council of Canada (NSERC), funding reference DIS-2022-568580. Research at Perimeter Institute is supported in part by the Government of Canada through the Department of Innovation, Science and Economic Development Canada and by the Province of Ontario through the Ministry of Colleges and Universities. 
\end{acknowledgments}

\software{The Einstein Toolkit (\citealt{loffler2012Einstein}; \href{http://einsteintoolkit.org}{http://einsteintoolkit.org}), \texttt{LORENE} \citep{gourgoulhon2001Quasiequilibrium},  \texttt{GRMHD\_con2prim} (\citealt{siegel2018Recovery}, \citealt{siegel2018soft}), \texttt{Skynet} \citep{lippuner2017SkyNet}, \texttt{Paramo} \citep{paramosoft},  \texttt{afterglowpy}  (\citealt{ryan2020GammaRay}), \texttt{WhiskyTHC} (\citealt{radice2012THC}, \citealt{radice2014HighOrder}, \citealt{radice2014Secondorder}),  \texttt{PyCactus} (\citealt{kastaun2021numerical}, \url{https://github.com/wokast/PyCactus}), \texttt{Matplotlib} \citep{hunter2007Matplotlib}, \texttt{NumPy} \citep{harris2020Array}, \texttt{SciPy} \citep{virtanen2020SciPy}, and \texttt{hdf5} \citep{hdf5}.}

\appendix

\section{Convergence of ejecta}
\label{app:convergence_tests}

In BNS merger simulations it is challenging to obtain a satisfactory convergence level of the total amount of dynamically ejected mass at finite (typical) numerical resolutions employed. This is in part due to non-linear effects involved in the ejection process (see, e.g., \citealt{radice2018Binary}) and the requirement for and treatment of an artificial atmosphere floor in grid-based simulations; another role is played by the methods of measuring the amount of unbound ejecta in a given simulation due to finite numerical resolution available for the numerical extraction techniques and ambiguities in unboundedness criteria. In this appendix, we explore how the properties of the dynamical ejecta change with resolution and other parameters such as the extraction radius of outflow detectors. We focus on the \sfho{} case and we compare results from simulations with a grid spacing of $\Delta x =180, 220, 250$\,m, which we label as high, medium, and low resolution, respectively.

\begin{figure}[tb!]
  \centering
  \includegraphics[width=0.45\columnwidth]{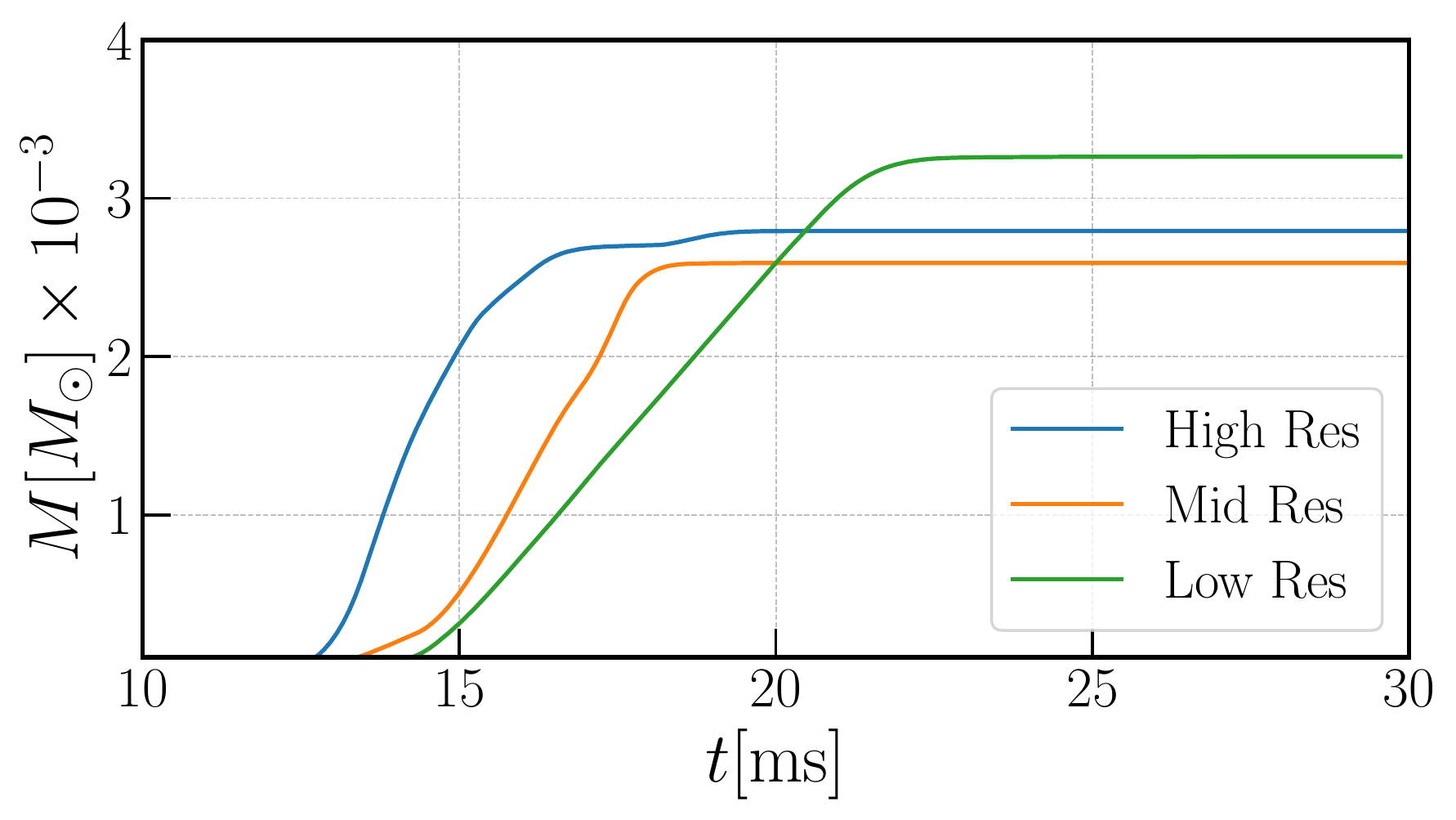}
   \includegraphics[width=0.45\columnwidth]{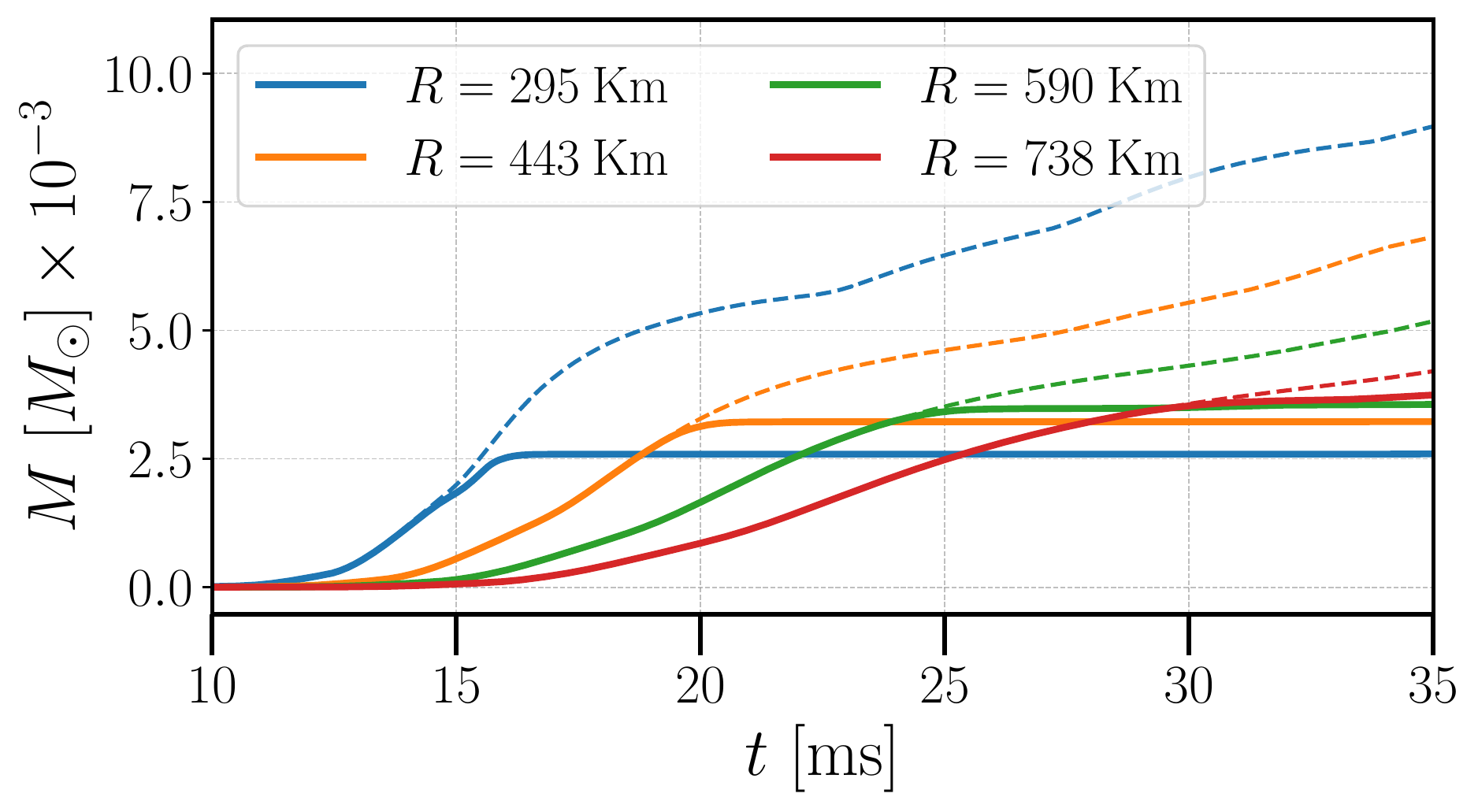}
  \caption{Convergence of total ejecta mass as a function of resolution and extraction radius. Left: Cumulative unbound ejecta mass as computed from the mass flux through a spherical detector surface at $\approx\!300$ km using the SFHo EOS simulations at different resolutions. Right: Cumulative unbound ejecta mass according to the geodesic criterion (thick lines) and the Bernoulli criterion (dashed lines) for \sfho{} at high resolution ($\Delta x = 180$\,m) and different detector radii.}
  \label{fig-conv-res}
\end{figure}

\begin{figure}[tb!]
  \centering
  \includegraphics[width=1\columnwidth]{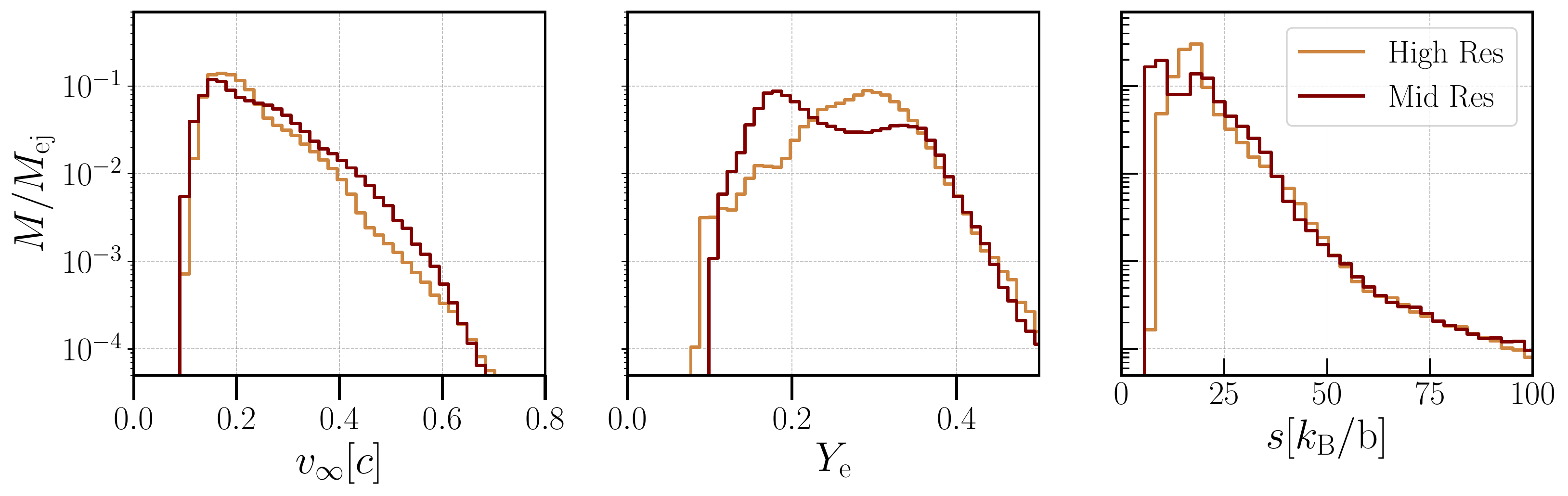}
  \caption{Convergence of detailed dynamical ejecta distributions in terms of asymptotic velocity (left), electron fraction (center), and specific entropy (right) based on the SFHo EOS simulations at different resolutions.}
  \label{fig-his-conv-res}
\end{figure}

The left panel of Fig.~\ref{fig-conv-res} shows a convergence test regarding the total amount of dynamically ejected mass using the geodesic criterion at a detector distance of 300\,km. We observe convergence at the 10\% level at the highest resolution employed. A comparison of the corresponding detailed ejecta mass distributions of the high and medium resolution runs according to asymptotic velocity, electron fraction, and specific entropy are shown in Fig.~\ref{fig-his-conv-res}. Except for moderate differences in the $Y_{\rm e}$ distribution around $Y_{\rm e}\approx0.25$, which we attribute to better or worse resolved shock heating, we find good agreement in the detailed ejecta distributions. \review1{The value $Y_{\rm e}\approx0.25$ being a threshold for lanthanide and actinide production (e.g., \citealt{lippuner2015RProcess}), these results emphasize the importance of well-resolved shock dynamics during the merger process.}

Finally, we compare the amount of unbound mass according to the Bernoulli and geodesic criteria for different detector radii based on the high-resolution run in the right panel of Fig.~\ref{fig-conv-res}. As discussed in Sec.~\ref{sec:ejecta_dynamics}, the unbound mass according to the geodesic criterion saturates at earlier times, because it only captures the relatively faster fluid elements that are dynamically unbound during the merger phase, while the Bernoulli criterion additionally captures slower material of high specific enthalpy unbound through winds during the post-merger phase, giving rise to a steadily growing unbound ejecta mass in the early post-merger phase. Early on, both criteria lead to identical ejecta masses, highlighting the dynamical nature of the underlying ejection process. We find almost indistinguishable variations in the total ejecta mass for detectors at radii larger than $\approx\!450$\,km using both the Bernoulli and the geodesic criterion, once accounting for a shift in arrival times of the outflow at the various detector locations.

\section{Tracer sampling}
\label{app:tracer_sampling}

\begin{figure}[tb!]
  \centering
  \includegraphics[width=0.45\columnwidth]{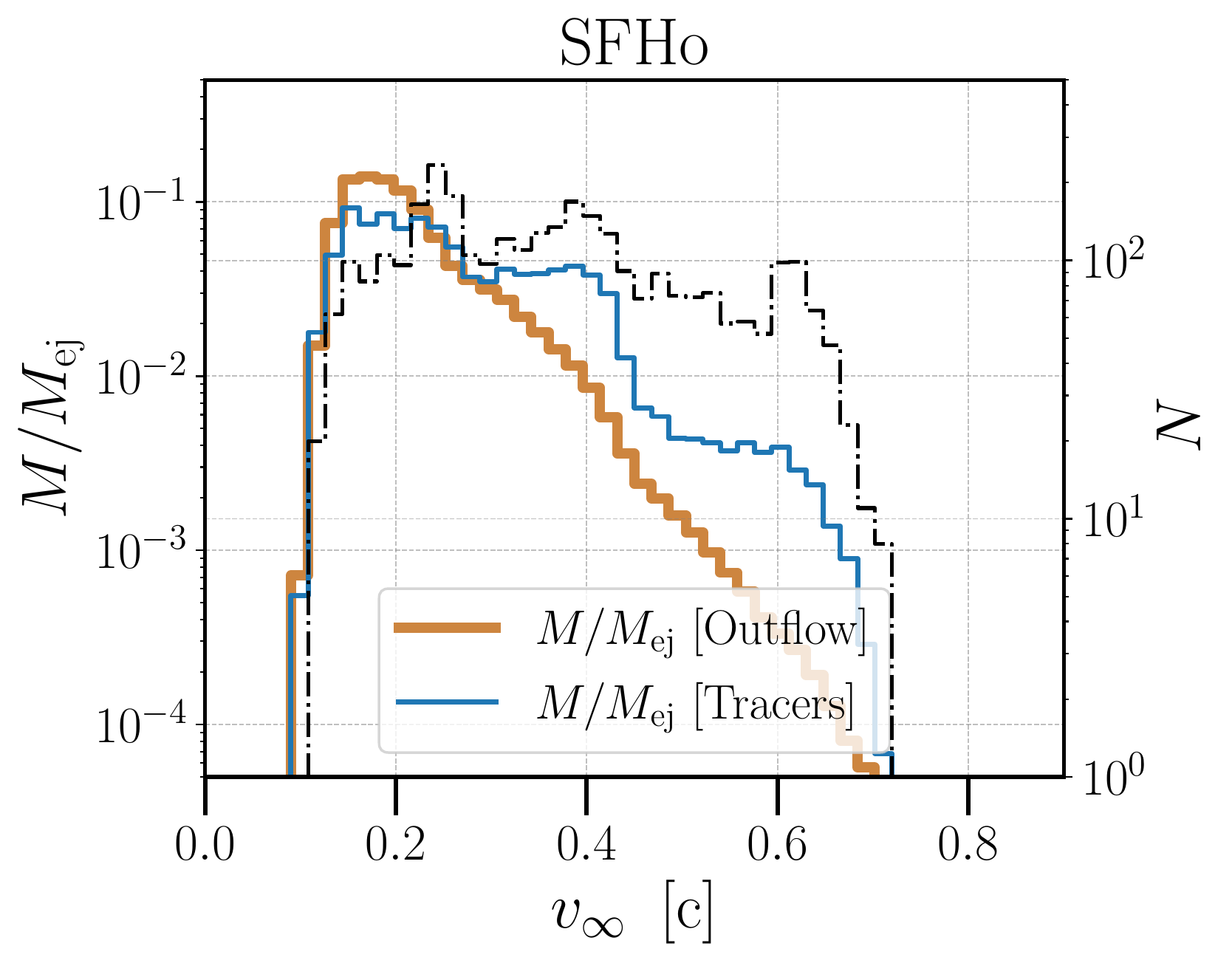}
  \includegraphics[width=0.45\columnwidth]{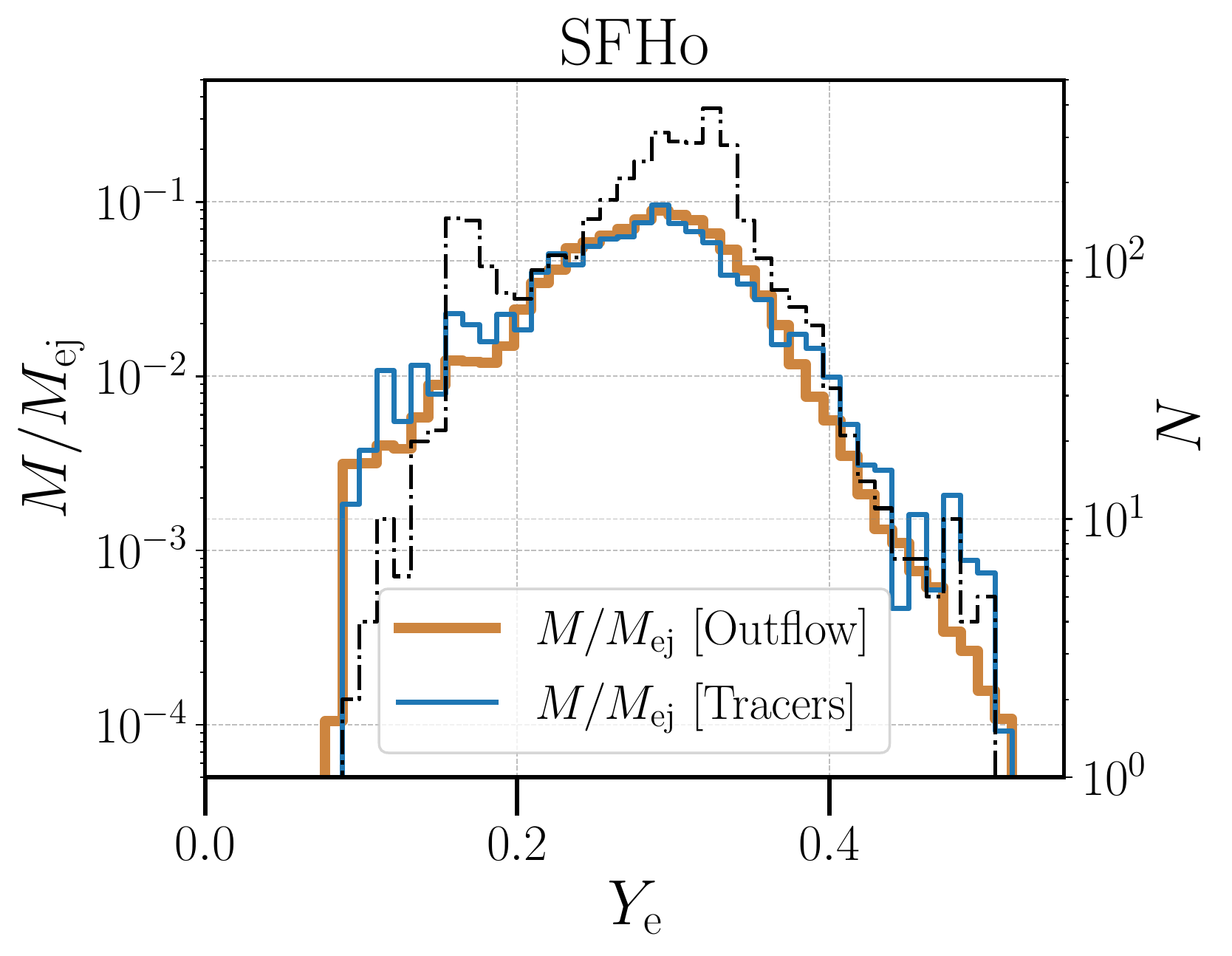}
    \includegraphics[width=0.45\columnwidth]{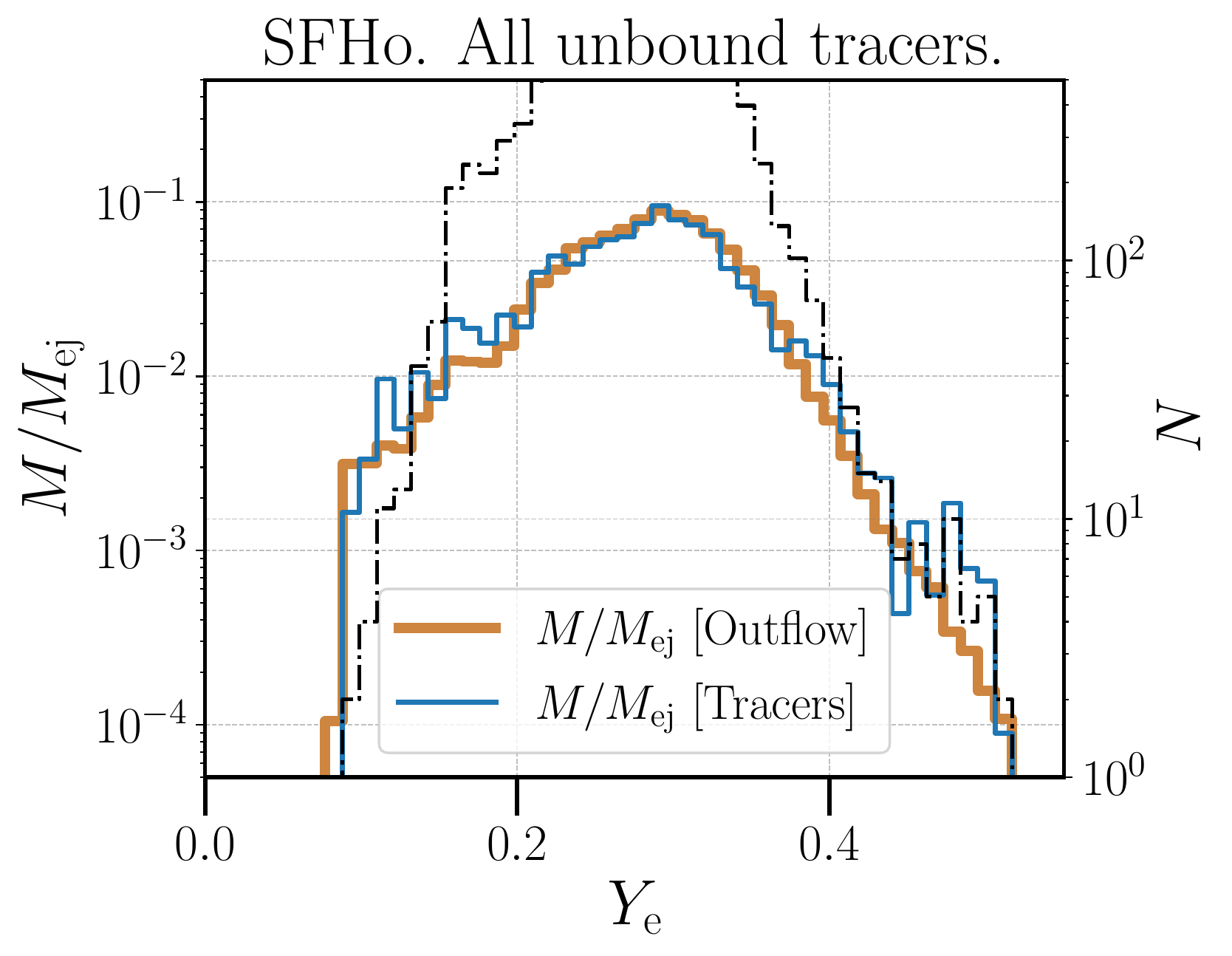}
  \caption{Mass distribution of electron fraction for dynamical ejecta material in our \sfho{} run, measured separately at $440$\,km by analyzing the mass flux on the Cartesian grid using a spherical detector surface (`Outflow'; Sec.~\ref{sec:ejecta_properties}) and by analyzing the properties recorded by passive tracers advected with the flow (`Tracers'; Sec.~\ref{sec:tracers}). The total number of tracer particles with scale on the right is shown as dot-dashed black lines. There is moderate oversampling of high-velocity ejecta (first panel), which we fix in subsequent analyses by excluding low-mass tracers that we consider sampling `noise'. Other distributions that are key to nucleosynthesis analyses, such as $Y_{\rm e}$ and specific entropy, show excellent agreement and are not affected by such oversampling, regardless of whether the full (third panel) or reduced (second panel) set of tracers is used. See the text for details.}
  \label{fig-tracers-ye}
\end{figure}

In Fig.~\ref{fig-tracers-ye}, we compare the distributions of dynamical ejecta mass extracted from our \sfho{} simulation using tracer particles (Sec.~\ref{sec:tracers}) and computing the unbound mass flux through a spherical detector surface (Sec.~\ref{sec:ejecta_properties}). Overall, we find excellent agreement, specifically in thermodynamic quantities such as electron fraction and specific entropy, which is key for accurate nucleosynthesis calculations. We find a moderate oversampling of high-velocity ejecta using the tracer approach, which is due to our method of placing a large number of tracers in the high-entropy, under-dense regions of shock heated ejecta during the merger (Sec.~\ref{sec:tracers}). We thus discard tracers below a certain mass-threshold 
as `noise' and exclude them from further analyses. In doing so, we improve the velocity distributions shown in Fig.~\ref{fig-tracers-ye} and obtain an overall good agreement and sampling of ejecta according to asymptotic velocity, with still more than 500 tracers resolving the high-velocity tail $v_\infty > 0.6c$. Other distributions, such as those in $Y_{\rm e}$ are insensitive to such oversampling (compare the second and third panel of Fig.~\ref{fig-tracers-ye}).

\bibliography{astrograv}
\bibliographystyle{aasjournal}



\end{document}